\documentclass[twocolumn,aps,prd,longbibliography]{revtex4-2}
\usepackage{dcolumn}
\usepackage{enumerate}
\usepackage{natbib}
\usepackage{amsmath}
\usepackage{amssymb}
\usepackage{graphicx}
\usepackage{graphics}
\usepackage{epstopdf}
\usepackage{lineno}

\begin{document}
\title{Nonlinear evolution of magnetorotational instability in a magnetized Taylor-Couette flow: Scaling properties and relation to upcoming DRESDYN-MRI experiment}

\author{Ashish Mishra}
\email{a.mishra@hzdr.de}
\affiliation{Helmholtz-Zentrum Dresden-Rossendorf, Bautzner Landstr.
400, D-01328 Dresden, Germany}
\affiliation{Center for Astronomy and Astrophysics, ER 3-2, TU Berlin, Hardenbergstr. 36, 10623 Berlin, Germany}
\author{George Mamatsashvili}
\affiliation{Helmholtz-Zentrum Dresden-Rossendorf, Bautzner Landstr.
400, D-01328 Dresden, Germany}
\affiliation{Abastumani Astrophysical Observatory, Abastumani 0301, Georgia}

\author{Frank Stefani}
\affiliation{Helmholtz-Zentrum~Dresden-Rossendorf, Bautzner Landstr.
400, D-01328 Dresden, Germany}


\begin{abstract}

Magnetorotational instability (MRI) is considered as the most likely mechanism driving angular momentum transport in astrophysical disks. However, despite many efforts, a direct and conclusive experimental evidence of MRI in laboratory is still missing. Recently, performing 1D linear analysis of the standard version of MRI (SMRI) in a Taylor-Couette (TC) flow between two rotating coaxial cylinders threaded by an axial magnetic field, we showed that SMRI can be detected in the upcoming DRESDYN-MRI experiment based on a cylindrical magnetized TC flow of liquid sodium. In this follow-up study, being also related to the DRESDYN-MRI experiments, we focus on the nonlinear evolution and saturation properties of SMRI and analyze its scaling behavior with respect to various parameters of the basic TC flow using a pseudo-spectral code. We conduct a detailed analysis over the extensive ranges of magnetic Reynolds number $Rm\in [8.5, 37.1]$, Lundquist number $Lu\in[1.5, 15.5]$ and Reynolds number,  $Re\in[10^3, 10^5]$. We focus on small magnetic Prandtl number, $Pm \ll 1$, regime down to $Pm\sim 10^{-4}$, aiming ultimately for those very small values typical of liquid sodium used in the experiments. In the saturated state, the magnetic energy of SMRI and associated normalized torque due to perturbations exerted on the cylinders, which characterizes angular momentum transport, both increase with $Rm$ for fixed $(Lu, Re)$, while for fixed $(Lu, Rm)$, the magnetic energy decreases and torque increases with increasing $Re$. We also study the scaling of the magnetic energy and torque in the saturated state as a function of $Re$ and find a power law dependence of the form $Re^{-0.6...-0.5}$ for the magnetic energy and $Re^{0.4...0.5}$ for the torque at all sets of $(Lu, Rm)$ and sufficiently high $Re\geq 4000$. We also explore the dependence on Lundquist number and angular velocity of the cylinders. The scaling laws derived here will be instrumental in the subsequent analysis of more realistic finite-length TC flows and comparison of numerical results with those obtained from the DRESDYN-MRI experiments in order to conclusively and unambiguously identify SMRI in laboratory. 
\end{abstract}
	
	\maketitle
	\section{Introduction}
	
The advent of magnetorotational instability (MRI) as the most probable mechanism to transport angular momentum allowing accretion of matter in astrophysical disks has received a great deal of attention in the scientific community. Although studied first in the context of cylindrical Taylor-Couette (TC) flow by Velikhov \cite{Velikhov_1959} and later more extensively by Chandrashekhar \cite{Chandrasekhar1961}, the importance of MRI to astrophysics remained hidden for almost three decades. It was in their seminal paper where Balbus and Hawley \cite{Balbus_Hawley_1991} showed that even a minimal seed magnetic field can lead to an effective transport of angular momentum in otherwise hydrodynamically stable accretion disks. Later Goodman \& Ji \cite{goodman_ji_2002} and R\"udiger \& Shalybkov \cite{Ruediger_Shalybkov2002} investigated the linear evolution of axisymmetric MRI again in TC flow with an imposed constant axial magnetic field (mimicking a disk flow) in the viscous and resistive regime corresponding to liquid metals that are normally used in experiments. These pioneering papers, first demonstrating theoretically the plausibility of MRI under experimental conditions, had spawned multiple commendable efforts to realize this instability in the laboratory \cite{Sisan_etal2004,Nornberg_Ji_etal_2010_PhysRevLett, Roach_etal2012PhRvL, Hung_etal2019CmPhy}. However, despite these efforts, the direct and conclusive evidence of MRI in the lab is presently still elusive. The main challenge is that due to extremely small magnetic Prandtl numbers, $Pm=\nu/\eta\sim 10^{-6}-10^{-5}$, of liquid metals in those experiments ($\nu$ is viscosity and $\eta$ resistivity), rather high Reynolds numbers $Re \gtrsim 10^6$ of cylinders' rotation are required for the onset of MRI. In this regard, recent claims by the Princeton group having detected MRI \cite{Wang_etal2022a,Wang_etal2022b} with an imposed purely axial magnetic field are still under scrutiny. In their experiments, both axisymmetric \cite{Wang_etal2022a} and non-axisymmetric \cite{Wang_etal2022b} modes were identified surprisingly close to each other in a similar range of rotation speed and magnetic field strength. However, a key problem is that both the rotation speed and field strength from this range are still several times smaller than their critical values for the onset of MRI as dictated by global linear analysis (\cite{Ruediger_Schultz2023} and Fig. 8 in \cite{Wang_etal2022b}).

The discovery of other versions of MRI such as helical MRI (HMRI, \cite{Hollerbach_Rudiger_2005}) and azimuthal MRI (AMRI, \cite{Hollerbach_Rudiger_2010}) occurring already at much smaller $Re \sim 10^3$ and their successful detection in the PROMISE experiment \cite{Stefani_Gundrum_Gerbeth_etal_2006PhRvL, Seilmayer_etal2014} has paved the way for more sophisticated large-scale experimental endeavors such as the DRESDYN-MRI experiment \cite{Stefani_etal2019}. Recently, we (Mishra et al.  \cite{Mishra_mamatsashvili_stefani_2022PRF}, hereafter Paper I) conducted a one-dimensional (1D) linear stability analysis of the Standard version of MRI (SMRI) in the setting of the upcoming DRESDYN-MRI experiment consisting of a big cylindrical TC machine filled with liquid sodium and showed that SMRI can be detected within the parameter regimes achievable in this experiment. While linear stability analysis has been successful in determining the onset regimes and growth rates of the instability in previous experiments, particularly in the case of HMRI \cite{Hollerbach_Rudiger_2005,Stefani_Gundrum_Gerbeth_etal_2006PhRvL} and AMRI \cite{Hollerbach_Rudiger_2010, Mishra_etal2021_JFM, Seilmayer_etal2014}, the study of nonlinear evolution of these instabilities revealed their characteristic behaviour in the saturation state and transition to turbulent regime \cite{Guseva_etal_2015NJPh, Mamatsashvili_etal2018}.
	
Since the rediscovery of SMRI, its nonlinear evolution and saturation properties have been the subject of great interest. It has been extensively investigated theoretically both in astrophysical and in laboratory settings. Hawley \& Balbus \cite{Hawley_Balbus_1991ApJ_nonlinear} conducted the first nonlinear simulation of axisymmetric SMRI in Keplerian accretion disks suggesting that MRI may approach saturated state by inducing radial motions which provide a favourable condition for magnetic reconnection. Knobloch \& Julien \cite{Knobloch_Julien_2005PhFl} studied the nonlinear evolution of SMRI mode analytically and suggested that MRI, at first, evolves on the rotational time scale of the TC flow, while the asymptomatic approach to saturated state occurs much slower, on the resistive time scale. Although they associated the different time evolution scales to a reconnection process, they did not take into account the effect of radial boundaries which are  important in experimental TC devices. 

Subsequent studies of the nonlinear evolution of MRI in magnetized, viscous and resistive TC flows allowed to better understand its saturation, dynamics and transport properties. Early works by Liu et al. \cite{Liu_Goodman_Ji_NonlinearMRI_2006ApJ} and later by Gellert et al. \cite{Gellert_etal2012} carried out numerical simulations of MRI in infinitely long TC geometry for a limited set of parameters far from experimental values (with the exception of few low resolution cases at $Rm\sim 20$ and $Re\sim 10^4$ in Liu et al.). Interesting results from this study include the observation of a jet-like outflow at the mid-height of the cylinder, the reconnection layer and scaling relations of the torque (angular momentum transport parameter) with respect to moderate and large magnetic Reynolds and Lundquist numbers. 

Umurhan et al. \cite{Umurhan_Menou_Regev_2007PhRvL} conducted a weakly nonlinear analysis of axisymmetric SMRI in a thin channel geometry at $Pm\ll 1$ closer to  experimental conditions and found that the saturation amplitude scales with $Pm^{1/2}$ while the angular momentum transport scales with $Pm$. By contrast, in the thin gap TC setup the saturation amplitude and angular momentum transport was found to scale with $Pm^{2/3}$ and $Pm^{4/3}$, respectively, differing from the previous results due to the development of a radial boundary layer \cite{Umurhan_Regev_Menou_2007PhRvE}. 
Clark and Oishi \cite{Clark_Oishi_2017ApJ_local_geometry} conducted also a weakly nonlinear analysis of SMRI in the local thin gap TC geometry at small $Pm$ and found corroborative scaling properties with those of thin gap TC setup by Umurhan et al. \cite{Umurhan_Regev_Menou_2007PhRvE}. However, their global analysis of a wide gap cylindrical TC flow \cite{Clark_Oishi_2017ApJ_TC_geometry} obtained steeper scalings of saturation amplitude and transport with $Pm$, different from that in the thin gap TC setup analysis due to the formation of an asymmetric radial boundary layer in the wide gap TC setup. In all these cases, the saturation of MRI occurs through the reduction of the background shear and rearrangement and strengthening of the background axial magnetic field. 

Recent papers \cite{Gissinger_Goodman_Ji_2012PhFl, Wei_etal2016, Choi_etal2019, Winarto_etal2020} studied the saturation properties and nonlinear evolution of MRI in a finite-height TC flow in the context of the Princeton MRI experiment, where the situation is complicated by Ekman circulations due to endcaps (see also a recent review \cite{Ji_Goodman2023}).  It is well-known from linear theory that at high enough Reynolds numbers $Re$ (i.e., at $Pm \ll 1$), the central parameters governing SMRI are Lundquist and magnetic Reynolds numbers such that the instability is almost insensitive to $Re$ \cite{Kirillov_Stefani_2010ApJ, Ruediger_etal_2018_PhysRepo}, while dependence on $Re$ is important in the nonlinear regime. Those studies analyzed the dependence of the saturated amplitude on these numbers and especially on $Re$ under insulating or conducing endaps, showing that the scaling with $Re$ (or $Pm$) is different for these two endcap types.  
	
The above-discussed studies indicate that the conclusive parametric dependence of nonlinear SMRI in a magnetized TC flow appears to be system-dependent (geometry, material of cylinders and endcaps, etc.) and is still not well understood. In this paper, we analyze the nonlinear saturation and dynamics of SMRI and seek to generalize its behavior with the flow parameters (Reynolds, magnetic Reynolds and Lundquist numbers), specifically in the context of the upcoming DRESDYN-MRI experiment. As distinct from the above studies, we do not limit ourselves to the weakly nonlinear regime of SMRI, which is valid only near the marginal stability of the flow, but consider the fully nonlinear problem. This study further extends our previous linear study of SMRI in DRESDYN-TC device (Paper I) where we showed that MRI should be detectable for the parameter regimes achievable in the DRESDYN experiment for all relevant rotation profiles in the Rayleigh-stable regime including the astrophysically important Keplerian profile. In the present study, adopting the same base TC flow setup with the parameter regimes characteristic to DRESDYN-MRI device, though without endcaps, we aim to infer various aspects of nonlinear development of MRI, including time evolution of magnetic energy and torque, saturation mechanism and scaling laws as a function of different system parameters. We consider again the small $Pm\ll 1$ regime and pay particular attention to the dependence of the saturated SMRI state on $Pm$ (i.e., on $Re$ at a given $Rm$), aiming to extrapolate it down to very small values $Pm\sim 10^{-6}-10^{-5}$ of liquid metals. 

The paper is organized as follows. The basic equations, setup and numerical scheme are discussed in section \ref{sec_2_math_setting}. Direct numerical simulations of the nonlinear evolution of SMRI, saturation mechanisms, dependence on the system parameters and relevance to the upcoming DRESDYN-MRI experiment are presented in section \ref{sec_3_results}. Summary and conclusion are given in section \ref{sec_4_conclusion}.

\section{Mathematical Setting} \label{sec_2_math_setting}
	
\begin{figure*}
\centering
\includegraphics[width=0.35\textwidth]{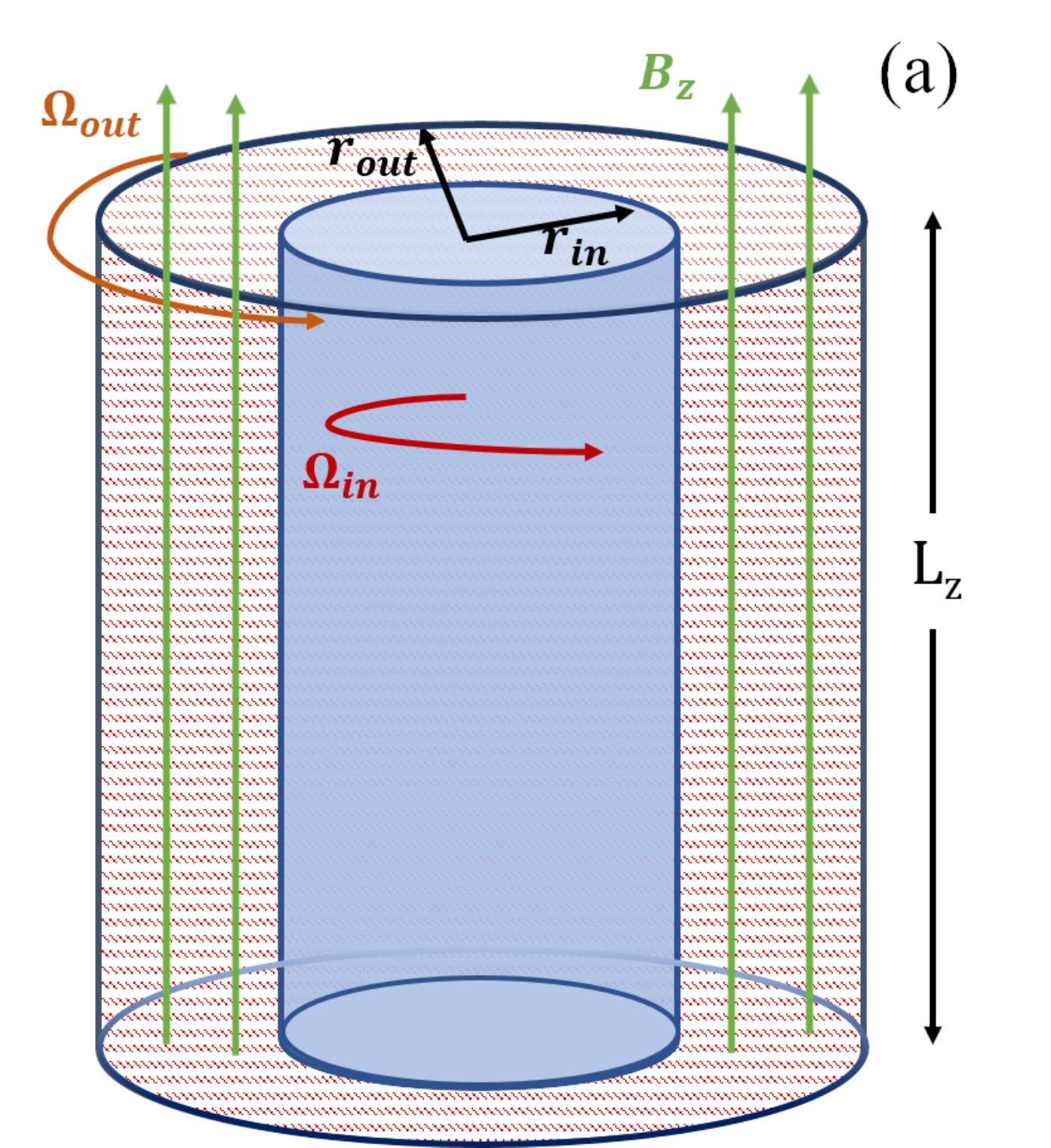}
\includegraphics[width=0.5\textwidth]{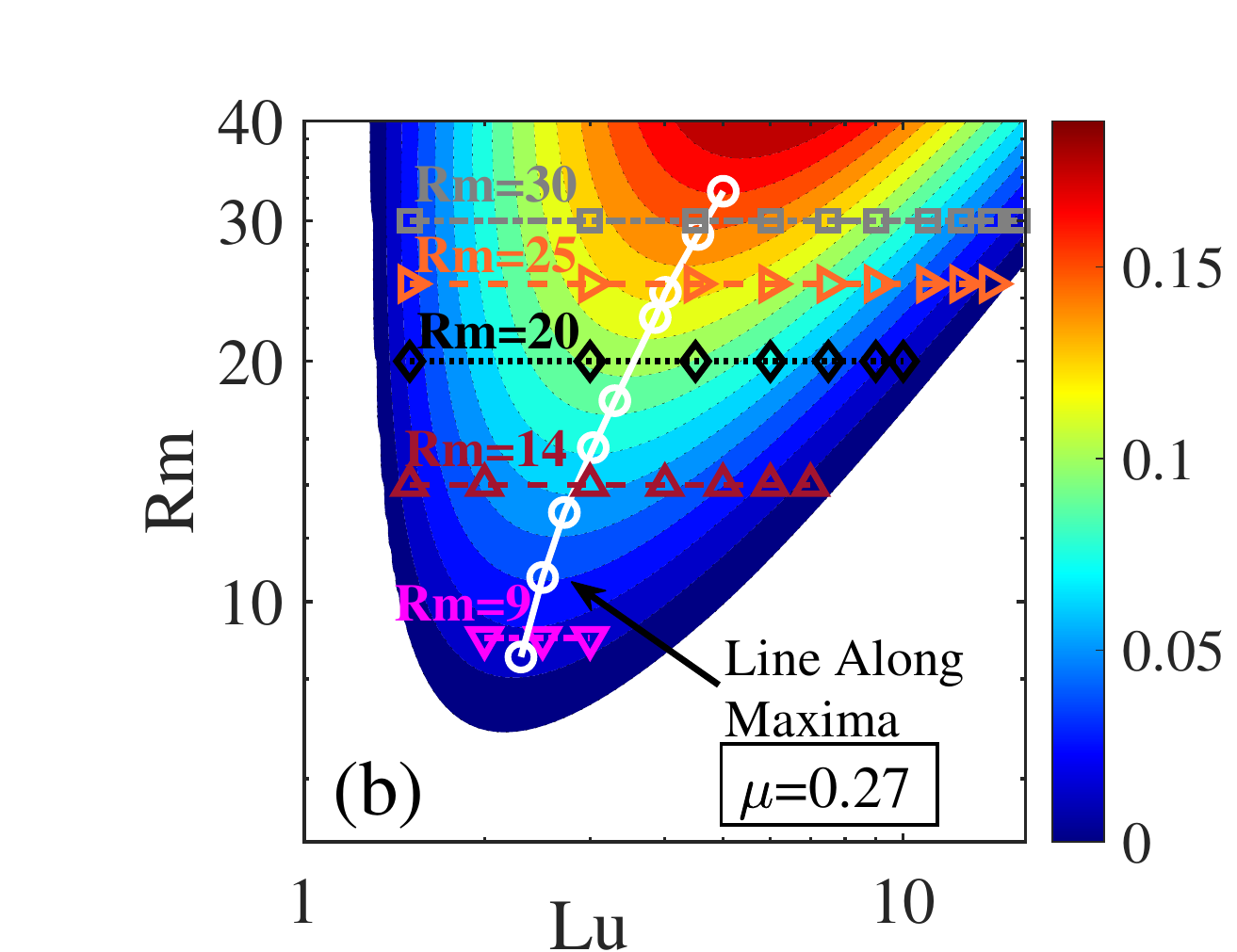}
\caption{(a) Taylor-Couette setup with an imposed axial magnetic field $B_{0z}$. (b) The growth rate of SMRI in the $(Lu,Rm)$-plane obtained from 1D linear stability analysis for $\mu=0.27$, $k_{z}\geq k_{z,min}=2\pi/L_z$ and $Pm=7.77\times 10^{-6}$ typical of liquid sodium \cite{Mishra_mamatsashvili_stefani_2022PRF}. In this map, $(Lu, Rm)$ pairs for which nonlinear simulations are performed are denoted with different symbols and lie on the horizontal lines with fixed $Rm \in \{9, 14, 20, 25, 30\}$ and varying $Lu \in [1.5, 15.5]$. White line traces the maxima of the growth rate in the $(Lu,Rm)$-plane obtained from 1D linear stability analysis and circles mark the points of nonlinear simulations.}
\label{fig:Analysis_points}
\end{figure*}

We work in a cylindrical TC flow setup containing viscous and resistive liquid metal threaded by an imposed constant axial magnetic field [Fig. \ref{fig:Analysis_points}(a)]. In the DRESDYN-MRI experiment, a current-carrying solenoid surrounds the outer cylinder which imposes a constant axial magnetic field $\boldsymbol{B}_0=B_{0z}\boldsymbol{\hat{z}}$ inside it. In the classical TC flow, the inner and outer cylinders with radii $r_{in}$ and $r_{out}$ rotate at angular velocities $\Omega_{in}$ and $\Omega_{out}$, respectively, and give rise to an equilibrium radial profile of the angular velocity $\Omega(r)$ of the fluid between the cylinders,
\begin{equation}\label{TC_profile}
\Omega(r) = \frac{\Omega_{out}r_{out}^2 -\Omega_{in}r_{in}^2}{r_{out}^2-r_{in}^2} +\frac{(\Omega_{in}-\Omega_{out})}{r_{out}^2-r_{in}^2}\frac{r_{in}^2r_{out}^2}{r^2}.
\end{equation}

The basic set of non-ideal MHD equations governing the motion of an incompressible conducting fluid is
\begin{equation} \label{moment}
\frac{\partial \boldsymbol{u}}{\partial t} + (\boldsymbol{u\cdot\nabla})\boldsymbol{u}=-\frac{1}{\rho}\nabla P+\frac{\boldsymbol{J \times B}}{\rho}+\nu \nabla^2\boldsymbol{u},
\end{equation}
\begin{equation}\label{induc}
\frac{\partial\boldsymbol{B}}{\partial t} =\nabla\times(\boldsymbol{u}\times\boldsymbol{B})+\eta\nabla^2\boldsymbol{B},
\end{equation}
\begin{equation}\label{non_compres}
\nabla\cdot {\boldsymbol{u}}=0,~~~\nabla\cdot {\boldsymbol{B}}=0,
\end{equation}
where  $\boldsymbol{u}$ is the velocity, $\textit{P}$ is the thermal pressure, $\boldsymbol{B}$ is the magnetic field and $\boldsymbol{J}=\mu_0^{-1} \nabla \times \boldsymbol{B}$ is the current density with $\mu_0$ being the magnetic permeability of vacuum. Density $\rho$, kinematic viscosity $\nu$ and magnetic diffusivity $\eta$ of the conducting fluid between the cylinders are all spatially constant. 

It is well known that in experiments with finite-length cylinders the theoretical rotation profile (\ref{TC_profile}) of TC flow does not hold near the top and bottom boundaries of the cylinders due to the presence of endcaps \cite{Kageyama_etal_2004JPSJ, Szklarski_Rudiger_2007PRE}. Nevertheless, for the sake of understanding the basic nonlinear evolution and saturation properties of SMRI, in the first approximation we ignore those effects due to endcaps and impose periodic boundary conditions along the cylinders, since the DRESDYN-MRI experiment has a large vertical aspect ratio $L_z/r_{in}=10$, where $L_z$ is the cylinder length (height), and also it is designed such that to minimize the effects of endcaps by introducing split rings at the top and bottom boundaries at an optimal radius $r_0=1.4r_{in}$ (see e.g., \cite{Szklarski_Rudiger_2007PRE,Stefani_Gerbeth_Gundrum_Etal_2009PhysRevE}). 

This approximation of infinite cylinders should be, however, viewed with caution, because the endcaps, breaking the translational symmetry of infinite cylinders, can affect the bifurcation properties, nonlinear saturation and dynamics of SMRI \cite{Gissinger_Goodman_Ji_2012PhFl}. Understanding the influence of endcaps on the basic TC flow profile and overall stability is crucial for an unambiguous identification of MRI in experiments, but at the same time, can quite complicate numerical treatment in the high $Re\sim 10^6$ regime of SMRI in liquid metal TC flows. The endcaps induce poloidal Ekman circulations on top of background TC flow altering the background  velocity structure upon which MRI feeds. In addition, this circulation as well as thin, high shear Stewartson layers formed near the split rings can themselves be subject to non-axisymmetric hydrodynamic instabilities at high enough $Re$, interfering with MRI and complicating its identification in the experiments \cite{Gissinger_Goodman_Ji_2012PhFl, Choi_etal2019}. In this study instead, we neglect these effects by considering infinite cylinders as in Paper I and focus on the basic nonlinear dynamics of SMRI. For this reason, although some of the results obtained here may not directly carry over to the experimental TC flow, they will serve as an essential guide into more complex dynamics of the instability and its spatial structure in the finite-length DRESDYN-MRI machine with endcaps.  
	
We non-dimensionalize time by $\Omega_{in}^{-1}$, angular velocities by $\Omega_{in}$, length by the gap width between the cylinders $d=r_{out}-r_{in}$, velocity by $\Omega_{in}d$, kinetic energy density by $\rho_0\Omega_{in}^2r_{in}^2$, magnetic field by the background field $B_{0z}$ and current density by $B_{0z}/(\mu_0 d)$. The main parameters are Reynolds number $Re=\Omega_{in}d^2/\nu$, magnetic Reynolds number $Rm=\Omega_{in}d^2/\eta$, magnetic Prandtl number $Pm=\nu/\eta=Rm/Re$ and Lundquist number $Lu=V_Ad/\eta$, where $V_A=B_{0z}/\sqrt{\rho \mu_0}$ is the Alfv\'en speed. 
	
In the DRESDYN-MRI machine, the ratio $r_{in}/r_{out}$ is fixed to $0.5$ and the vertical aspect ratio $L_z/r_{in}$ to 10, while the ratio of angular velocities of the cylinders $\mu=\Omega_{out}/\Omega_{in}$ can be varied (see Table II of Paper I). In this study, we focus on the Rayleigh-stable regime with $\mu > (r_{in}/r_{out})^2=0.25$. This condition prevents hydrodynamic instabilities and ensures that only magnetic instabilities can develop. Here, we limit ourselves to axisymmetric perturbations with zero azimuthal wavenumber, $m=0$, which are the dominant ones for SMRI at large $Lu$ and $Rm$ discussed in Paper I. To find the optimal scaling relations for the detection of SMRI in terms of $\mu$, $Re$, $Rm$ and $Lu$ in the DRESDYN-MRI experiments, we consider the ranges of these parameters accessible in the experiments (Table II of Paper I). 

\subsection{Numerical Method}
	
We use the pseudo-spectral code from \cite{Guseva_etal_2015NJPh} to solve the basic nonlinear Eqs. (\ref{moment})-(\ref{non_compres}). In this code, the radial expansion is based on a high order finite-difference method while Fourier expansion is done in the  axial and azimuthal directions. Time discretization of second order is done using the implicit Crank–Nicolson method. Nonlinear terms are calculated using the pseudo-spectral method and are dealiased using the 2/3-rule. The boundary conditions are no-slip for velocity and insulating for magnetic field, as it is in the DRESDYN-MRI machine, with periodic boundary conditions for both these quantities in the axial direction. More details about this code and its validation tests can be found in \cite{Guseva_etal_2015NJPh}. 
The cylindrical flow domain is $(r, \phi, z) \in [r_{in}, r_{out}] \times [0, 2\pi] \times [0, L_z]$. For our setup, in the non-dimensional units, $r_{in}=1, r_{out}=2$ and $L_z=10$. Chebyshev collocation method is used to distribute the points along radius in order to achieve higher resolution near the boundaries. We conduct a high resolution study with $N_r=500$ finite difference points in the radial direction and $N_z=800$ Fourier modes in the axial direction, which, as we checked, appear to be sufficient to resolve the main small-scale features -- viscous boundary layers near the cylinder walls and magnetic reconnection region -- of the dominant large-scale nonlinear SMRI state at high $Re$ up to the maximum $Re=10^5$ (see section III). Since SMRI is dominated by $m=0$ modes, our simulations are axisymmetric, not including non-axisymmetric $|m| \geq 1$ modes, with $N_{\phi}$ set to 1. We have also done several test runs including non-axisymmetric modes up to $N_{\phi}=20$ and found that the energy of non-axisymmetric modes always remains orders of magnitude smaller than that of the axisymmetric SMRI mode for the considered ranges of $Lu$, $Rm$, $Re$ and $\mu$, which validates our axisymmetric approach. The minimum axial wavelength is set to $k_{z,min}=2\pi/L_z$, so that at least one full axial wavelength fits in the domain.

\begin{figure*}
\centering
\includegraphics[width=0.31\textwidth]{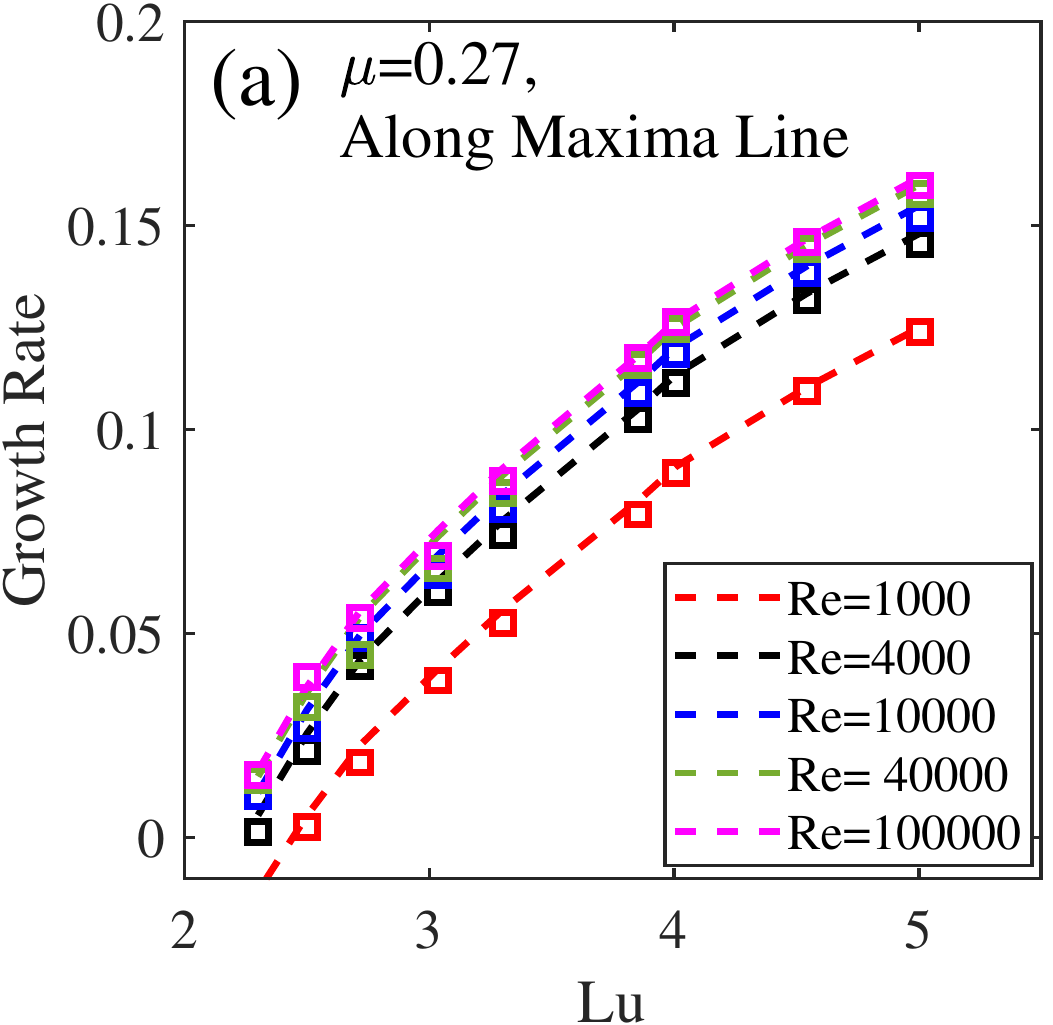}
\hspace{1em}
\vspace{1em}
\includegraphics[width=0.31\textwidth]{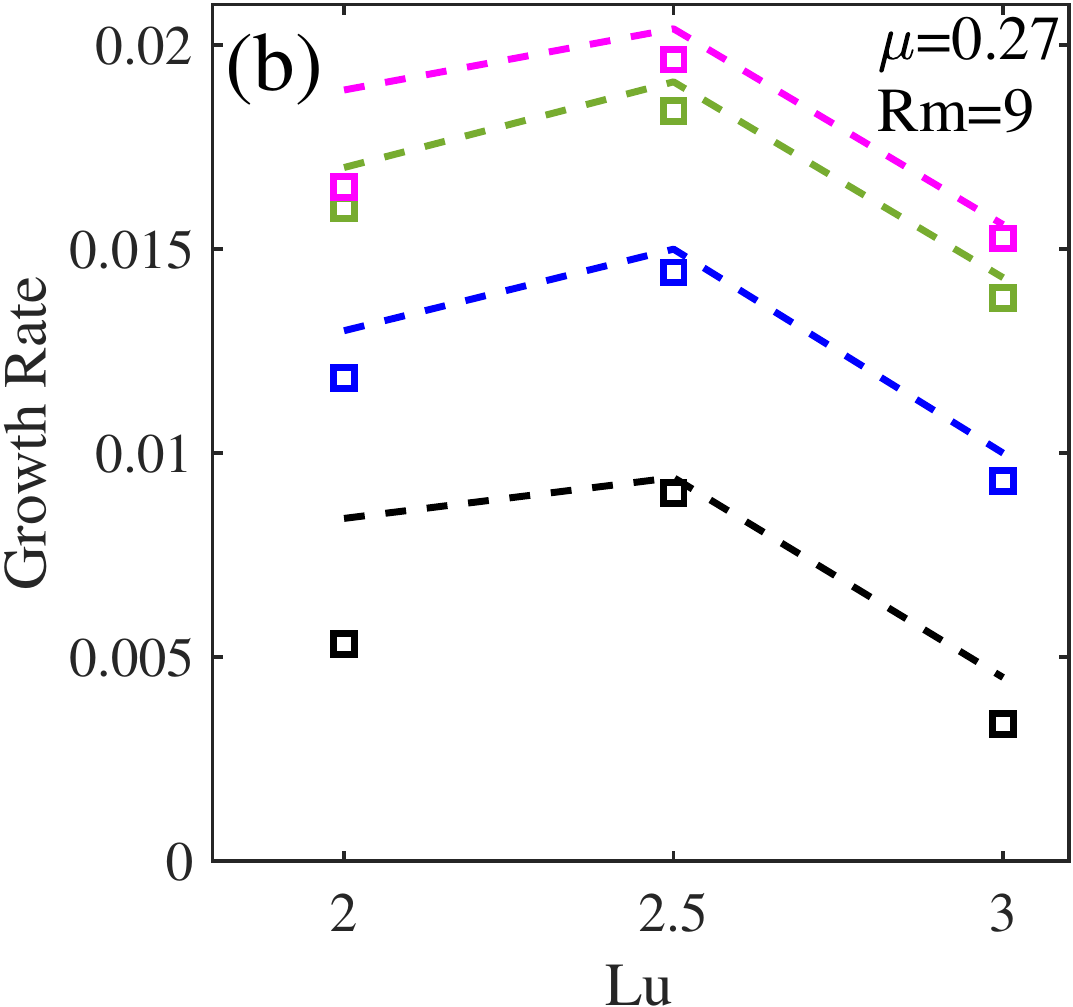}
\hspace{1em}
\includegraphics[width=0.31\textwidth]{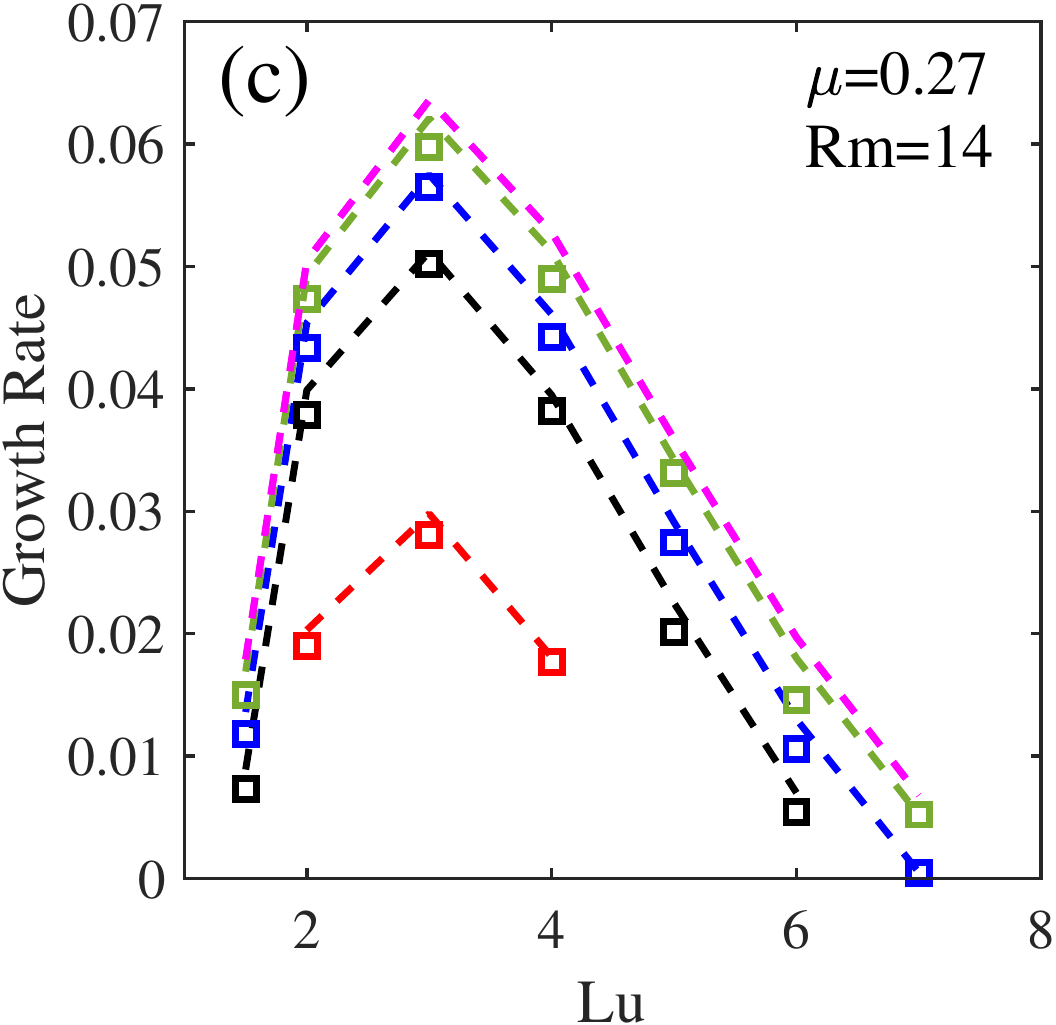}
\hspace{1em}
\includegraphics[width=0.31\textwidth]{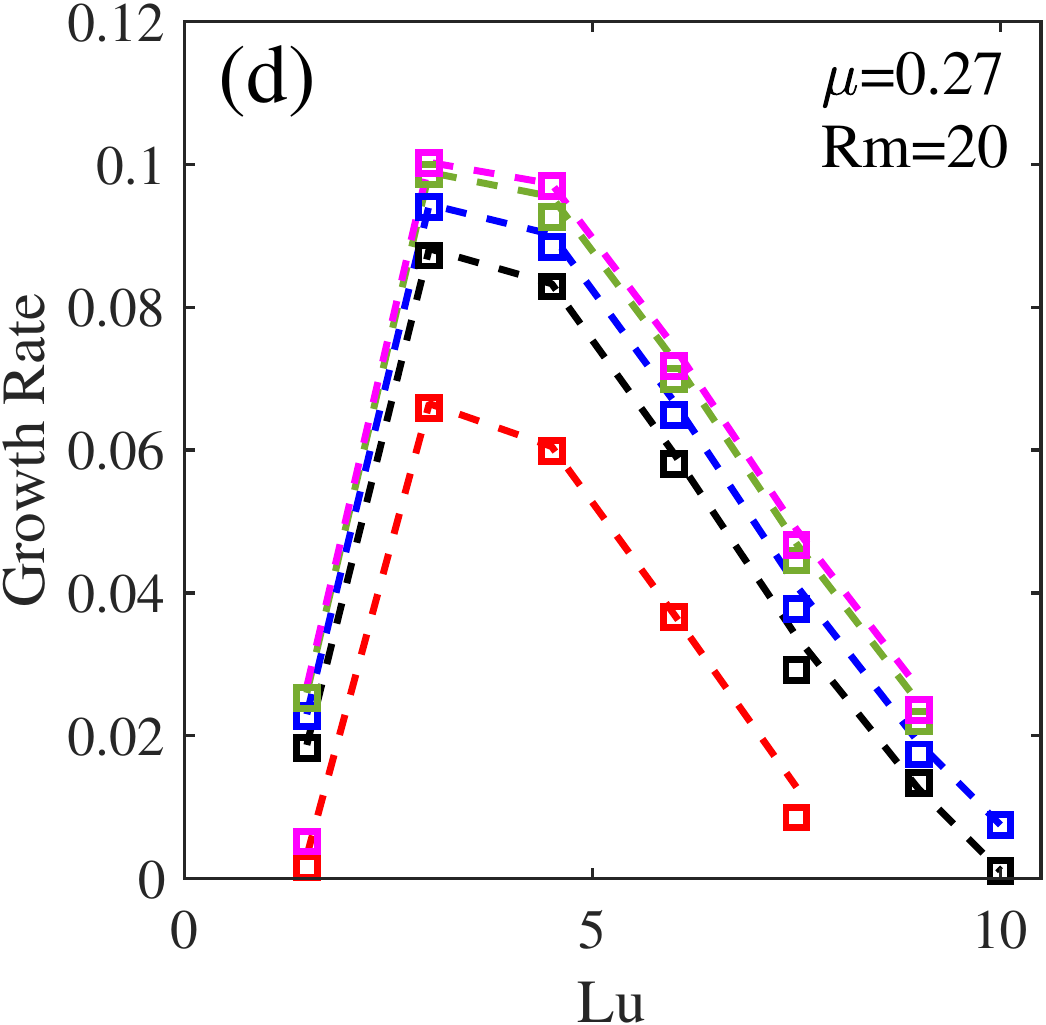}
\hspace{1em}
\includegraphics[width=0.31\textwidth]{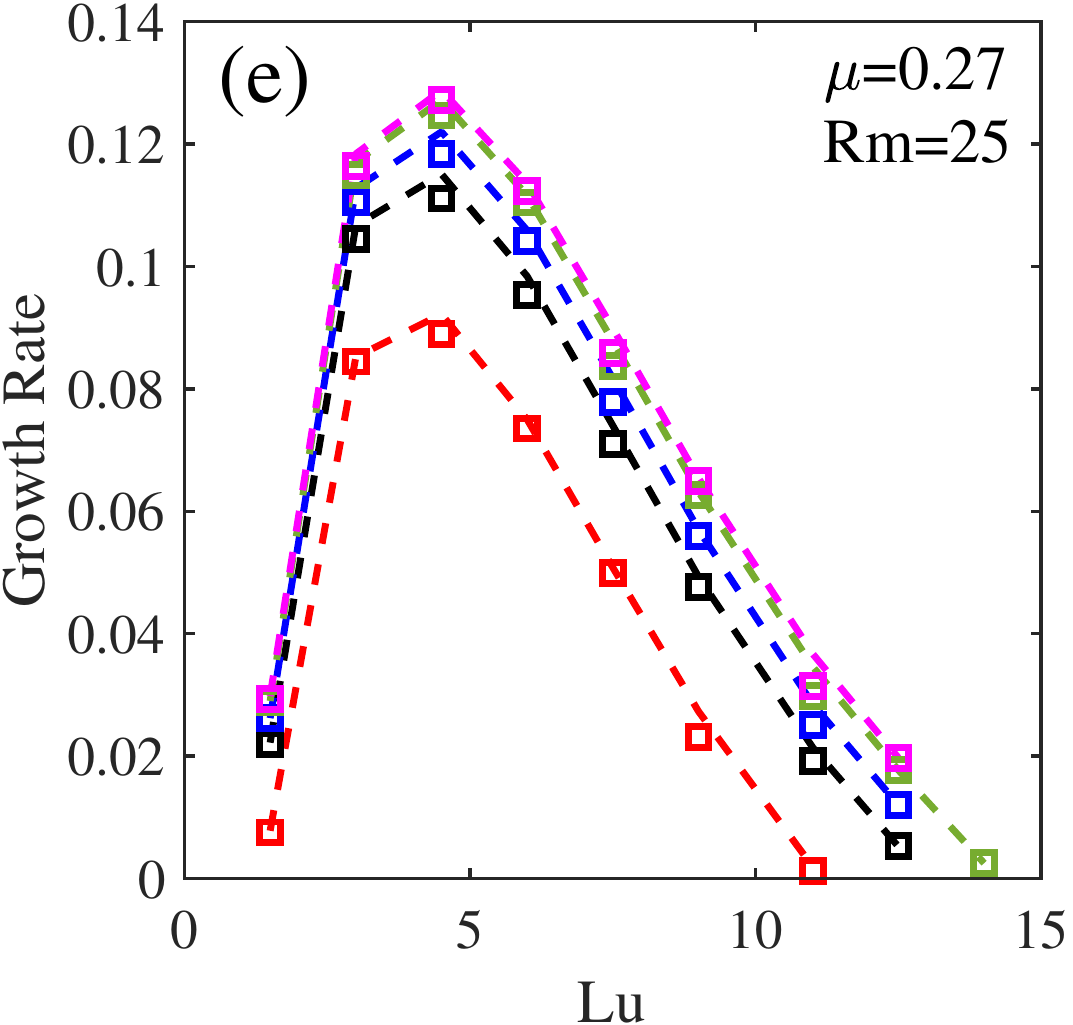}
\hspace{1em}
\includegraphics[width=0.31\textwidth]{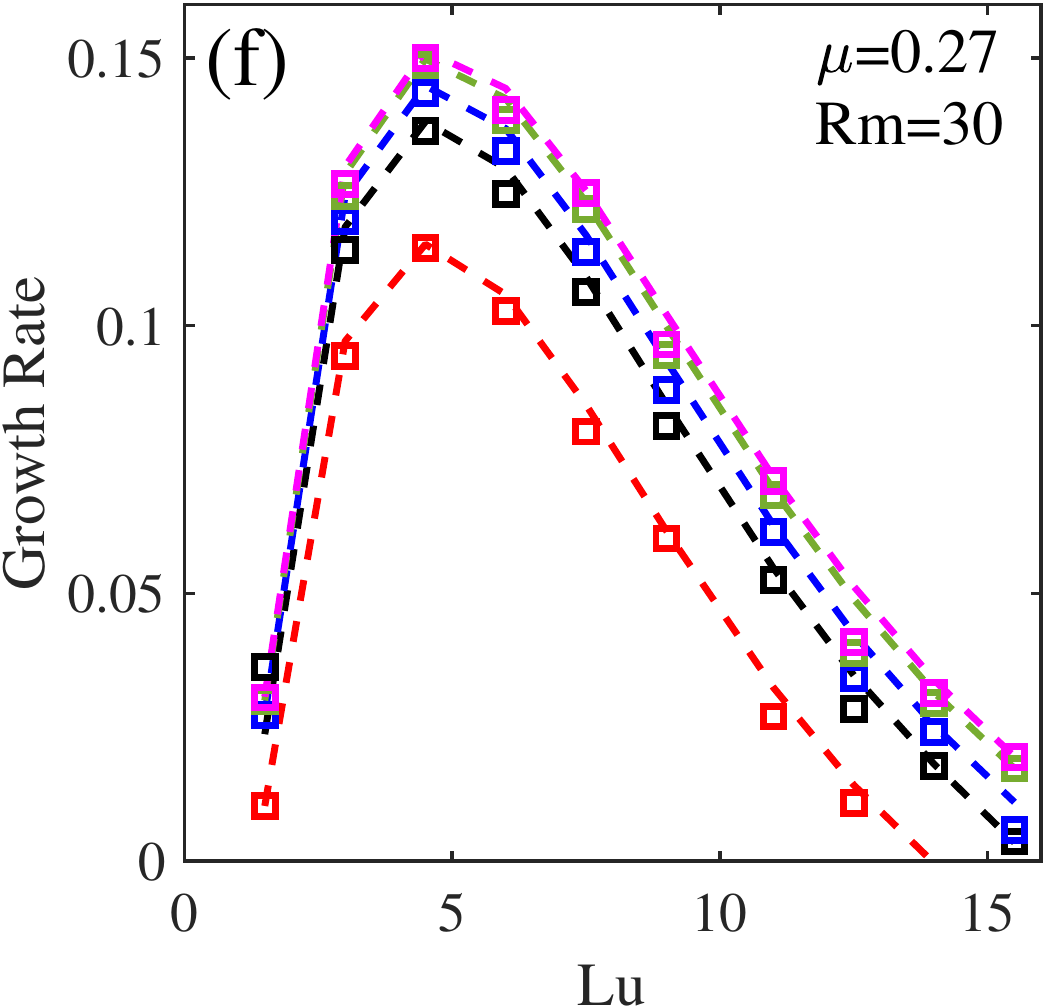}
\caption{Growth rate comparison at different $Re$ and the simulation $(Lu, Rm)$ pairs from Fig. \ref{fig:Analysis_points}. Dashed lines show the growth rates obtained from the 1D linear stability analysis, while  square markers show the growth rates from the nonlinear simulations for $Re=10^3$ (red), $Re=4\cdot 10^3$ (black), $Re=10^4$ (blue), $Re=4\cdot 10^4$ (green), $Re=10^5$ (magenta). The growth rate (a) along the white line of the maximum growth in Fig. \ref{fig:Analysis_points}(b) and (b)-(f) for $Rm\in\{9, 14, 20, 25, 30\}$.}\label{fig:growth_rate}
\end{figure*}

\section{Results}\label{sec_3_results}

In this section, we first compare the SMRI growth rates obtained from our nonlinear analysis in the initial, exponential growth stage of evolution with that of 1D linear analysis. This comparative approach ensures that our nonlinear simulations well reproduce the linear analysis results. Then we move to the main subject of this study -- nonlinear saturation, dynamics and transport of SMRI and its dependence on the parameters of the TC flow.
 
Figure \ref{fig:Analysis_points}(b) shows the positive growth rate of SMRI obtained previously from 1D linear analysis for $Pm=7.77\times 10^{-6}$ typical of liquid sodium \cite{Mishra_mamatsashvili_stefani_2022PRF}. We start with the $\mu=0.27$ case, because it is already Rayleigh-stable and also have relatively low critical $Lu$ and $Rm$ for the onset of SMRI, while larger $\mu$ will be considered in the subsections III.C and III.D. We conduct the nonlinear analysis along lines with fixed $Rm \in \{9, 14, 20, 25, 30\}$ and varying $Lu \in [1.5, 15.5]$ in the $(Lu,Rm)$-plane, as shown in Fig. \ref{fig:Analysis_points}(b). To diversify and generalize the analysis, we also conduct nonlinear simulations along the line of the largest growth rate obtained from the 1D linear analysis results [white curve in Fig. \ref{fig:Analysis_points}(b)]. Along this maxima line $Rm$ increases approximately linearly with $Lu$ according to $Rm\approx A\cdot Lu+B$, where $A=8.86$ and $B=-11.43$ with the minimum and maximum pairs of $(Lu, Rm)$ being (2.3, 8.5) and (5.0, 32.7), respectively.
Further, to obtain the scaling laws for various parameters in the saturated state, we analyse each pair $(Lu,Rm)$ for different $Re \in \{1, 4, 7, 10, 20, 30, 40, 100\} \times 10^3$. Since the minimum and maximum $Rm$ chosen here are, respectively, 8.5 and 32.7 (the endpoints of the line of the maximum growth), the range of $Pm\in [8.5\times10^{-5}, 0.0327]$. This, in turn, enables us to perform a detailed study of the nonlinear dynamics of SMRI and its saturation properties at very small $Pm$ approaching experimental ones. 
	
\begin{figure*}
\centering
\includegraphics[width=0.47\textwidth]{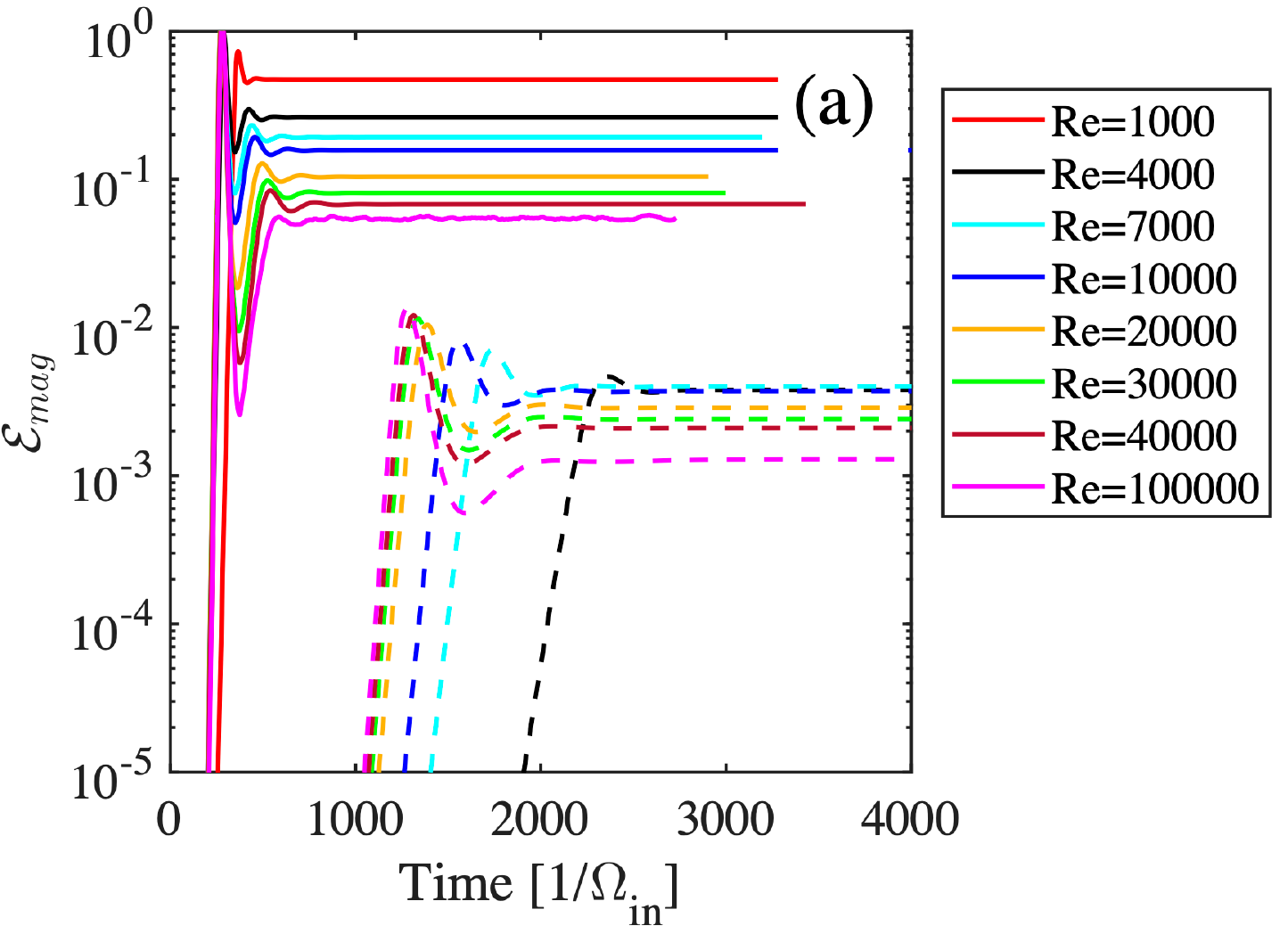}
\includegraphics[width=0.5\textwidth]{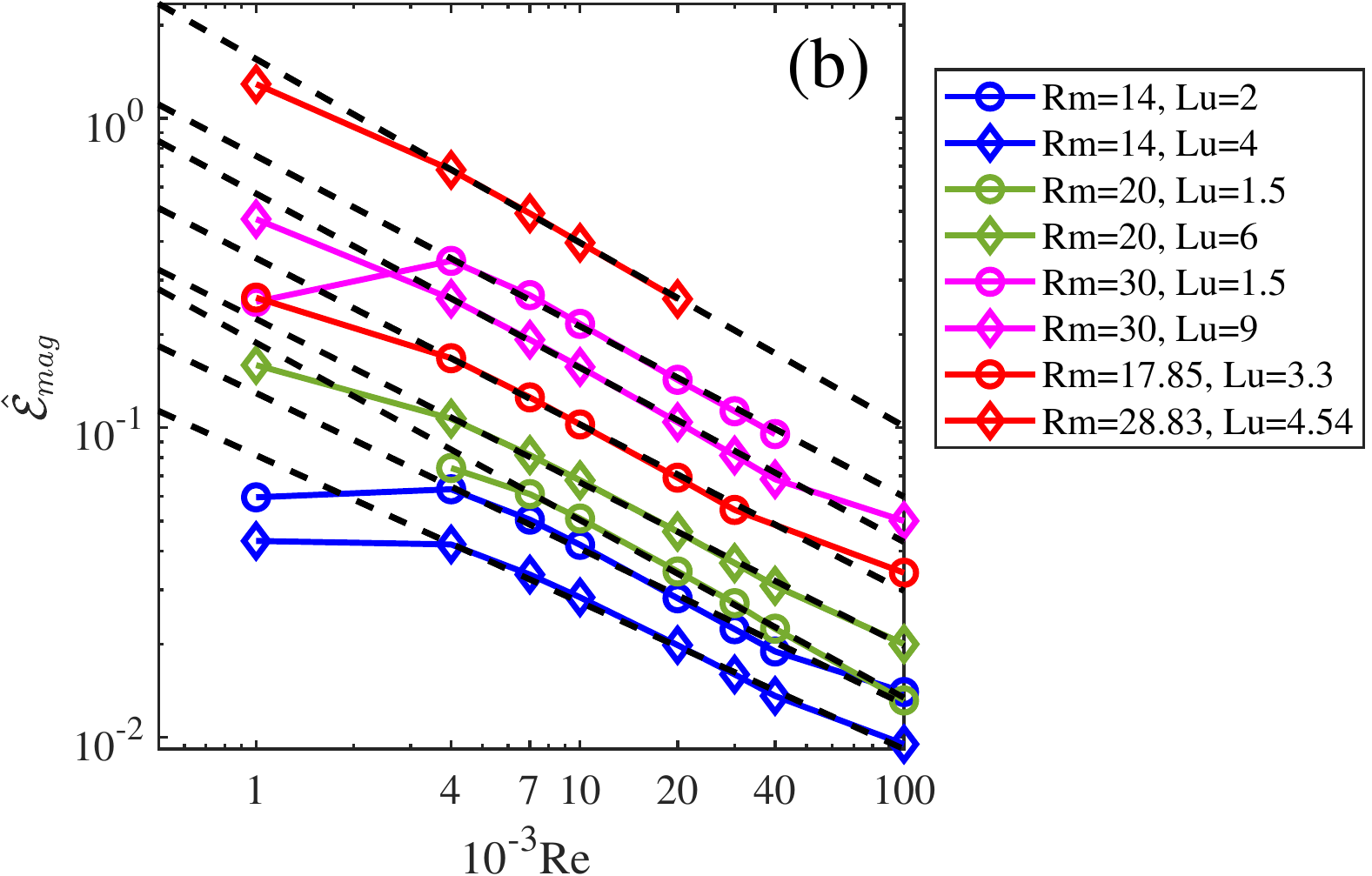}
\caption{(a) Time evolution of the volume-integrated magnetic energy, $\mathcal{E}_{mag}$, for $\mu=0.27$, different $Re$ and two pairs of $(Lu, Rm) = (2.5, 9)$ (dashed) and (9, 30) (solid). (b) Magnetic energy in the saturated state, $\mathcal{\hat{E}}_{mag}$, vs. $Re$ for different pairs $(Lu, Rm)$. Black dashed lines are the power law fits of the form $Re^a$ with an average value of the exponent $a\approx-0.5$.}\label{fig:Scaling_energy}
\end{figure*}
	
\subsection{Instability growth rate in the linear regime}
	
In the nonlinear simulations, the growth rate $\gamma$ at the initial linear stage of the instability evolution can be obtained by assuming that at these times the volume-integrated energy of magnetic field perturbations $\boldsymbol{b}=\boldsymbol{B}-\boldsymbol{B}_0$, $\mathcal{E}_{mag}=(1/2)\int_V \boldsymbol{b}^2dV$, where $V$ is the volume of the TC device, increases exponentially, $\mathcal{E}_{mag} \propto \exp(2\gamma t)$, and therefore $\gamma=(1/2)dlog(\mathcal{E}_{mag})/dt$. Figure \ref{fig:growth_rate} compares the growth rates from the 1D linear stability analysis and from the initial, linear stage of nonlinear simulations as a function of $Lu$ at  $\mu=0.27$ and different $Rm$ and $Re$. Dashed lines correspond to the growth rate maximized over axial wavenumbers $k_z \geq k_{z, min}$ from the 1D linear analysis \cite{Mishra_mamatsashvili_stefani_2022PRF}, while square markers to the instability growth rates obtained from the nonlinear simulations for $Re \in \{1, 4, 7, 10, 20, 30, 40, 100\} \times 10^3$. Figure \ref{fig:growth_rate}(a) shows that along the  line of maximum instability, $Rm\approx A\cdot Lu+B$,  the growth rates from the 1D linear and nonlinear analysis agree quite well with each other for all $Re$ considered. It is seen in Fig. \ref{fig:growth_rate}(a) also that at larger $Re$ (smaller $Pm$) these growth rates tend to converge which is a typical feature of SMRI, as already mentioned above. Figures \ref{fig:growth_rate}(b)-\ref{fig:growth_rate}(f) further show the same comparison of the growth rates as a function of $Lu$ for the same parameters as above but fixed $Rm \in \{9, 14, 20, 25, 30\}$. For $Rm=9$ in Fig. \ref{fig:growth_rate}(b), it can be seen that the difference is somewhat larger for smaller $Lu$, because the growth rates are very small for smaller $Lu$ and the nonlinear simulations cannot capture them quite accurately, while at larger $Lu$ this difference is quite small. (The case with $Re=1000$ is stable for $Rm=9$ and therefore not shown in this plot.) Figures \ref{fig:growth_rate}(c)-\ref{fig:growth_rate}(f) show the growth rates as a function of $Lu$ for several higher $Rm \in \{14, 20, 25, 30\}$, respectively, which further confirms much better correspondence between the growth rates obtained from 1D linear analysis and the nonlinear simulations. Hence, the above analysis conducted for a broad set of $(Lu, Rm)$ and each for different $Re$ demonstrates that our nonlinear setup works quite well for a wide range of $Pm$ and its results in the linear regime of evolution are in a good correspondence with those of the 1D linear stability analysis.

\subsection{Nonlinear evolution of SMRI}
		
To analyze the growth and subsequent nonlinear saturation of axisymmetric MRI modes, in Fig. \ref{fig:Scaling_energy}(a) we follow the evolution of the volume-integrated magnetic energy of perturbations for varying Reynolds numbers $Re$ and two pairs of $(Lu, Rm)=(9, 30)$ and $(2.5, 9)$. In all the cases, the magnetic energy initially grows exponentially with the growth rate predicted by the 1D linear analysis and eventually saturates with the energies and stresses remaining nearly constant in time. This behavior is consistent with previous nonlinear studies of SMRI both at infinite \cite{Knobloch_Julien_2005PhFl,Liu_Goodman_Ji_NonlinearMRI_2006ApJ,Gellert_etal2012} and finite \cite{Gissinger_Goodman_Ji_2012PhFl, Wei_etal2016, Winarto_etal2020} height of the cylinders. Since the magnetic energy of SMRI depends mostly on $Rm$ (see also Paper I), the exponential growth and saturation energy is smaller for smaller $Rm$. 

In Fig. \ref{fig:Scaling_energy}(b), we plot the magnetic energy values in the nonlinear saturated state, $\mathcal{\hat{E}}_{mag}$, as a function of $Re$ for various pairs $(Lu, Rm)$, the black dashed lines show the power law fit. Evidently, for all sets of ($Lu, Rm$), the saturated magnetic energy decreases with increasing $Re$, forming a family of lines with nearly the same slope in the logarithmic scale at $Re\geq 4000$, where the asymptotic regime appears to be already reached. We numerically fit this dependence with the power law $Re^a$ shown by black dashed lines in Fig. \ref{fig:Scaling_energy}(b), which are more or less parallel to each other with the common average exponent $a \approx -0.5$. This implies that the magnetic energy in the saturated state follows a power law scaling $Re^{-0.5}$ or, $Pm^{0.5}$ as $Rm$ is fixed, for all considered pairs $(Lu,Rm)$. This result is consistent with a previously reported scaling law for SMRI in a wide gap TC geometry with endcaps \cite{Gissinger_Goodman_Ji_2012PhFl}. This result will be important below for extrapolating the values of the magnetic energy down to even smaller, experimentally relevant $Pm \sim 10^{-6}-10^{-5}$.  

\subsubsection{Transition from the linear phase to nonlinear saturated state -- a closer look}
	
\begin{figure*}
\centering
\includegraphics[width=0.9\textwidth]{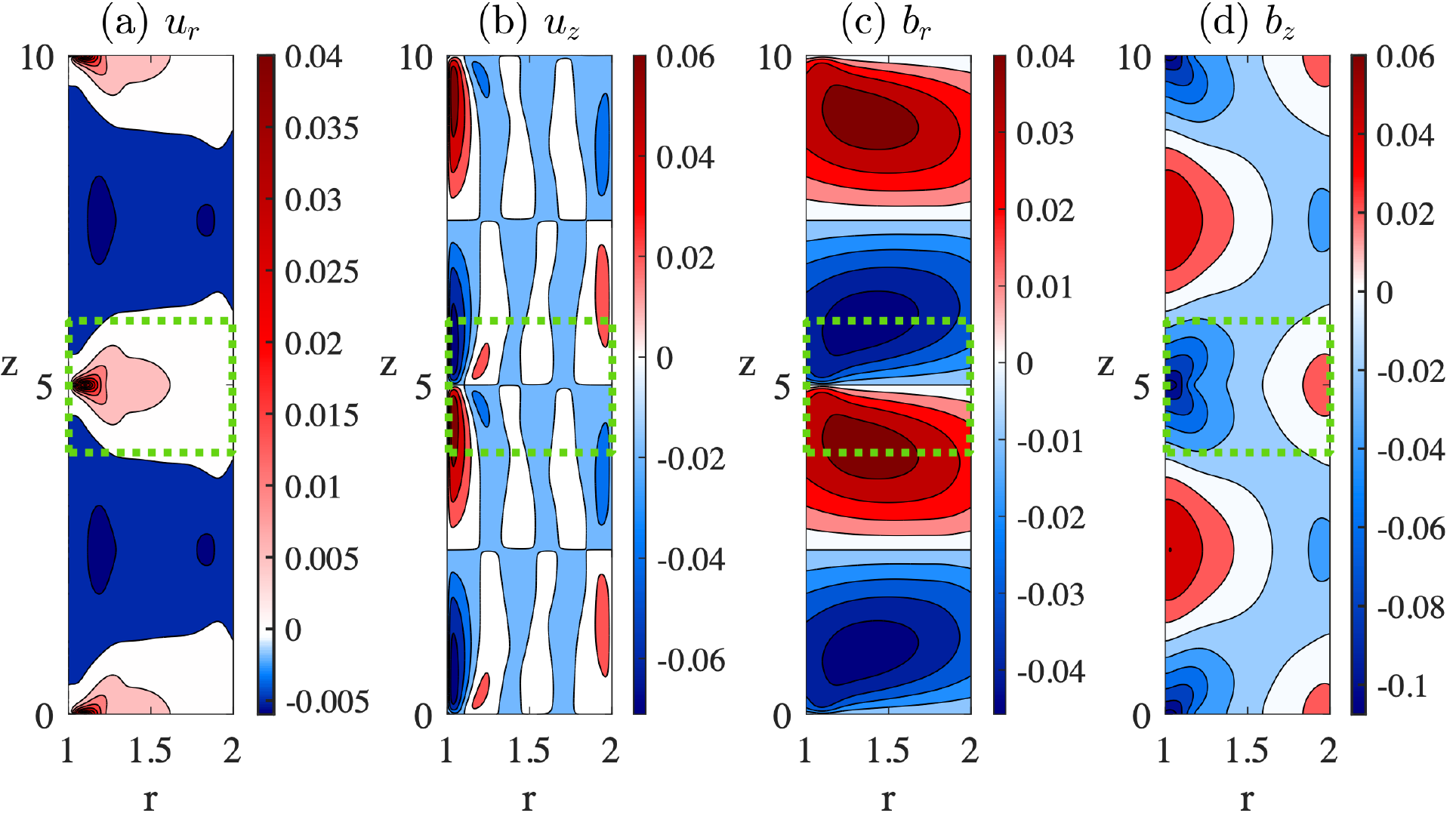}
\caption{Structure of (a) the radial velocity, (b) axial velocity, (c) radial magnetic field and (d) axial magnetic field in $(r,z)$-plane in the saturated state at $\mu=0.27, Lu=9, Rm=30$ and $Re=40000$. A region marked with green dashed rectangle encompasses the reconnection site near $z=5$ and is used in Fig. \ref{fig:reconnection}.}
\label{fig:r_z_slice}
\end{figure*}

Some of the earliest studies of the nonlinear dynamics of SMRI suggested  magnetic reconnection as a primary mechanism of its saturation \cite{Hawley_Balbus_1991ApJ_nonlinear}. The role of reconnection in the saturation of SMRI  was later analytically \cite{Knobloch_Julien_2005PhFl} and numerically \cite{Liu_Goodman_Ji_NonlinearMRI_2006ApJ} analyzed in greater detail for a magnetized TC flow. For the Princeton MRI experiment, Liu et. al. \cite{Liu_Goodman_Ji_NonlinearMRI_2006ApJ} showed that the transition of SMRI from the exponential growth phase to the nonlinear saturated state occurs via magnetic reconnection and discussed related jet formation and its scaling behaviour. In this subsection, we aim to explore SMRI saturation mechanism in the context of the DRESDYN-MRI experiment.

Figure \ref{fig:r_z_slice} shows the radial and axial velocity and magnetic field perturbations in the $(r,z)$-plane in the nonlinear saturated state of SMRI for $Lu=9, Rm=30$ and $Re=40000$. It is seen from this figure that the spatial distribution of these variables has evolved significantly from the eigenfunctions at the linear stage (Paper I) to almost steady structures in the saturated state. The radial velocity $u_r$ exhibits the formation of outflowing jets distributed periodically along the height of the cylinder [the one at $z=5$ is marked with green dashed rectangle in Fig. \ref{fig:r_z_slice}(a)]. At the same time, the axial velocity $u_z$ has the form of a standing wave, predominantly concentrated near the inner cylinder. In the vicinity of the jet (at $z=5$), two narrow regions, an upper one with $u_z<0$ (blue) and a lower one with $u_z>0$ (red), are seen in Fig. \ref{fig:r_z_slice}(b), which indicate the axial inflow into the jet. Radial magnetic field $b_r$ in the jet region [at $z=5$ inside green dashed rectangle in Fig. \ref{fig:r_z_slice}(c)] shows the field lines with opposite polarities approaching each other, while the axial magnetic field perturbations $b_z$ in the same region are dominated by a comparatively large-scale structure [Fig. \ref{fig:r_z_slice}(d)]. The radial jet with the associated velocity and radial magnetic field structures represents, as we demonstrate below, a favorable site for magnetic reconnection. 

\begin{figure*}[t!]
\centering
\includegraphics[width=0.3\textwidth]{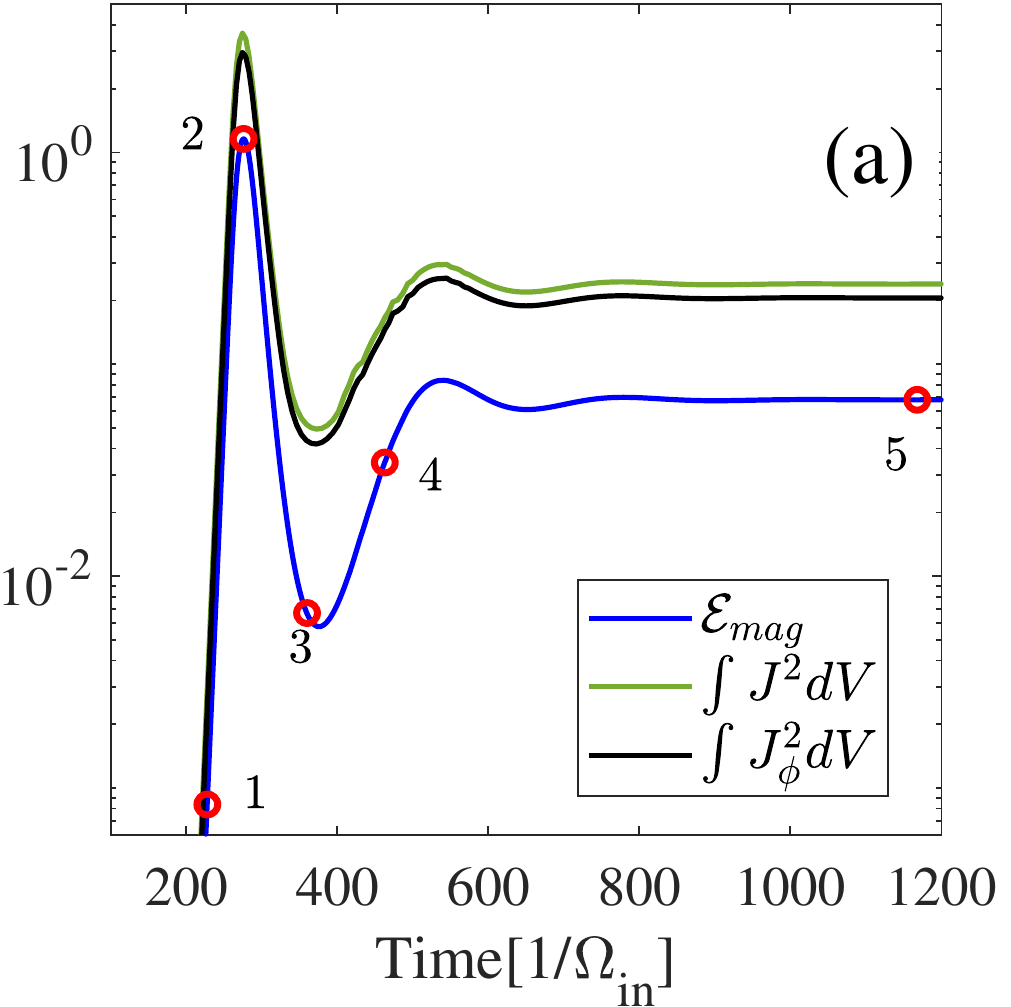}
\includegraphics[width=0.32\textwidth]{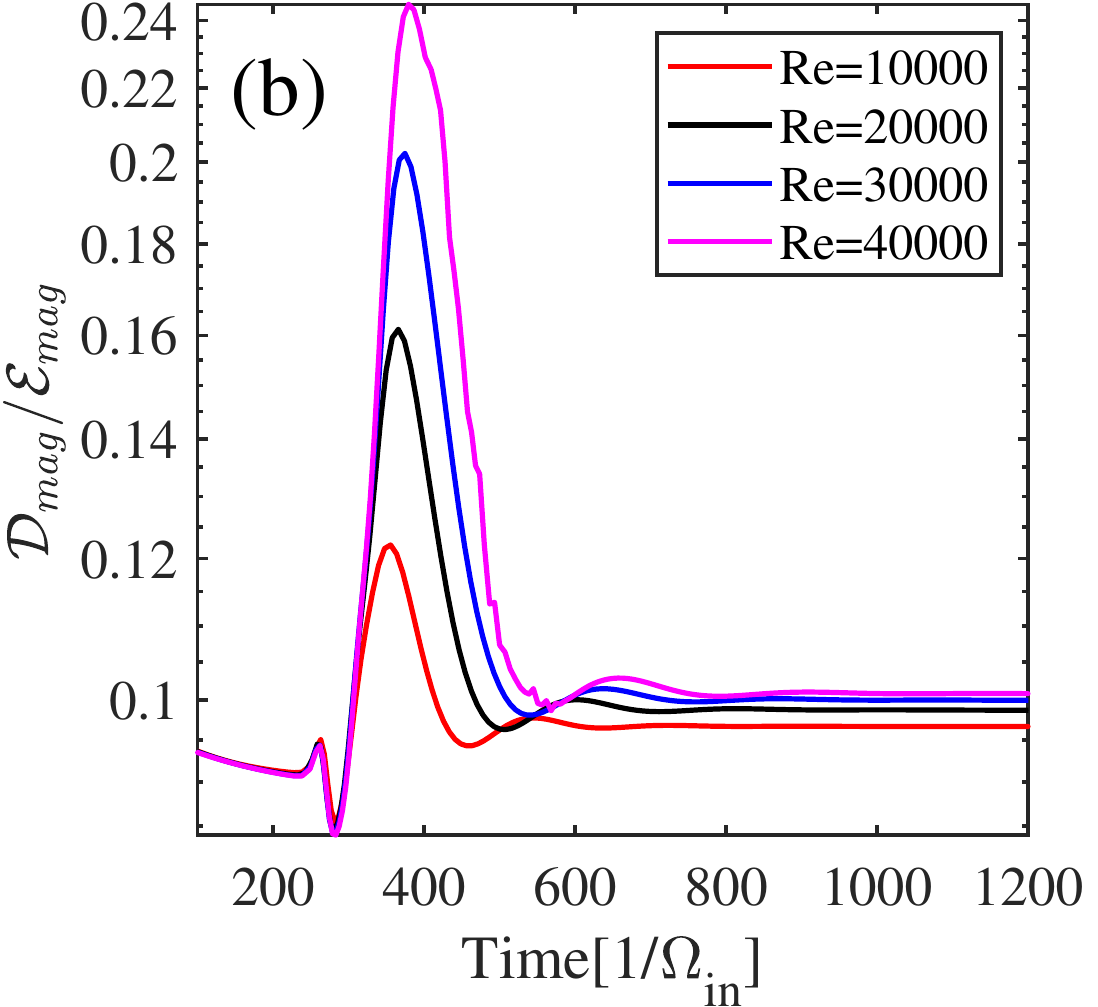}
\includegraphics[width=0.31\textwidth]{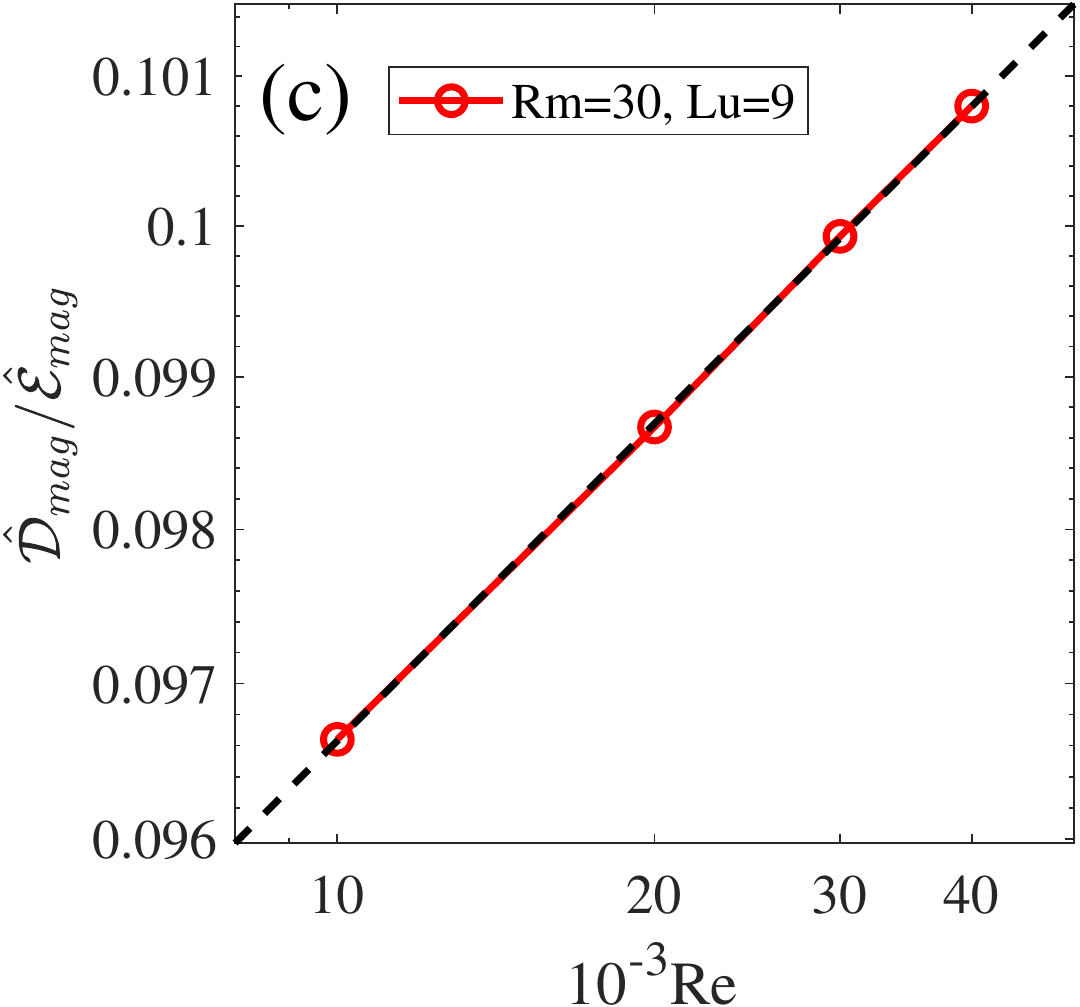}
\caption{(a) Time evolution of the volume-integrated magnetic energy $\mathcal{E}_{mag}$ (blue), total current density $\int J^2 dV$ (green) and azimuthal current density $\int J_\phi^2 dV$ (black) at $\mu=0.27, Lu=9, Rm=30$ and $Re=40000$. Red circles on the magnetic energy curve mark the time moments at which flow states are analysed in Fig. \ref{fig:reconnection}. (b) Evolution of the ratio of the volume-integrated magnetic dissipation to magnetic energy, i.e., magnetic dissipation rate, $\mathcal{D}_{mag}/ \mathcal{E}_{mag}$, for the same parameters as in (a) but different $Re \in \{1, 2, 3, 4 \} \times 10^4$. (c) This ratio in the saturated state, $\hat{\mathcal{D}}_{mag}/\hat{\mathcal{E}}_{mag}$, follows the power law $Re^c$, albeit with small $c\approx 0.03$ (dashed line). From (b) and (c) it is seen that the magnetic dissipation rate increases with $Re$ for a given $Rm$, more strongly during the transition phase [point 3 in (a)].} \label{fig:current_density}
\end{figure*}

To understand the reconnection process and associated formation of the jet, we analyze the evolution of the current density together with magnetic field. Earlier investigations have shown that current density $J$ is a proxy of the site of magnetic reconnection, where it exhibits current sheet structures \cite{Bodo_etal_2022MNRAS, Rosenberg_Ebrahimi_2021ApJL}. 
Figure \ref{fig:current_density}(a) shows the time evolution of the volume-integrated magnetic energy $\mathcal{E}_{mag}$, current density squared, $\int J^2 dV$, where $J=\nabla \times \boldsymbol{b}$, and azimuthal current density squared, $\int J_{\phi}^2 dV$, where $J_{\phi}=(\partial b_r/\partial z - \partial b_z/\partial r)$, for $Lu=9, Rm=30$ and $Re=40000$. The total current density $J$ is dominated by the azimuthal component $J_{\phi}$ as seen in Fig. \ref{fig:current_density}(a). On a comparative note, the evolution of the magnetic energy follows the current density $J$, or for that matter, $J_\phi$, at all times. It is clear from Fig. \ref{fig:current_density}(a) that the magnetic energy at all stages of evolution, i.e., growth (denoted with 1), peak (2), transitional phase consisting of decrease (3) and restructuring of fields (4) and the final saturated state (5) match well with those of the current density. Figure \ref{fig:current_density}(b) shows the time evolution of the magnetic dissipation rate, i.e., the ratio of magnetic dissipation $\mathcal{D}_{mag}\, (=Rm^{-1}\int J^2 dV)$ to the volume-integrated magnetic energy, $\mathcal{E}_{mag}$, at different $Re=\{1, 2, 3, 4\}\times 10^4$, while Fig. \ref{fig:current_density}(c) shows this ratio versus $Re$ in the saturated state, which exhibits a weak power law dependence $Re^c$ with small $c\approx 0.03$. Comparing  Figs. \ref{fig:current_density}(a) and \ref{fig:current_density}(b) at $Re=40000$, it is seen that during the transition phase, when $\mathcal{E}_{mag}$ decreases with time and reaches a minimum, $\mathcal{D}_{mag}/ \mathcal{E}_{mag}$, by contrast, reaches a peak due to increased resistive dissipation at these times. This ratio increases with increasing $Re$ both in the saturated and even stronger in the transition phases, implying that higher Reynolds numbers result in higher resistive dissipation rates due to the formation of smaller-scale structures \cite{Riols_etal17,Mamatsashvili_etal20, Held_Mamatsashvili22}.  Specifically, we will see below that a strong concentration of the current -- current sheet -- forms at the beginning of the transition phase, which implies a key role of $J$ (or for that matter $J_\phi$) in the saturation mechanism. This current density also increases with increasing $Re$, causing the increase in the resistive dissipation rate and hence decrease in the magnetic energy.

\begin{figure*}
\includegraphics[width=0.9\textwidth,height=4.2cm]{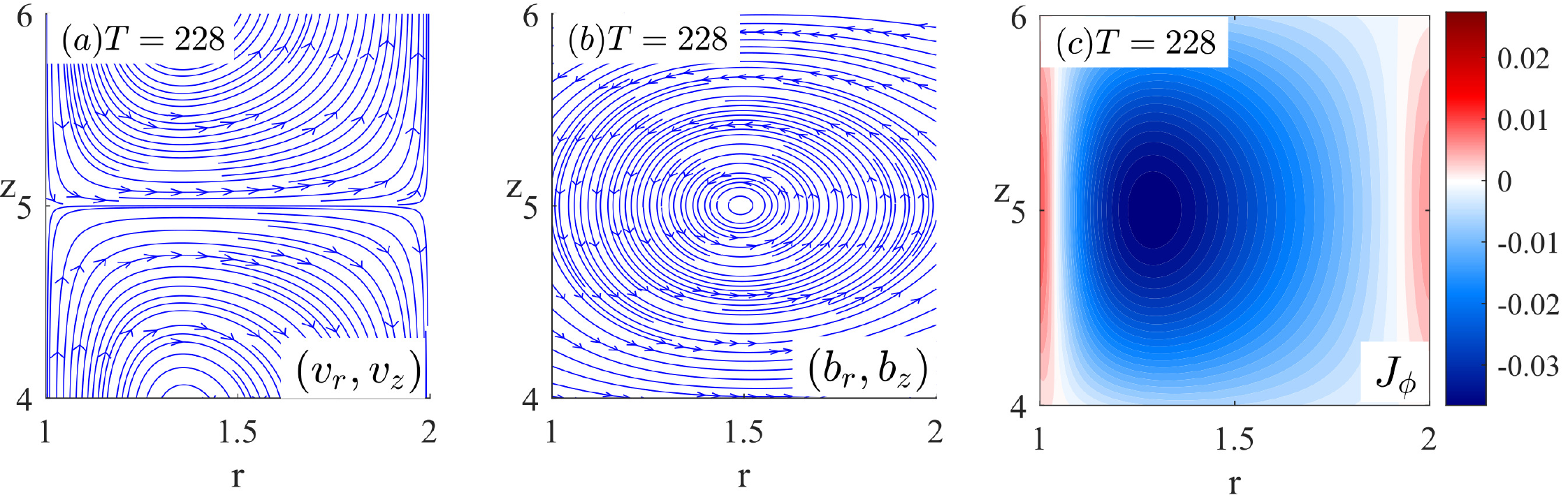}
\includegraphics[width=0.9\textwidth,height=4.2cm]{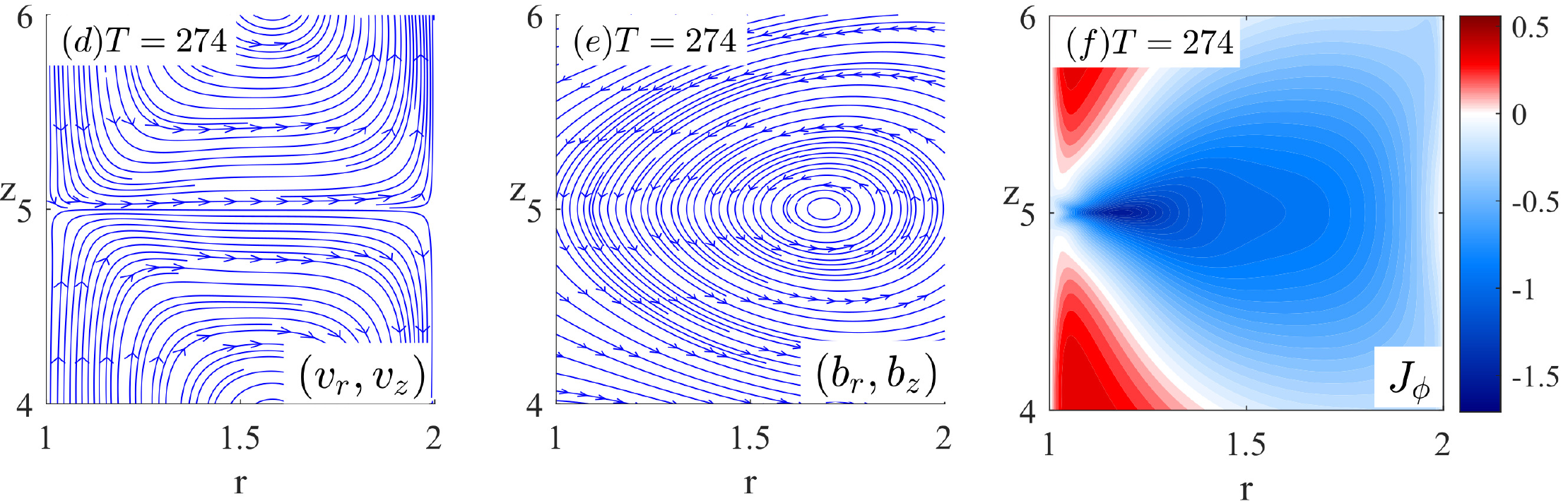} 
\includegraphics[width=0.9\textwidth,height=4.2cm]{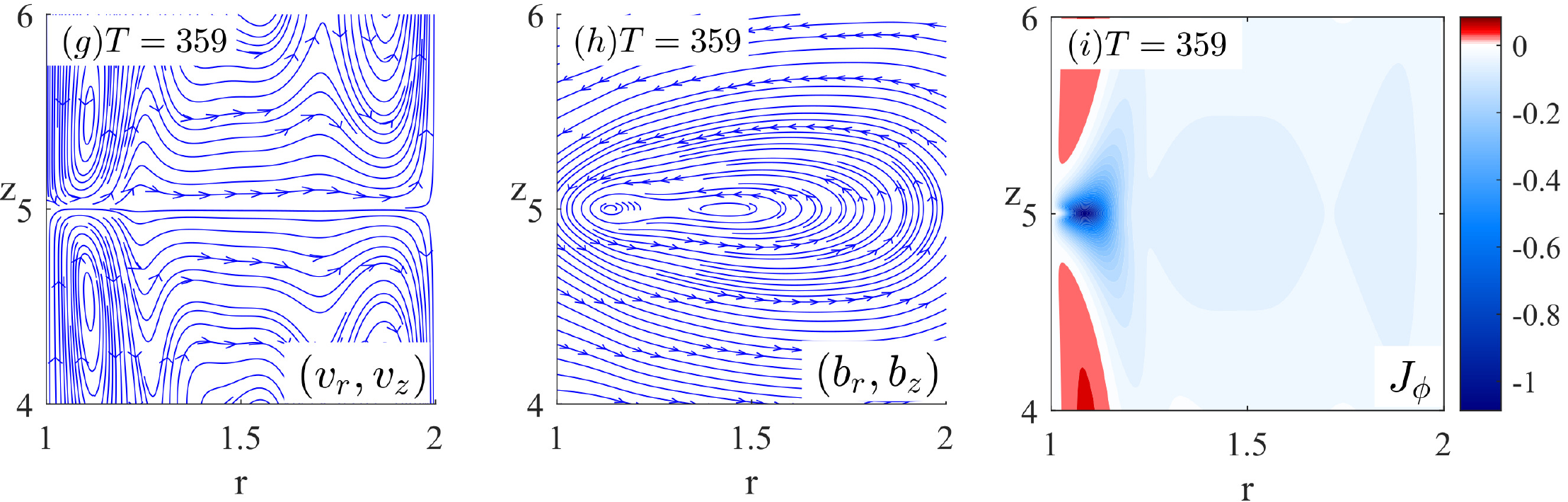}
\includegraphics[width=0.9\textwidth,height=4.2cm]{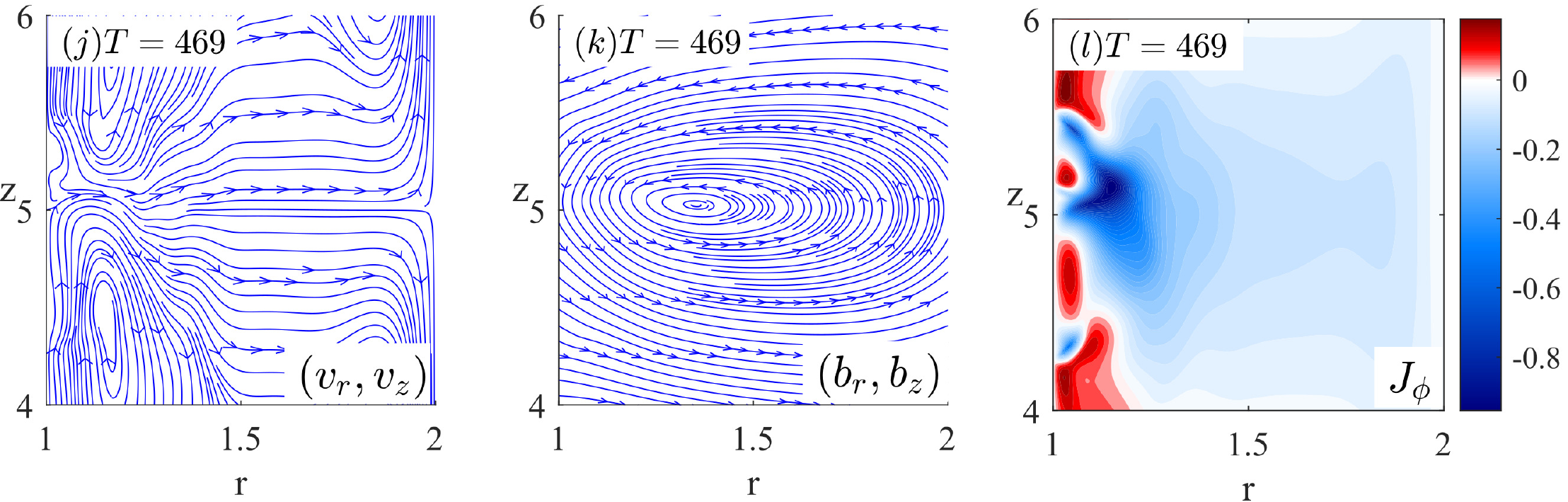}
\includegraphics[width=0.9\textwidth,height=4.2cm]{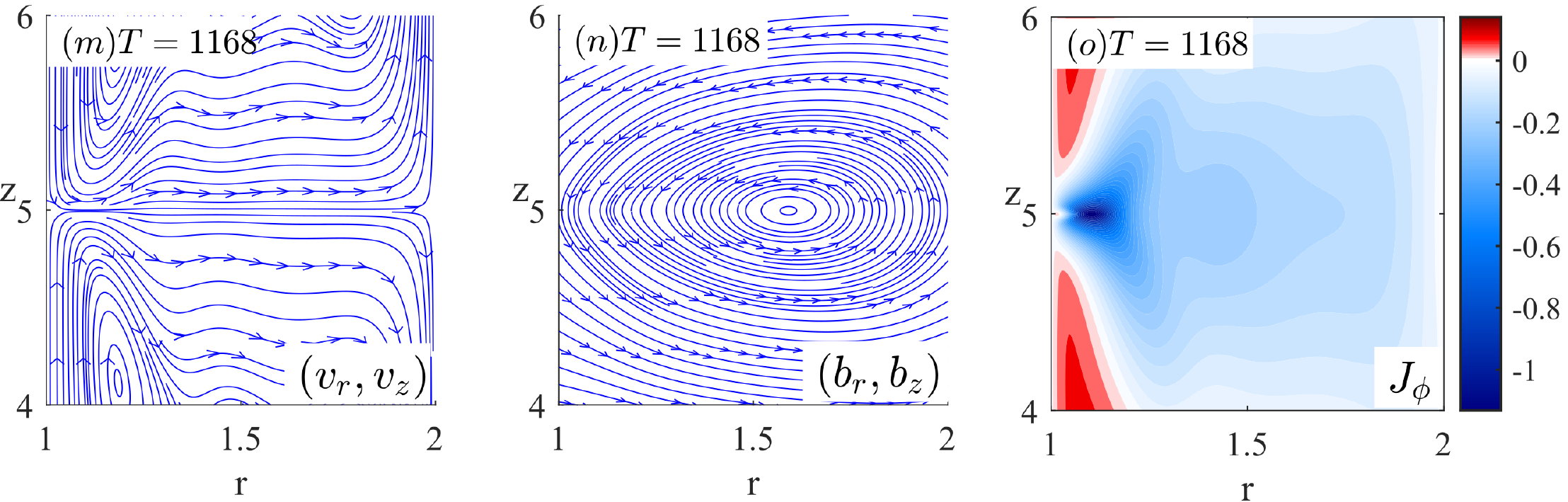}
\caption{Left column depicts the streamlines obtained from the radial $u_r$ and axial $u_z$ velocity components, middle column depicts the field lines obtained from the radial $b_r$ and axial $b_z$ magnetic field components, while right column depicts the associated distribution of $J_\phi$ in the $(r,z)$-plane for $\mu=0.27, Lu=9, Rm=30$ and $Re=40000$ at different characteristic time moments during the saturation process: $T=228$ (a,b,c), $T=274$ (d,e,f), $T=359$ (g,h,i), $T=469$ (j,k,l) and $T=1168$ (m,n,o), corresponding, respectively, to the red points 1-5 in Fig. \ref{fig:current_density}(a).}\label{fig:reconnection}
\end{figure*}

\begin{figure}
\includegraphics[width=0.9\columnwidth]{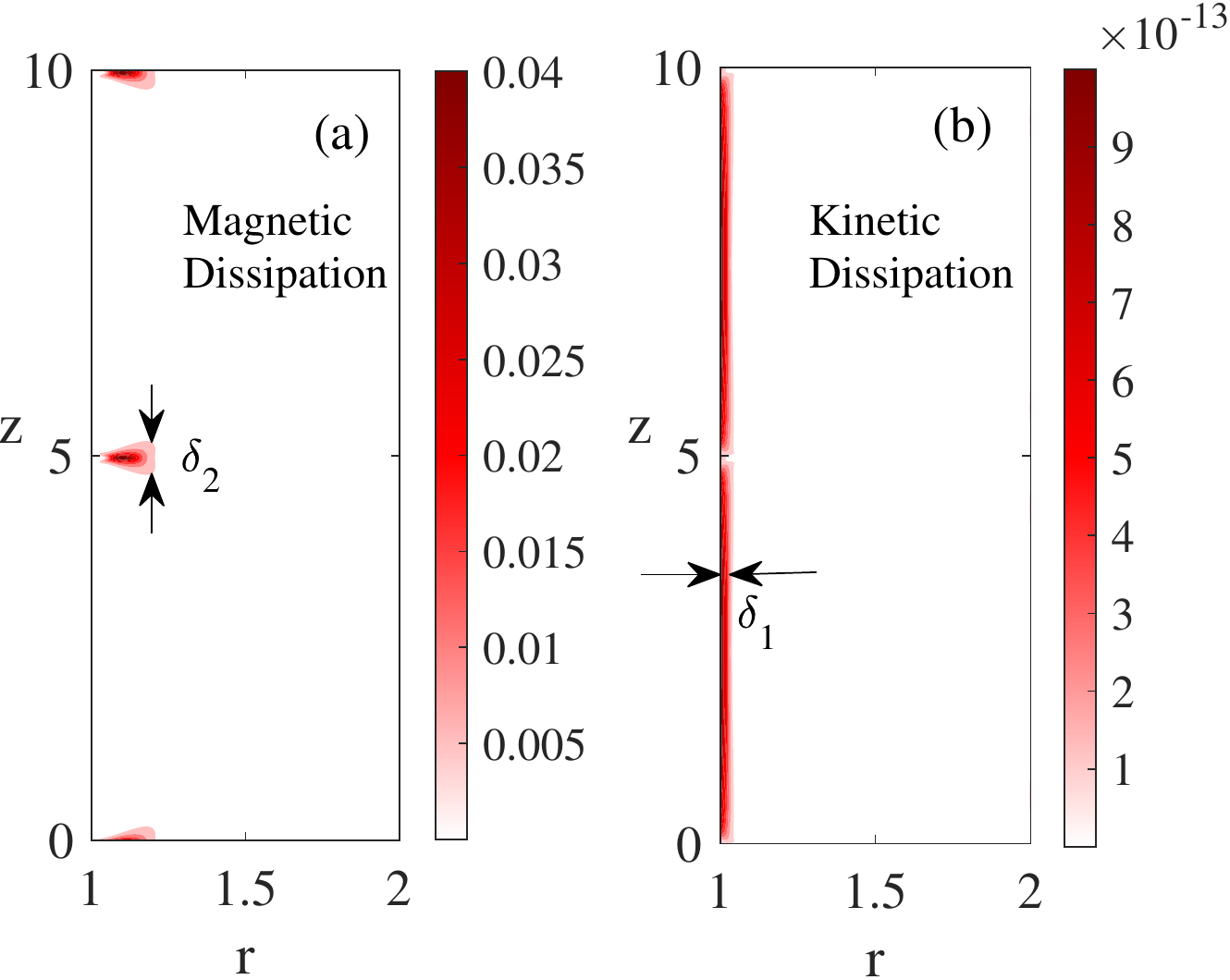}
\caption{Distribution of (a) magnetic and (b) kinetic dissipation in the $(r,z)$-plane in the saturated state for the same parameters as in Fig. \ref{fig:r_z_slice}. Comparing with that figure, magnetic dissipation is concentrated in the reconnection region (with size $\delta_2$), while kinetic dissipation in the boundary layer (with thickness $\delta_1$) near the inner cylinder wall.}\label{fig:mag_kin_dissipation}
\end{figure}

\begin{figure*}
\centering
\includegraphics[width=0.3\textwidth]{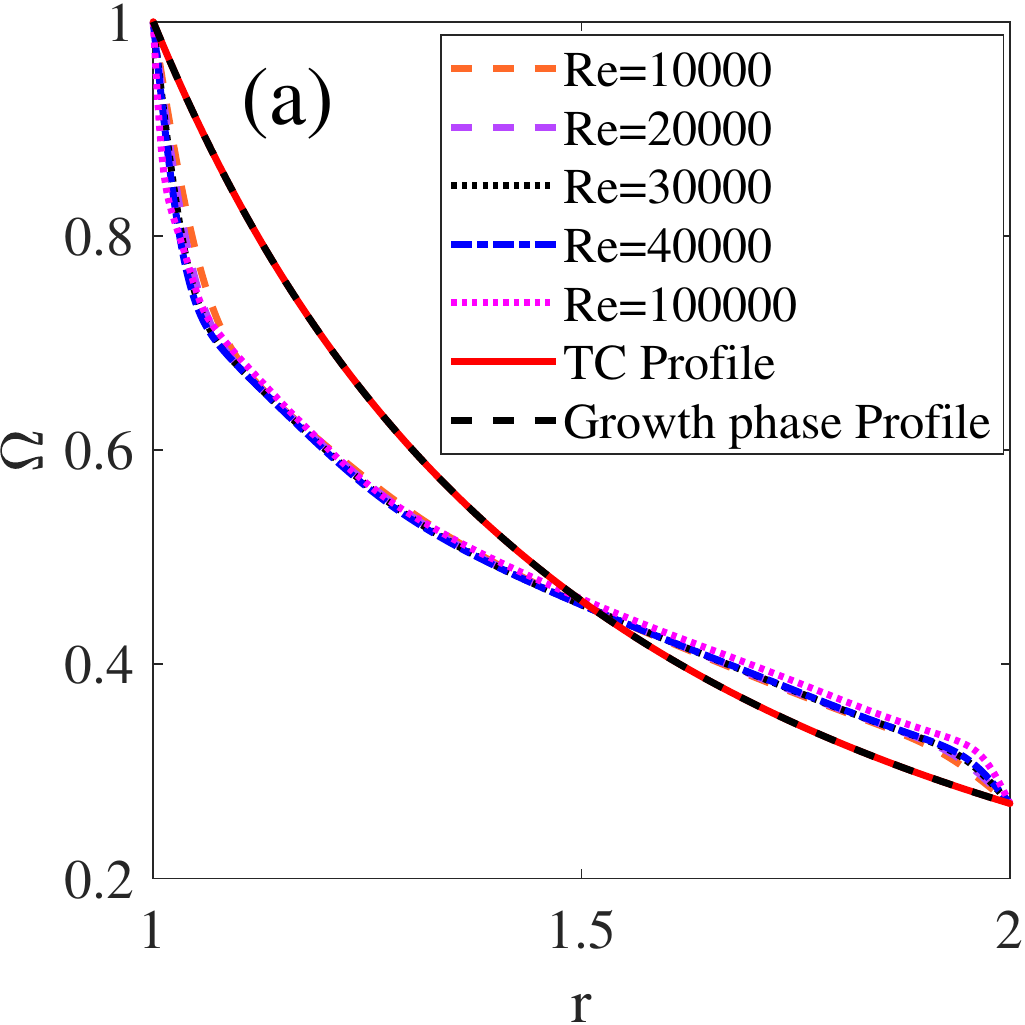}
\hspace{1em}
\includegraphics[width=0.31\textwidth]{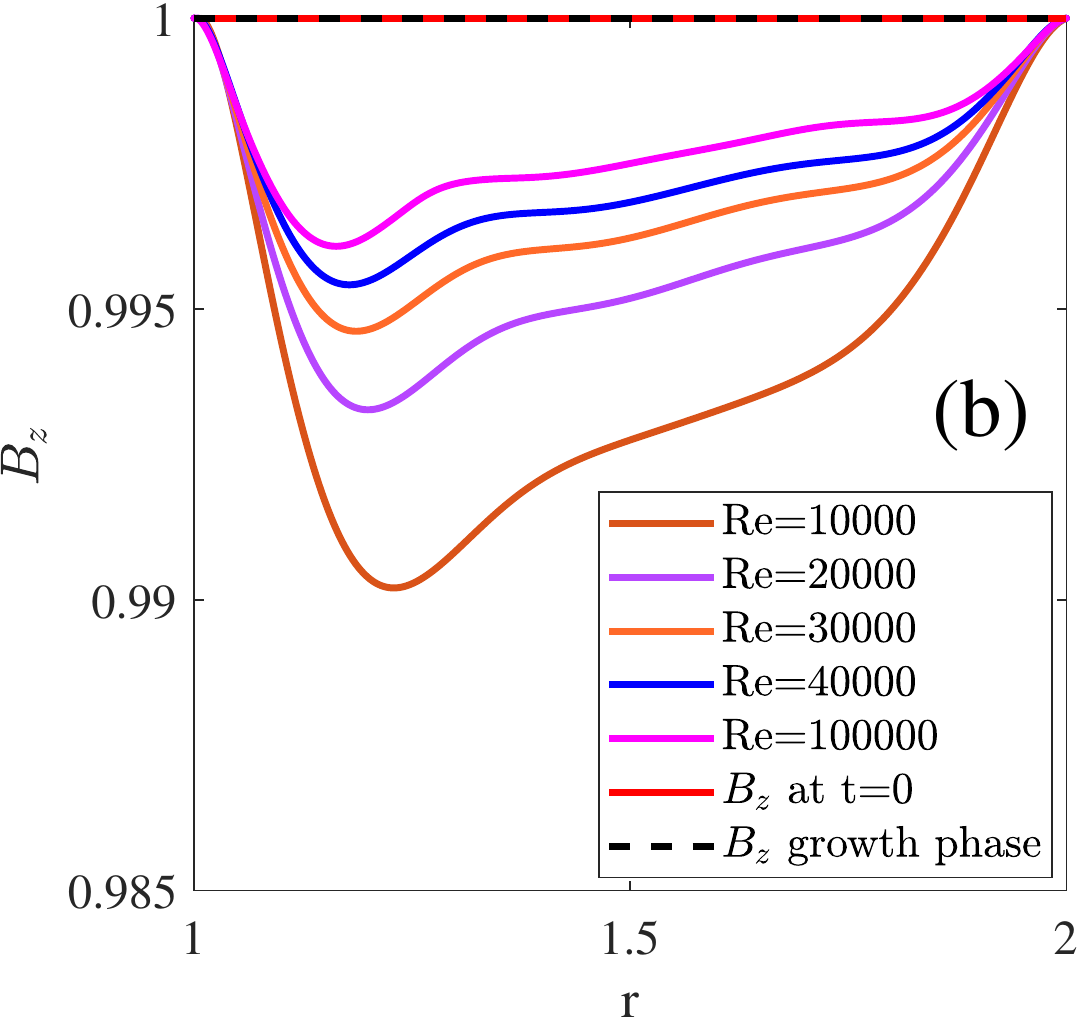}
\hspace{1em}
\includegraphics[width=0.29\textwidth]{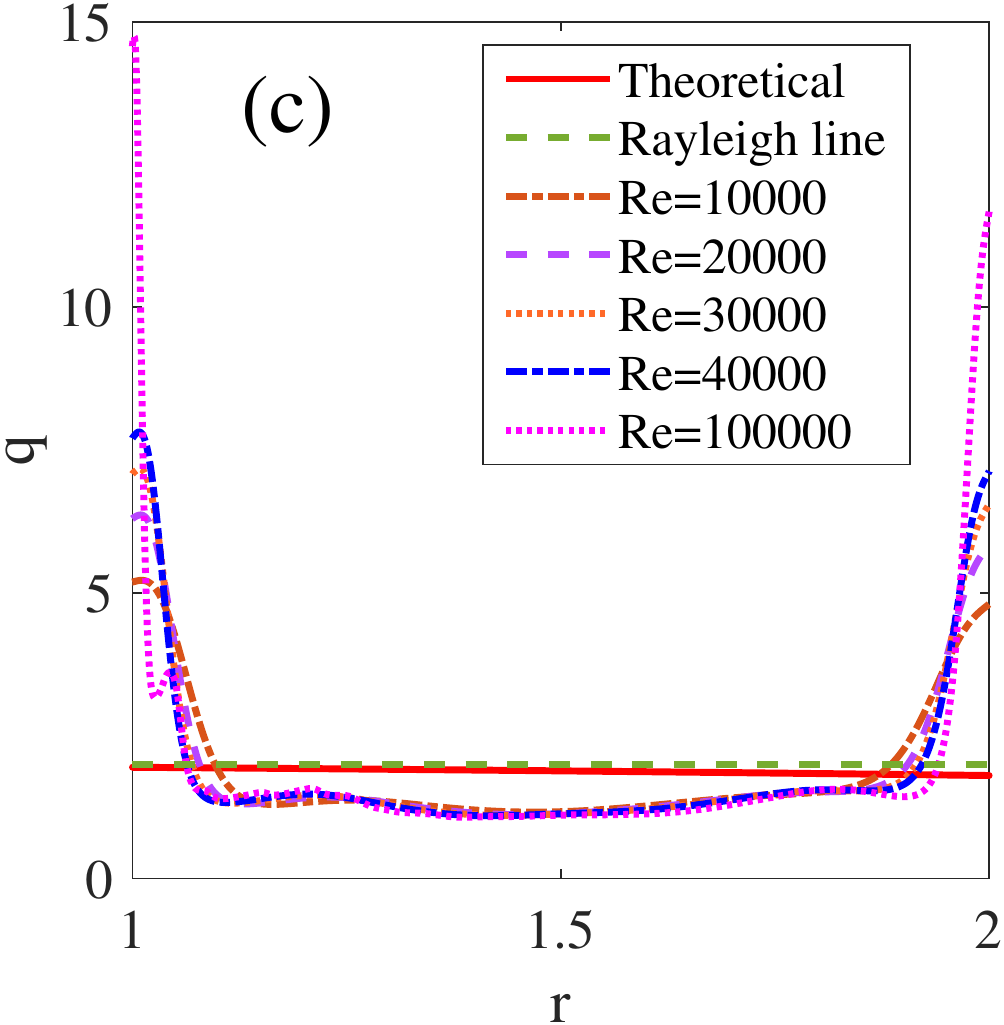}
\caption{(a) Angular velocity profiles in the linear regime, during the exponential growth phase  (dashed black) and in the saturated state at different $Re$. The latter profiles deviate from the classical TC profile (red) given by Eq. (\ref{TC_profile}), but seem to be almost independent of $Re$ in the bulk of the flow. (b) The total axial magnetic field in the linear stage (dashed black), which coincides with the imposed axial field $B_{0z}=1$ (red), and in the saturated state which is very slightly smaller than the imposed field. (c) The local shear parameter $q=-d{\rm ln}~\Omega/d{\rm ln}~r$ in the saturated state as a function of $r$ at different $Re$. The shear parameter for the classical TC flow profile (red) and Rayleigh stability threshold $q=2$ (dashed green) are shown for reference. It is seen that SMRI saturates through reduction of the shear in the mean angular velocity. Note, however, that near the cylinder walls the shear becomes quite large above the Rayleigh threshold value, which further increases with $Re$. In all the panels, $\mu=0.27$, $Lu=9$ and $Rm=30$ the same as in Fig. \ref{fig:reconnection}.}\label{fig:Omega_Bz_deviation}
\end{figure*}

In the above analysis, we showed that the time evolution of azimuthal current density $J_\phi$ agrees well with different phases of the SMRI evolution, especially during the transition to saturated state. Here, we exploit this feature of $J_\phi$ together with the perturbation velocity and magnetic field structure in order to examine the site and properties of magnetic reconnection in the present setup. It is seen from the structure of the nonlinear state in Fig. \ref{fig:r_z_slice}, that reconnection takes place around the mid height of the cylinder, so we choose an interval $4\leq z\leq 6$ (green dashed rectangle in Fig. \ref{fig:r_z_slice}) to better look into this process. Figure \ref{fig:reconnection} shows the velocity and magnetic field lines, which are constructed, respectively, from the radial and axial, ($u_r, u_z$) and $(b_r,b_z)$, components of these fields together with the distribution of $J_{\phi}$ in the $(r,z)$-plane for the same parameters and the characteristic time moments shown with red circles in Fig. \ref{fig:current_density}(a). Although the field lines are dominated by large-scale fields, together with $J_\phi$, they allow us to better look into the saturation mechanism. Figures \ref{fig:reconnection}(a)-\ref{fig:reconnection}(c) show the structure of the velocity and magnetic field lines as well as azimuthal current in the exponential growth phase at $T=228$ [point $1$ in Fig. \ref{fig:current_density}(a)]. As expected, these field lines have a regular round shape in the $(r,z)$-plane which is similar to that of eigenfunctions in the linear stability analysis from Paper I; $J_\phi$ also shows the concentration of current in the same region of the $(r,z)$-plane. Figures \ref{fig:reconnection}(d)-\ref{fig:reconnection}(f) show these quantities, respectively, at the peak of the exponential growth phase at $T=274$ [point $2$ in Fig. \ref{fig:current_density}(a)]. At this point nonlinearity comes into play and, as a consequence, the structure of $J_\phi$ changes (flattens) significantly as seen in Fig. \ref{fig:reconnection}(f), becoming more concentrated near the inner cylinder at $z=5$ and forming there a pinch-type structure -- current sheet. Accordingly, the velocity and magnetic field lines in the $(r,z)$-plane start to deform and also become more concentrated near the inner cylinder [Figs. \ref{fig:reconnection}(d) and \ref{fig:reconnection}(e)]. As a result,  magnetic reconnection occurs in the vicinity of the current sheet, which allows the system to release energy through the formation of the radially outflowing jet, as was also observed in Ref. \cite{Liu_Goodman_Ji_NonlinearMRI_2006ApJ}. This corresponds to the fast decay phase of increased resistive dissipation in Fig. \ref{fig:current_density}(a) lasting from point 2 to 3. Figures \ref{fig:reconnection}(g)-\ref{fig:reconnection}(i) show the situation near the end of the fast decay phase at $T=359$ [point $3$ in Fig. \ref{fig:current_density}(a)], where an X-type, or null point at $(r,z)\approx (1.3, 5)$ emerges in the magnetic field lines [Fig. \ref{fig:reconnection}(h)], which indicates that magnetic reconnection is happening. The corresponding velocity distribution in this region has the form of two axial streams near the inner cylinder wall, which meet at $z=5$, and then flow radially outwards forming a jet [Fig. \ref{fig:reconnection}(g)]. The azimuthal current $J_\phi$ in this reconnection region is highly concentrated with almost more than 95\% of the absolute value [Fig. \ref{fig:reconnection}(i)]. After the fast decay phase due to magnetic reconnection, fields start to restructure due to the excited radial motions (i.e., radial jet) near the inner cylinder. This restructuring phase is shown in Figs. \ref{fig:reconnection}(j)-\ref{fig:reconnection}(l) at $T=469$ [point $4$ in Fig. \ref{fig:current_density}(a)]. The newly formed strong radial jet [Fig. \ref{fig:reconnection}(j)] distorts and diffuses the current density distribution [Fig. {\ref{fig:reconnection}(l)] compared to that in the preceding reconnection phase depicted in Fig. \ref{fig:reconnection}(i). As a result, the structure of magnetic field lines also changes. Figures \ref{fig:reconnection}(m)-\ref{fig:reconnection}(o) show the final saturated state at $T=1168$ [point $5$ in Fig. \ref{fig:current_density}(a)]. After saturation has been reached, magnetic field, current density and velocity remain constant in time, that is, a stationary distribution of these fields has been established (see also Fig. \ref{fig:r_z_slice}), where the strong axial inflow velocity in the boundary layer formed near the inner cylinder [Fig. \ref{fig:reconnection}(m)] continually brings together field lines and causes them to reconnect [indicated by pointed shape of the field lines at $z=5$ in Fig. \ref{fig:reconnection}(n)]. The fluid flowing axially into the reconnection region is ejected in the form of the radially outflowing jet, which takes up part of the energy released during the reconnection process, while the other part is dissipated by resistivity. In the region of magnetic reconnection, the azimuthal current $J_\phi$ is accordingly highly concentrated, reaching a maximum near the inner cylinder  [Fig. \ref{fig:reconnection}(o)], and therefore magnetic energy dissipation is highest there [Fig. \ref{fig:mag_kin_dissipation}(a)]. Since this state is stationary, the strong dissipation of magnetic energy in the reconnection region is balanced by energy injection from the basic TC flow into perturbations due to SMRI. In this case, the majority of kinetic energy dissipation is concentrated in the viscous boundary layer near the inner cylinder wall [Fig. \ref{fig:mag_kin_dissipation}(b)].

Figures \ref{fig:Omega_Bz_deviation}(a) and \ref{fig:Omega_Bz_deviation}(b) show the angular velocity and total axial magnetic field averaged over $z$-axial coordinate in the growth and saturated stages at different $Re$, which basically remain the same in the instability growth phase (blacked dashed line) as long as the perturbations are still small. However, in the saturated state, the angular velocity profile noticeably deviates from the initial theoretical TC profile [Eq. \ref{TC_profile}, red line in Fig. \ref{fig:Omega_Bz_deviation}(a)], reducing its shear (see below), but does not change with $Re$ in the bulk of the flow. This behaviour is in agreement with other studies of nonlinear evolution of MRI in TC flow \cite{Knobloch_Julien_2005PhFl, Liu_Goodman_Ji_NonlinearMRI_2006ApJ, Mamatsashvili_etal2018}. As for the averaged axial field, its deviation from the imposed field remains very small ($0.5\%$ of the imposed field), however, its specific structure [see Fig. \ref{fig:r_z_slice}(d)], exhibiting concentration near the walls is due to the magnetic reconnection process discussed above. 

In Fig. \ref{fig:Omega_Bz_deviation}(c) we examine in more detail the modification of the local shear of the angular velocity in the saturated state in terms of the shear parameter $q=-d{\rm ln}~\Omega/d{\rm ln}~r >0$ at different $Re$. The flow is Rayleigh-stable for $q \leq 2$ and unstable for $q>2$. The marginal stability threshold $q=2$ is shown as a green dashed line. For $\mu=0.27$, the theoretical value of $q$ for TC profile (\ref{TC_profile}) is close to the Rayleigh line near the inner radius while it slightly decreases with radius. On the other hand, in the saturated state, $q$ is substantially modified, being higher and Rayleigh-unstable in the boundary layers adjacent to the inner and outer cylinders and further increasing as $Re$ increases. By contrast, in the bulk of the flow, the shear profile remains in the Rayleigh-stable regime with a nearly constant value of $q\approx 1.3$ independent of $Re$, which is smaller than its initial classical TC flow value at $\mu=0.27$. This indicates that the saturation of SMRI is accompanied by decrease in the shear of the mean angular velocity profile compared to the original TC flow.

\subsubsection{Magnetic Energy in the Saturated State: Scaling Properties}	

In Fig. \ref{fig:Sat_mag_energy_vs_Lu} we plot the magnetic energy in the saturated state, $\hat{\mathcal{E}}_{mag}$, as a function of $Lu$ at different $Re$ and $Rm$ as indicated in Fig. \ref{fig:Analysis_points}(b). Figure \ref{fig:Sat_mag_energy_vs_Lu}(a) shows this dependence along the line of the maximum growth. As expected, the magnetic energy increases with increasing $Lu$ for fixed $Re$. However, it saturates at much larger values for $Re=1000$ and gradually decreases with increasing $Re$, consistent with the behavior seen in Fig. \ref{fig:Scaling_energy}(a). Figures \ref{fig:Sat_mag_energy_vs_Lu}(b)-\ref{fig:Sat_mag_energy_vs_Lu}(f) show the saturated magnetic energy for different $Rm=\{9, 14, 20, 25, 30\}$ as a function of $Lu$, which, for a given $Re$, first increases with $Lu$, attains a maximum at a certain $Lu$ and then gradually decreases towards larger $Lu$, like the exponential growth rate in the linear regime in Fig. \ref{fig:growth_rate}. As noted above, $\hat{\mathcal{E}}_{mag}$ decreases with increasing $Re$, or decreasing $Pm$, since for fixed $Rm$, $Re$ is inversely proportional to $Pm$. 
	
Based on the power law scaling obtained from Fig. \ref{fig:Scaling_energy}(b), in Figs. \ref{fig:Scal_Sat_Mag_Energy_Vs_Lu}(a)-\ref{fig:Scal_Sat_Mag_Energy_Vs_Lu}(f) we plot the saturated magnetic energy $\hat{\mathcal{E}}_{mag}$ for several $Rm$ as a function of $Lu$ scaled by the power law relation $Re^{a}$, where exponent $a$ is obtained numerically. Figure \ref{fig:Scal_Sat_Mag_Energy_Vs_Lu}(a) shows $\hat{\mathcal{E}}_{mag}$ as a function of $Lu$ along the line of the maximum growth rate. It is evident from this figure that with this scaling all the curves at $Re \geq 4000$ collapse into a single curve for all $Lu$ for a scaling parameter $a\approx -0.6$. Although the curve for $Re=1000$ does not collapse with the other ones, since it is not yet in the truly asymptotic regime with respect to $Re$, it is still reasonably close to them. This figure convincingly shows that the magnetic energy in the saturated state at large $Re$ follows a scaling law, which can be extrapolated to even larger $Re\sim 10^5-10^6$ relevant to experiments (see section B.5) and beyond the scope of capabilities of current numerical simulations.   

Figures \ref{fig:Scal_Sat_Mag_Energy_Vs_Lu}(b)-\ref{fig:Scal_Sat_Mag_Energy_Vs_Lu}(f) provide a broader picture of the overall scaling behavior of $\hat{\mathcal{E}}_{mag}$ with $Re$ as a function of $Lu$ for chosen $Rm=\{9, 14, 20, 25, 30\}$, which generalizes the scalings from Fig. \ref{fig:Scaling_energy}(b) given only for several $(Lu, Rm)$ pairs. In this case, at smaller $Rm=9$ the scaling exponent becomes $a\approx -0.5$ for large $Re \geq 10000$ [Fig. \ref{fig:Scal_Sat_Mag_Energy_Vs_Lu}(b)]. This scaling does not, however, hold well for smaller $Re$, since $Rm$ is also relatively small and the system is close to marginal stability, but it improves a lot with increasing $Rm$ as seen in Figs. \ref{fig:Scal_Sat_Mag_Energy_Vs_Lu}(c)-\ref{fig:Scal_Sat_Mag_Energy_Vs_Lu}(f). Indeed, already for $Rm=14$, $\hat{\mathcal{E}}_{mag}$ well follows this scaling at $a\approx -0.55$ for large $Re \geq 4000$ with minor deviations for $Re=4000$ [Fig. \ref{fig:Scal_Sat_Mag_Energy_Vs_Lu}(c)]. Still, this scaling seems not to hold for lower $Re = 1000$, since it is not yet in the asymptotic regime. On further increasing $Rm=\{20, 25, 30\}$, we see that the scaling laws with $-0.6 \leq a\leq -0.55$ hold remarkably well for $Re \geq 4000$ while the  magnetic energy at $Re=1000$, although somewhat deviating from these scalings, is now reasonably close to them [Figs. \ref{fig:Scal_Sat_Mag_Energy_Vs_Lu}(d)-\ref{fig:Scal_Sat_Mag_Energy_Vs_Lu}(f)].
Thus, the above analysis demonstrates that the saturated magnetic energy plotted as a function of $Lu$ scales well with $Re^{-0.6...-0.5}$ for all $Re \geq 4000$. It should be noted that for a fixed $Rm$, this scaling is the stronger the smaller $Lu$ is.

\begin{figure*}[!htb]
\includegraphics[width=0.266\textwidth]{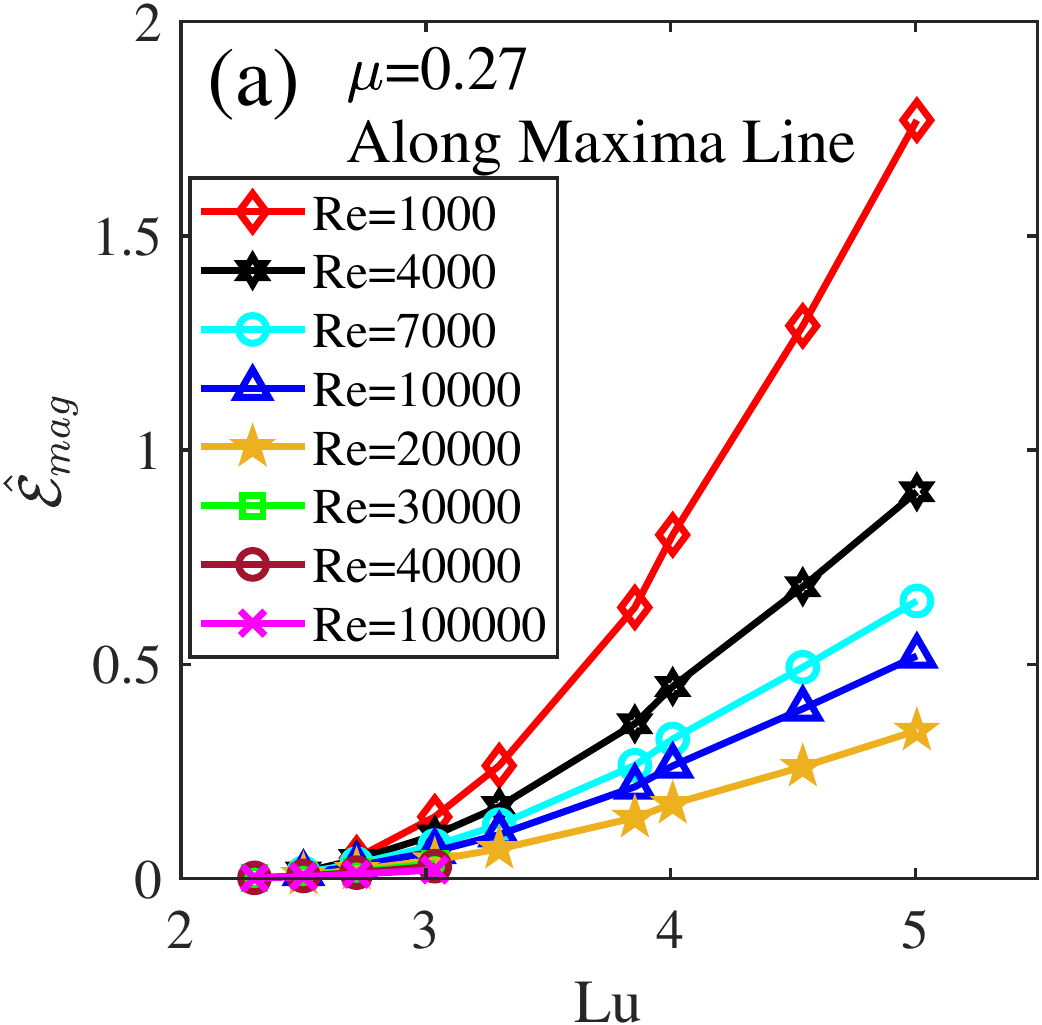}
\hspace{1.2em}
\includegraphics[width=0.26\textwidth]{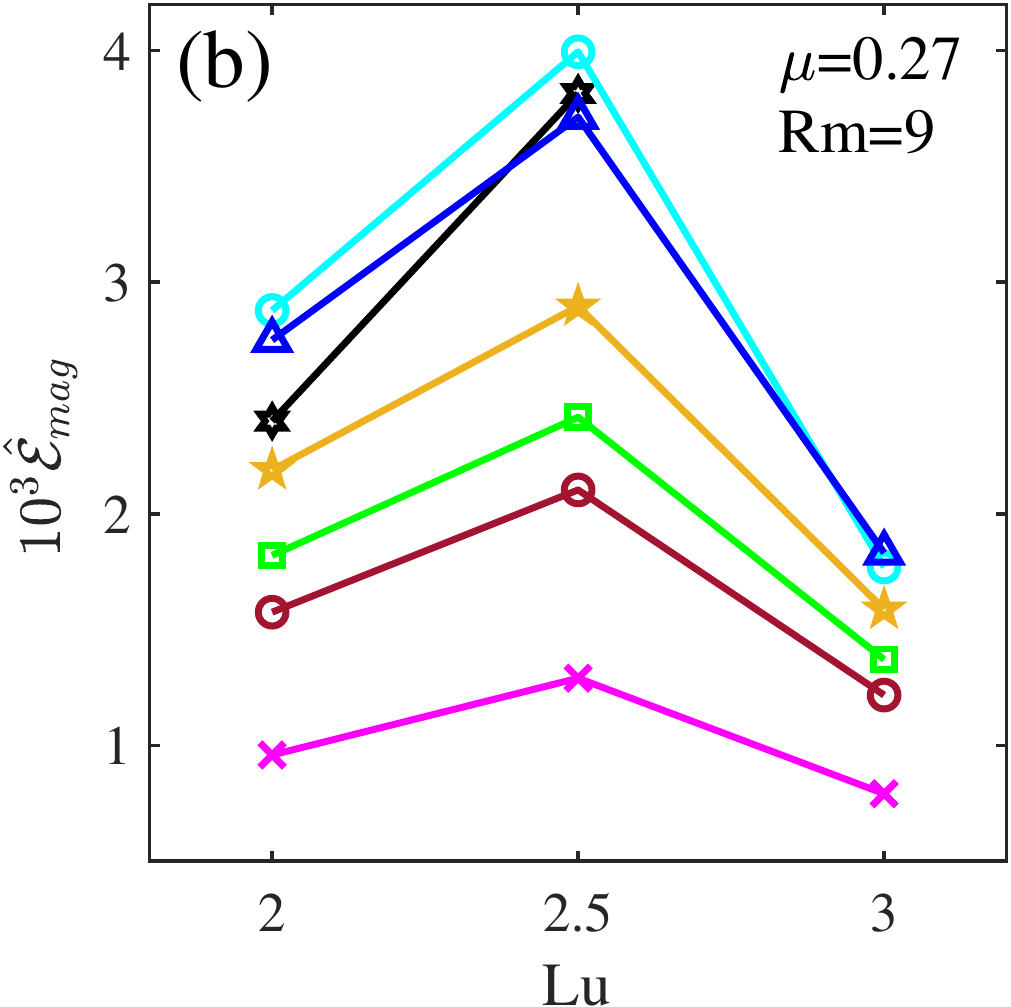}
\hspace{1em}
\includegraphics[width=0.27\textwidth]{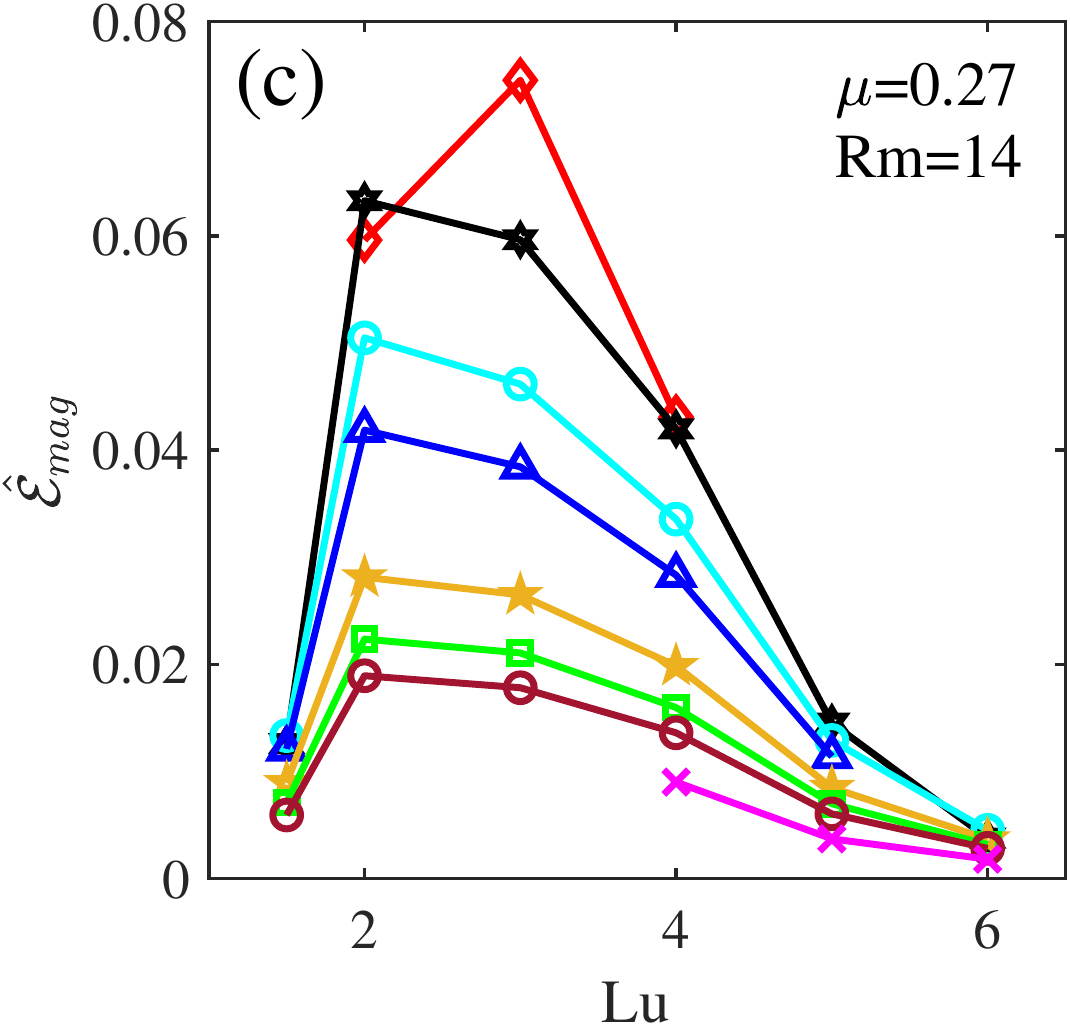}
\hspace{1em}
\includegraphics[width=0.274\textwidth]{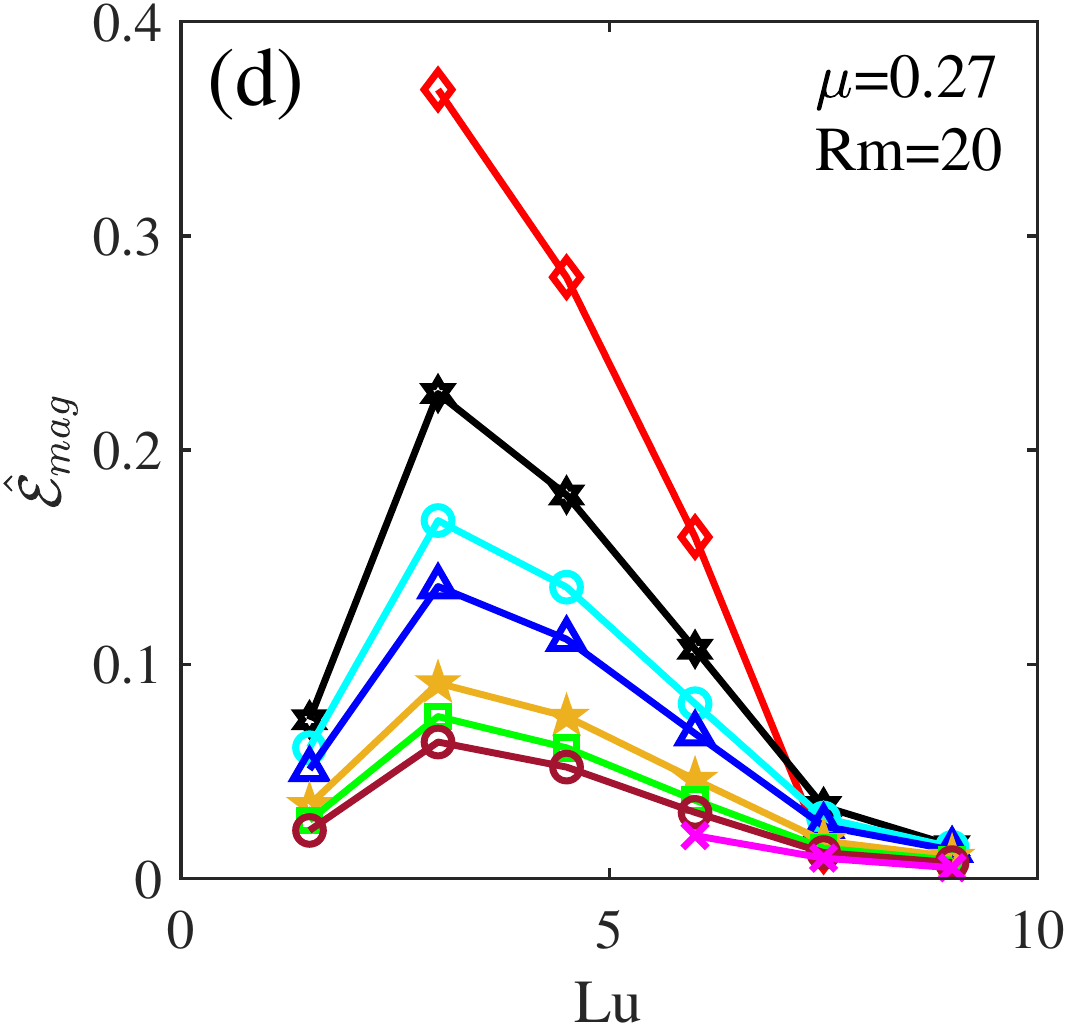}
\hspace{1em}
\includegraphics[width=0.27\textwidth]{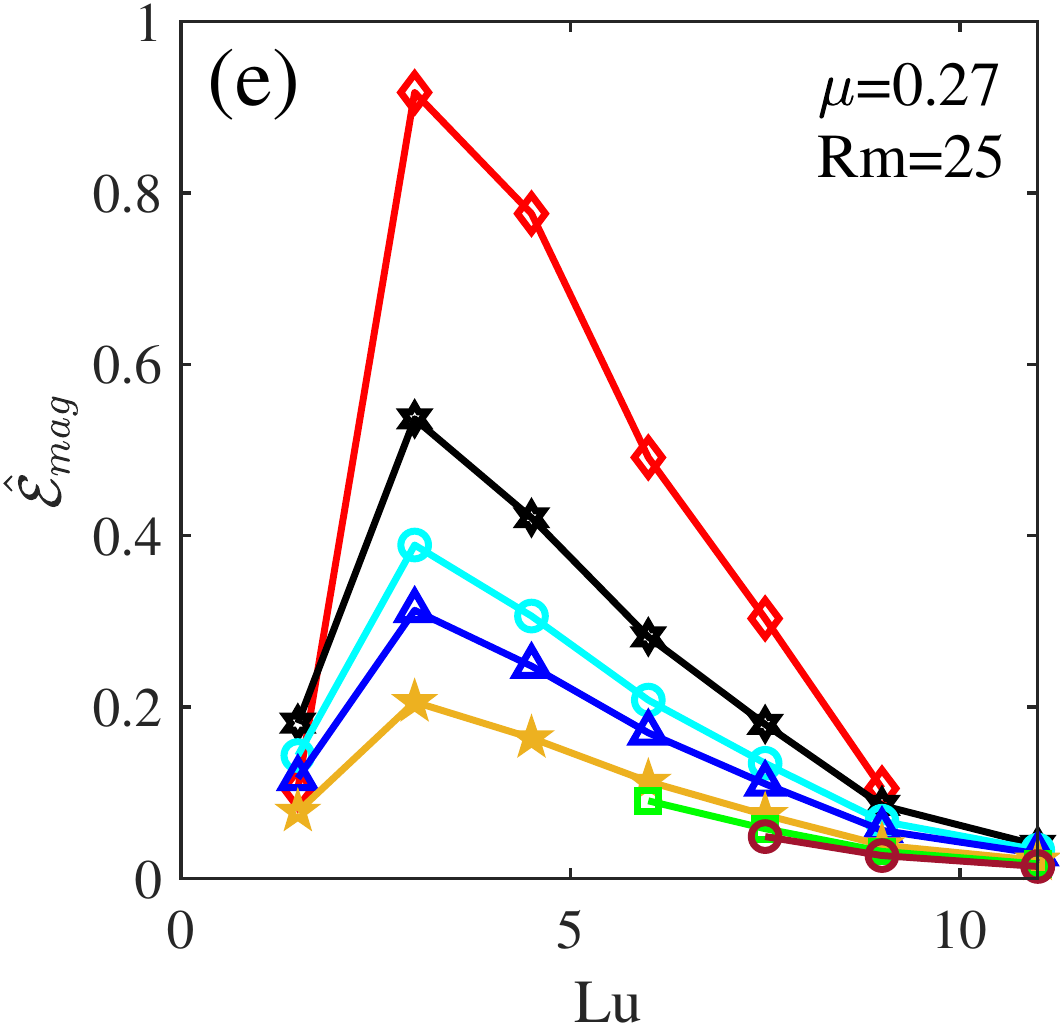}
\hspace{1em}
\includegraphics[width=0.265\textwidth]{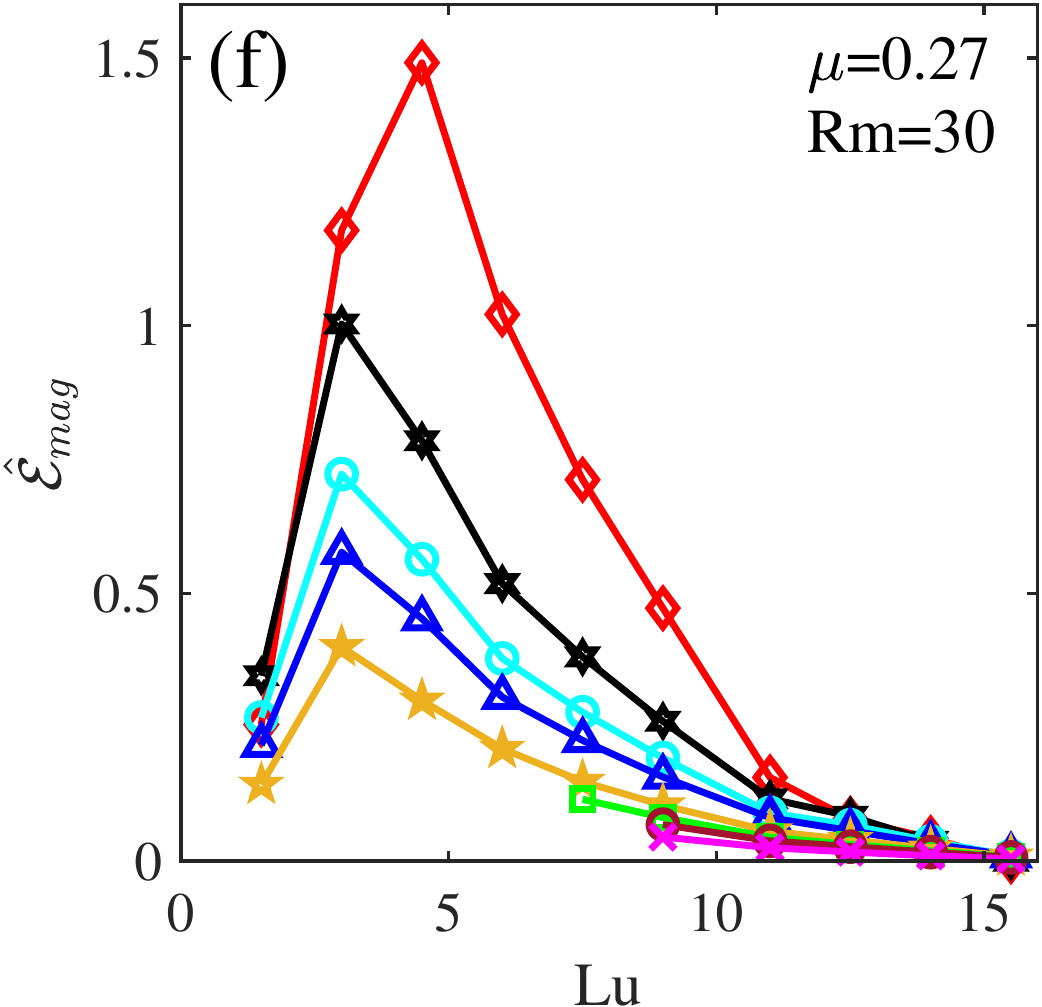}
\caption{Saturated magnetic energy, $\hat{\mathcal{E}}_{mag}$, as a function of $Lu$ for fixed $\mu=0.27$ and different $Re=1000$ (red), 4000 (black), 7000 (cyan), 10000 (blue), 20000 (yellow), 30000 (green), 40000 (brown), 100000 (magenta) (a) along the line of the maximum growth as shown in Fig. \ref{fig:Analysis_points}(b) and for different (b) $Rm=9$, (c) $Rm=14$, (d) $Rm=20$, (e) $Rm=25$, (f) $Rm=30$.}\label{fig:Sat_mag_energy_vs_Lu}
\end{figure*}

\begin{figure*}[!htb]
\hspace{-0.5em}
\includegraphics[width=0.267\textwidth]{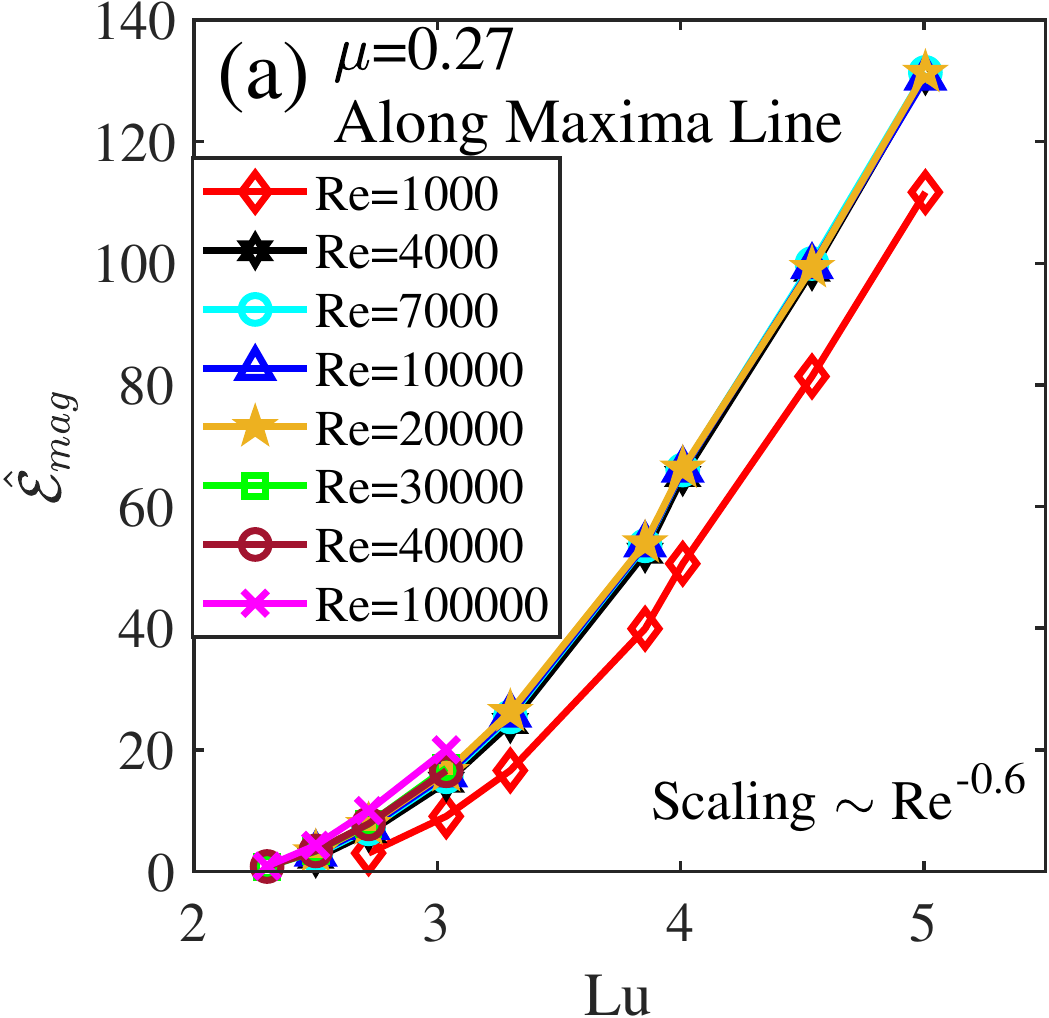}
\hspace{0.4em}
\includegraphics[width=0.275\textwidth]{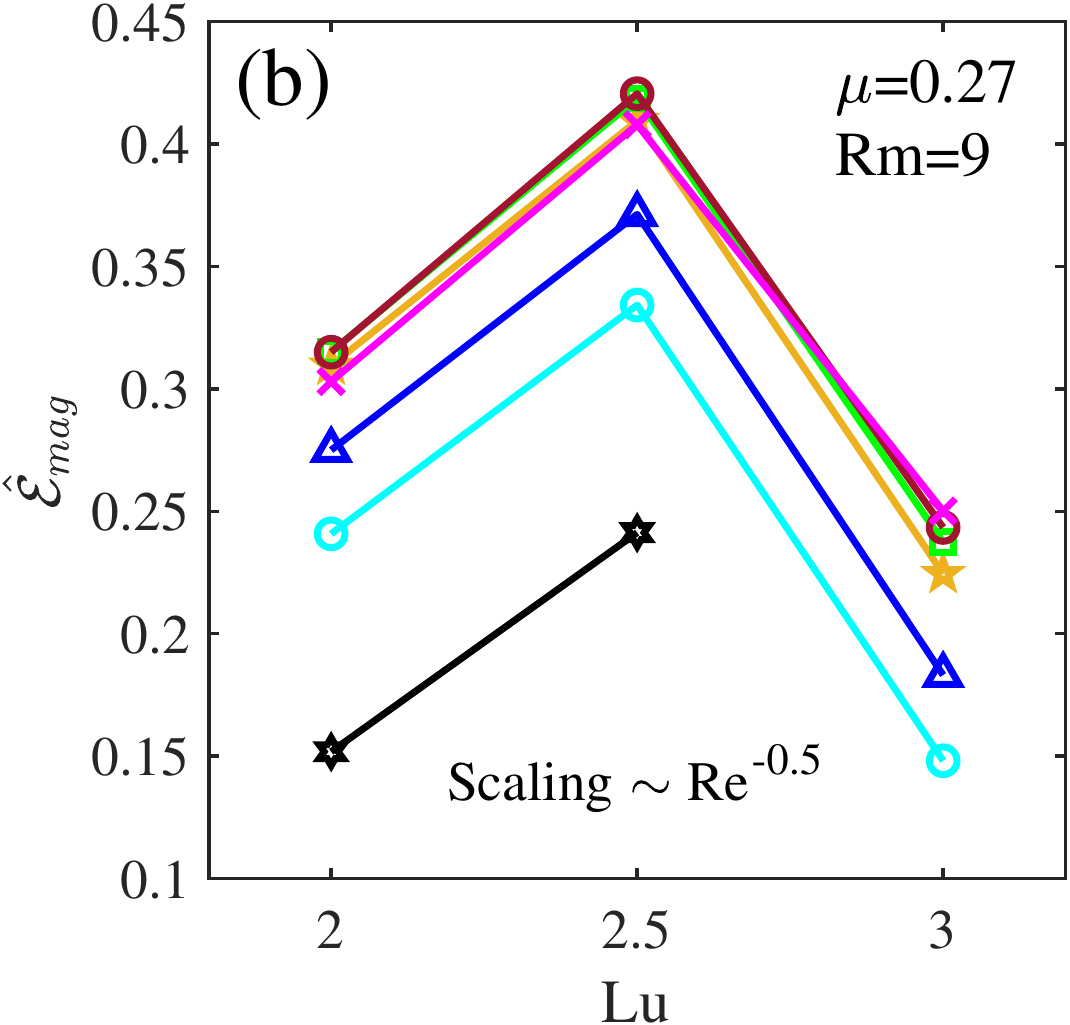}
\hspace{1.8em}
\includegraphics[width=0.257\textwidth]{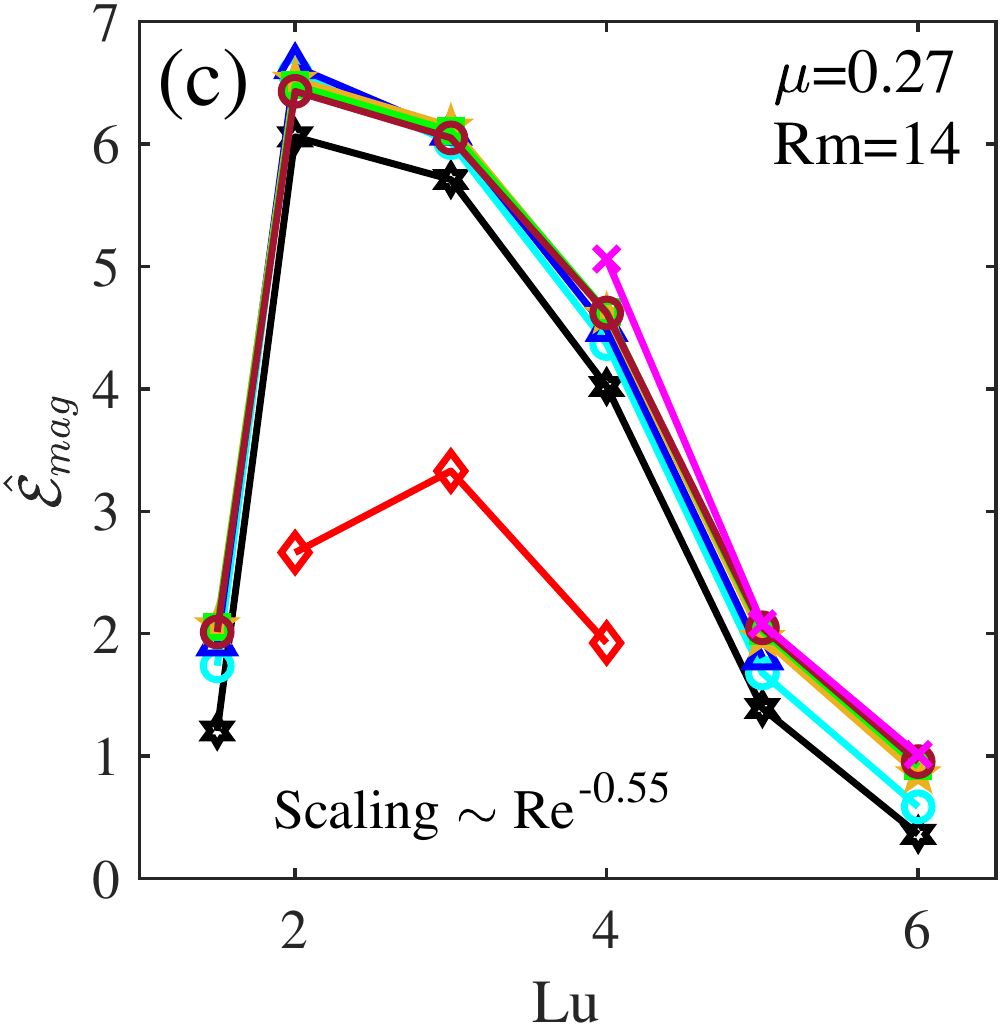}
\includegraphics[width=0.27\textwidth]{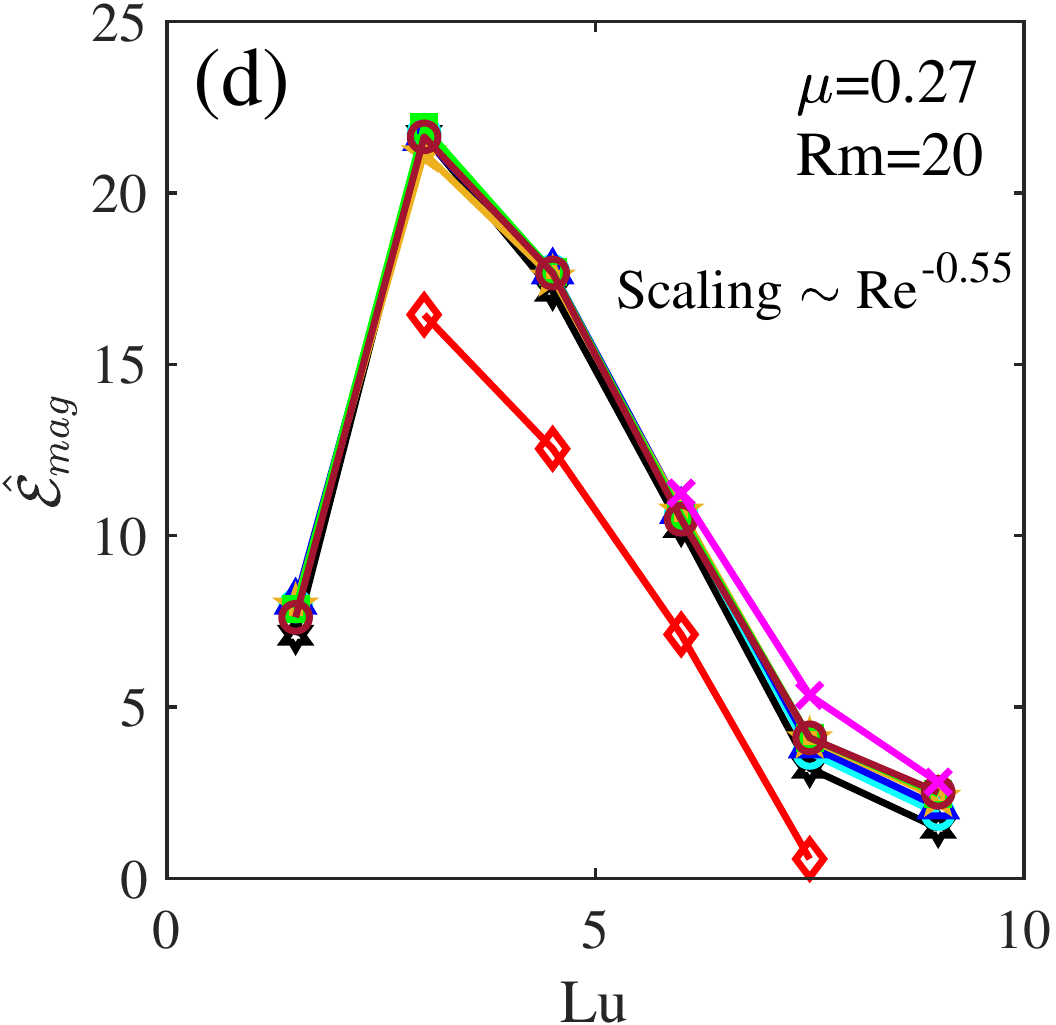}
\hspace{1em}
\includegraphics[width=0.27\textwidth]{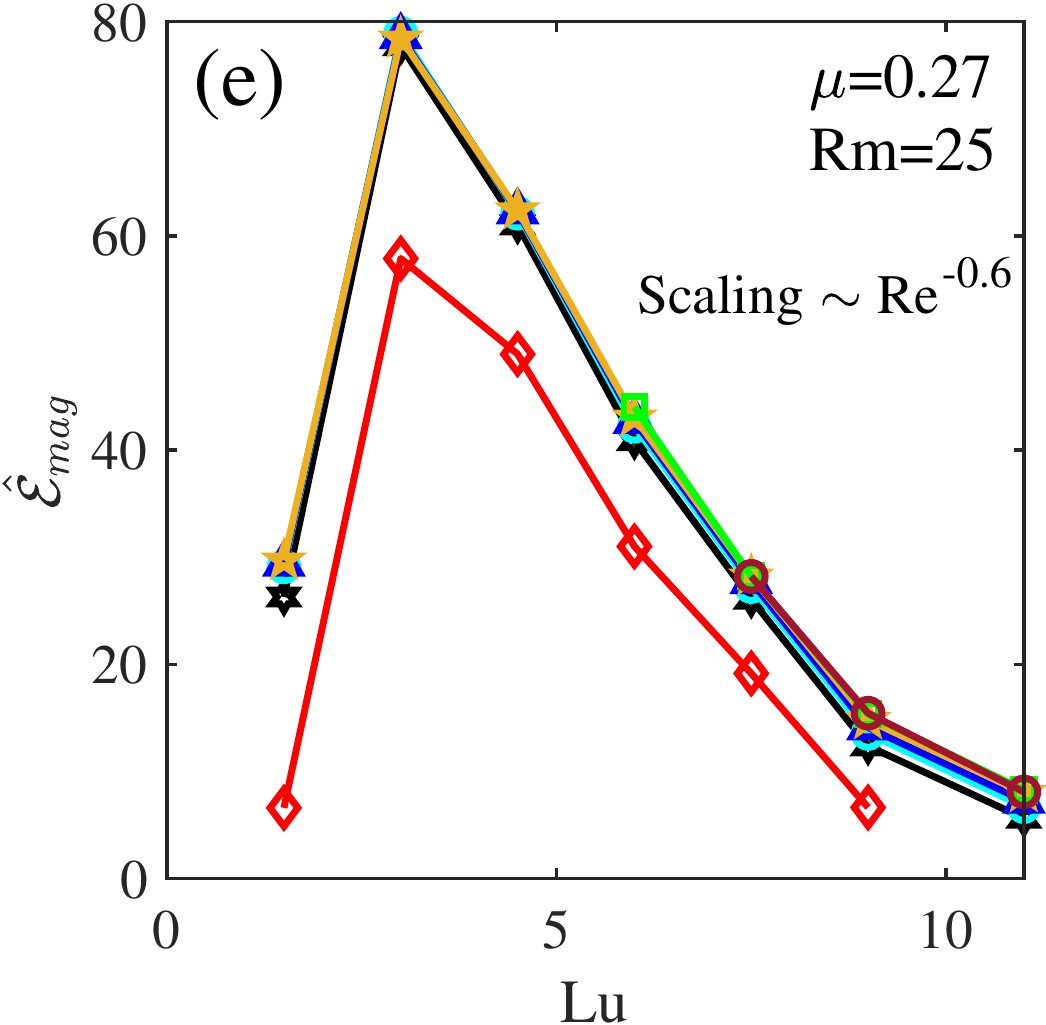}
\hspace{1em}
\includegraphics[width=0.27\textwidth]{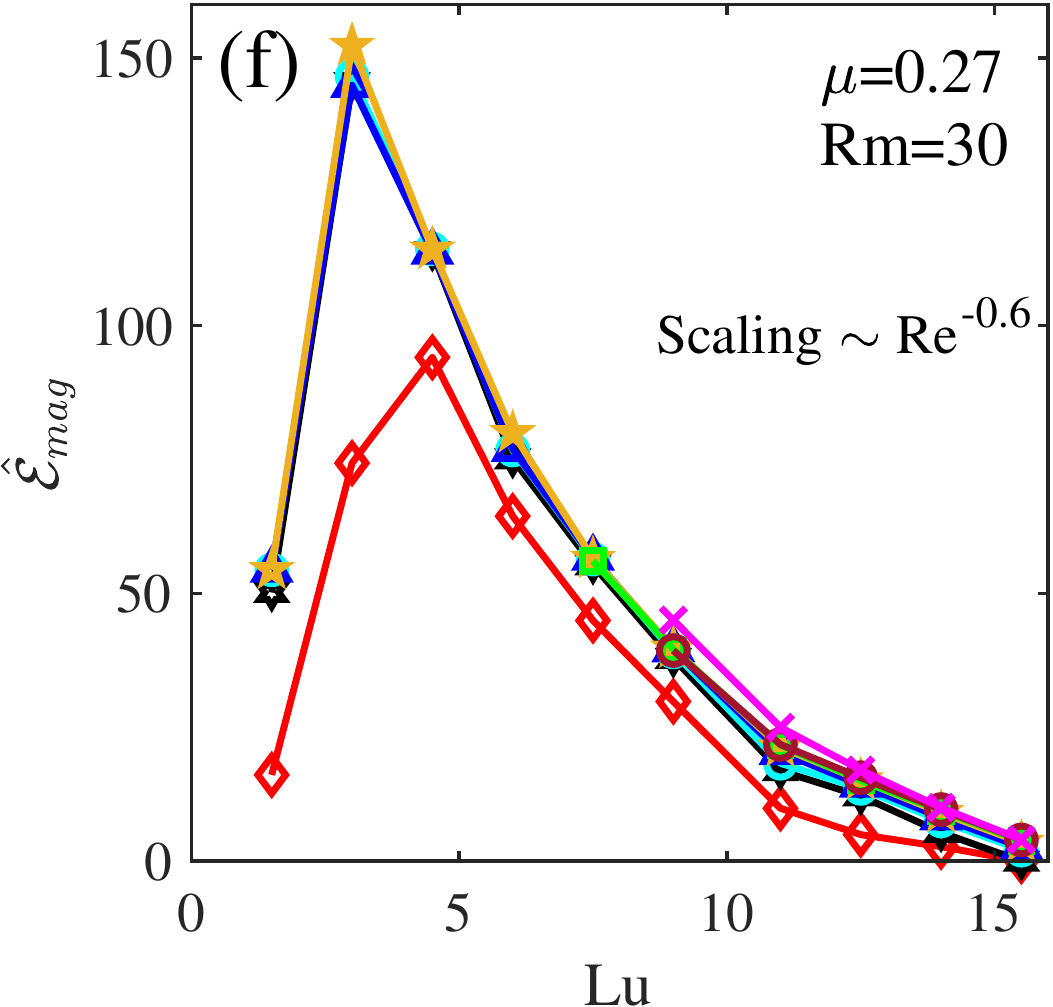}
\caption{Same as in Fig. \ref{fig:Sat_mag_energy_vs_Lu} except that the curves are scaled with the power law $Re^a$, where $a$ is the scaling exponent optimally found for each $Rm$ in (a)-(f) panels. As a result of this scaling, all the curves for $Re \geq 4000$ and $Rm \geq 14$ well collapse compared to those in Fig. \ref{fig:Sat_mag_energy_vs_Lu}.}
\label{fig:Scal_Sat_Mag_Energy_Vs_Lu}
\end{figure*}

\subsubsection{Angular Momentum Transport Evolution and Scaling Behavior}

Radial transport of angular momentum in a TC flow is mediated via torque. In this setup, there are three types of torque: advective torque due to Reynolds stress, magnetic torque due to Maxwell stress (Lorentz force) and viscous torque due to azimuthal velocity shear. As a result of the adopted no-slip and insulating boundary conditions, the volume integrated advective and magnetic torques vanish at the boundaries and thus do not contribute to the net transport of angular momentum. Hence, the change of the total angular momentum of the flow in the domain, $L=\int \rho ru_{\phi}dV$, with time is governed only by viscous torques $G_{in}$ and $G_{out}$ at the inner and out cylinders, respectively, which are given by \cite{Mamatsashvili_etal2018,Guseva_etal_2017ApJ}
\begin{equation}\label{ang_moment_from_torque}
\frac{dL}{dt}=G_{in}-G_{out}
\end{equation}
\noindent
and 
\begin{equation}\label{torque_at_cylinders}
G_{in, out}= -\frac{r_{in, out}^3}{Re}\int_0^{2\pi}\int_0^{L_z} \frac{d}{dr}\Big(\frac{u_\phi}{r} \Big)|_{r=r_{in, out}} d\phi dz.
\end{equation}
For the basic TC flow, torque at the inner and outer cylinders is the same $G_{lam}=-(2\pi L_z/Re)r_{in}^3d\Omega/dr|_{r=r_{in}}$ and is used here to normalize the total viscous torque, $G/G_{lam}$, in order to characterize the effective angular momentum transport in the nonlinear state.

\begin{figure*}
\centering
\includegraphics[width=0.44\textwidth]{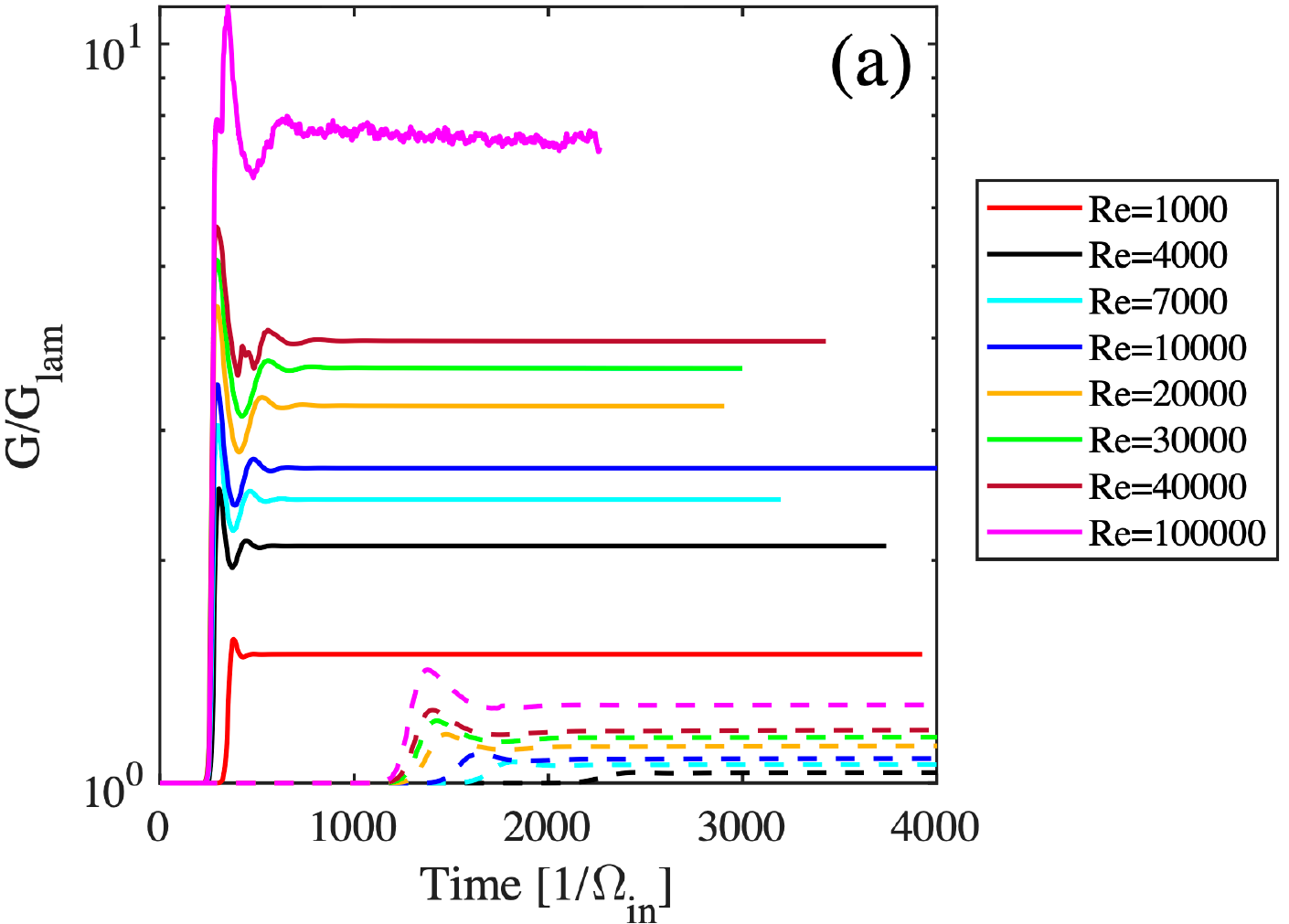}
\includegraphics[width=0.5\textwidth]{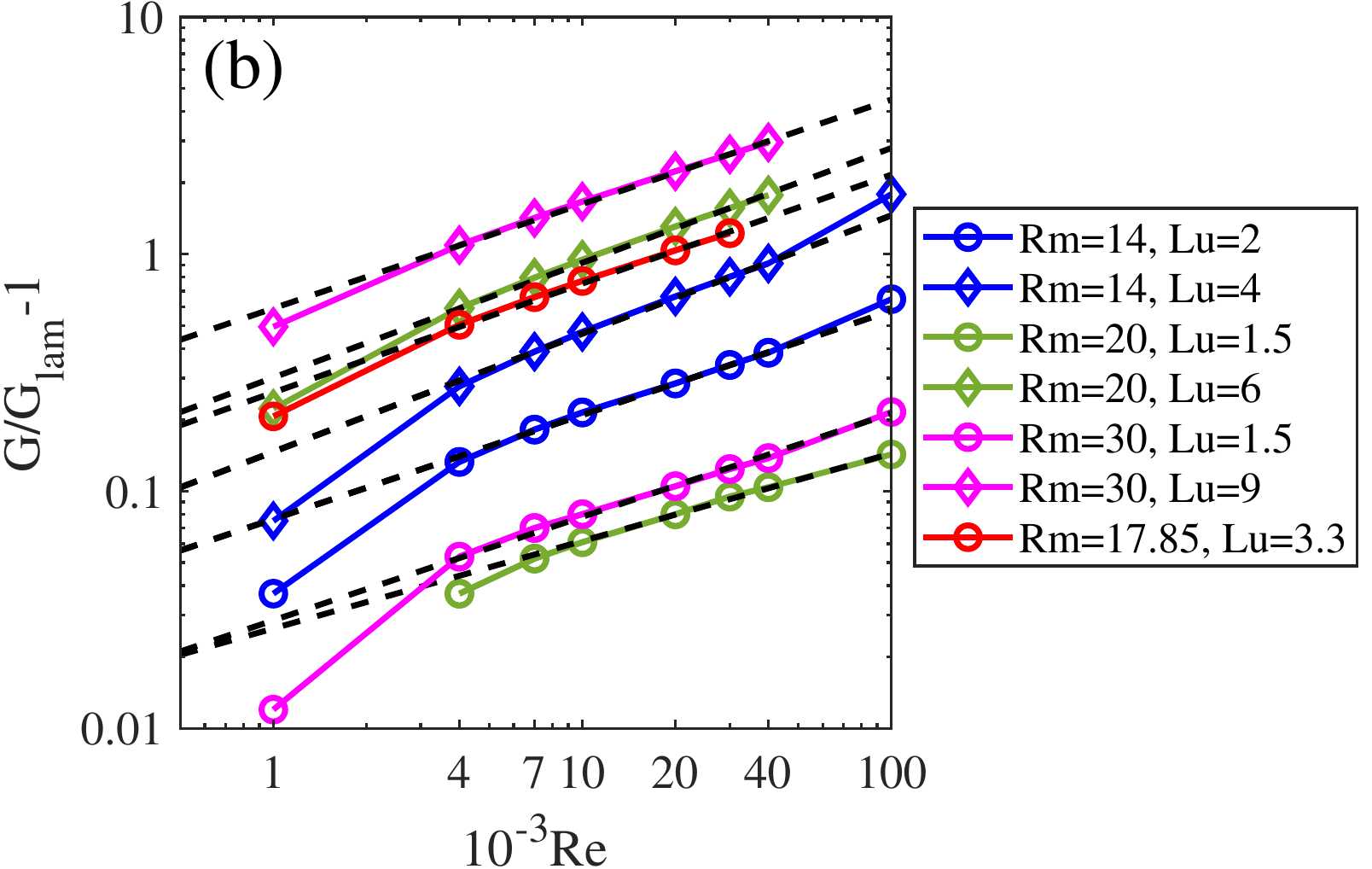}
\caption{(a) Evolution of the normalized torque $G/G_{lam}$ for different $Re$ and fixed $\mu=0.27$, $(Lu, Rm) = (2.5, 9)$ (dashed) and $(9, 30)$ (solid). (b) Normalized perturbation torque $G/G_{lam}-1$ in the saturated state as a function of $Re$ for several pairs of $(Lu, Rm)$. Black dashed lines show fitting with power law $Re^b$ with the average scaling exponent $b\approx 0.5$.}\label{fig:torque_energy}
\end{figure*}

Figure \ref{fig:torque_energy}(a) shows the time evolution of the normalized torque $G/G_{lam}$ measured at the inner cylinder for two pairs of $(Lu, Rm)=(9, 30)$ (solid) and $(2.5, 9)$ (dashed) and different Reynolds numbers $Re$. Since the torque measured in the simulations is normalized using the laminar torque, the minimum torque is equal to 1. It is seen in Fig. \ref{fig:torque_energy}(a) that, in contrast to the magnetic energy, $G/G_{lam}$ increases with $Re$ both during the exponential growth at the linear stage and in the saturated state. During the evolution, the torque at the inner cylinder $G_{in}$ increases the angular momentum of the flow while the torque at outer cylinder $G_{out}$ decreases it (Eqs. \ref{ang_moment_from_torque} and \ref{torque_at_cylinders}), eventually reaching an equilibrium in the saturated steady state when both torques become equal, $G_{in}\approx G_{out}$, on average in time.

As discussed above, the normalized torque $G/G_{lam}$ characterizes the effective angular momentum transport in the flow. We can obtain the contribution from perturbations to this torque by subtracting the equilibrium (laminar) one, $G/G_{lam} - 1$. This normalized torque solely due to perturbations is used everywhere below and referred to as ``torque'' for short. Figure \ref{fig:torque_energy}(b) shows $G/G_{lam} - 1$ at the cylinders in the saturated state as a function of $Re$ for different $Lu$ and $Rm$. The black dashed lines show the power law fitting of the form $Re^{b}$ where $b$ is the scaling exponent. Similar to the saturated magnetic energy, the torque also forms a family of parallel lines for different ($Lu, Rm$) but unlike the former, it increases with  increasing $Re$. The fitted power law exponent $b\approx 0.5$ indicates that $G/G_{lam} - 1 \propto Re^{0.5}$, that is, $\propto Pm^{-0.5}$ as $Rm$ is fixed. This is consistent with the similar $Re^{0.5}$ scaling of the torque for SMRI at large $Re$ and $Rm$ found by Liu et al. \cite{Liu_Goodman_Ji_NonlinearMRI_2006ApJ}.
   
Figure \ref{fig:Torque_Sat_vs_Lu} shows the torque in the saturated state as a function of $Lu$ for the same values of $Rm$ and $Re$ as for the saturated magnetic energy above (Fig. \ref{fig:Sat_mag_energy_vs_Lu}). 
Figure \ref{fig:Torque_Sat_vs_Lu}(a) shows its almost linear increase with $Lu$ along the line of the maximum growth at different $Re$. Figures \ref{fig:Torque_Sat_vs_Lu}(b)-\ref{fig:Torque_Sat_vs_Lu}(f) show that for fixed $Rm$ and $Re$, $G/G_{lam} - 1$ first increases with $Lu$, reaches a peak for a certain $Lu$ and then decreases gradually, while increasing with $Re$ at all $Lu$. This behavior is in contrast to that of the magnetic energy $\hat{\mathcal{E}}_{mag}$ in Fig. \ref{fig:Sat_mag_energy_vs_Lu} and implies that for fixed $Rm$ an increase in $Re$ (decrease in $Pm$) leads to the prevalence of velocity perturbations over magnetic ones and hence to an increase of the torque on the cylinders, which is due to the azimuthal velocity shear (Eq. \ref{torque_at_cylinders}).
    
After showing the scaling of the torque as a function of $Re$ for several pairs of $(Lu, Rm)$ in Fig. \ref{fig:torque_energy}(b) above, we take now a more general approach and show it as a function of $Lu$ optimally scaled again with the power law $Re^b$ in Fig. \ref{fig:Torque_Scaling}. In Fig. \ref{fig:Torque_Scaling}(a) we first plot $G/G_{lam} - 1$ in the saturated state as a function of $Lu$ along the line of the maximum growth for different $Re$. It exhibits a good scaling with $b\approx 0.4$ for all $Re \geq 4000$. In a similar manner, we further demonstrate the scaling of the torque with $Re$ as a function of $Lu$ for each value of $Rm$ in Figs. \ref{fig:Torque_Scaling}(b)-\ref{fig:Torque_Scaling}(e), respect. For $Rm=9$ the numerically obtained scaling exponent becomes $b\approx 0.5$ [Fig. \ref{fig:Torque_Scaling}(b)]. For larger $Re$ and smaller $Lu$, this scaling is stronger than for larger $Lu$ and smaller $Re$, which can be attributed to weak growth of SMRI and correspondingly very small saturation levels of the torque in the latter case at small $Rm$. At higher $Rm=14$, the scaling parameter is $b\approx 0.5$ [Fig. \ref{fig:Torque_Scaling}(c)]. Since now $Rm$ is relatively large, implying stronger growth of the instability and therefore higher saturation level, the scaling is better than that for $Rm=9$. Still, this scaling is stronger for smaller $Lu\lesssim 4$ and gets a bit weaker for larger $Lu$. A similar trend is seen for larger $Rm$ in Figs. \ref{fig:Torque_Scaling}(d)-\ref{fig:Torque_Scaling}(f) and the obtained scaling exponent  $b\approx 0.45$ is slightly smaller  than those for $Rm=9$ and $14$. Nevertheless, the scaling is overall stronger for larger $Rm$. 

\begin{figure*}
\centering
\includegraphics[width=0.265\textwidth]{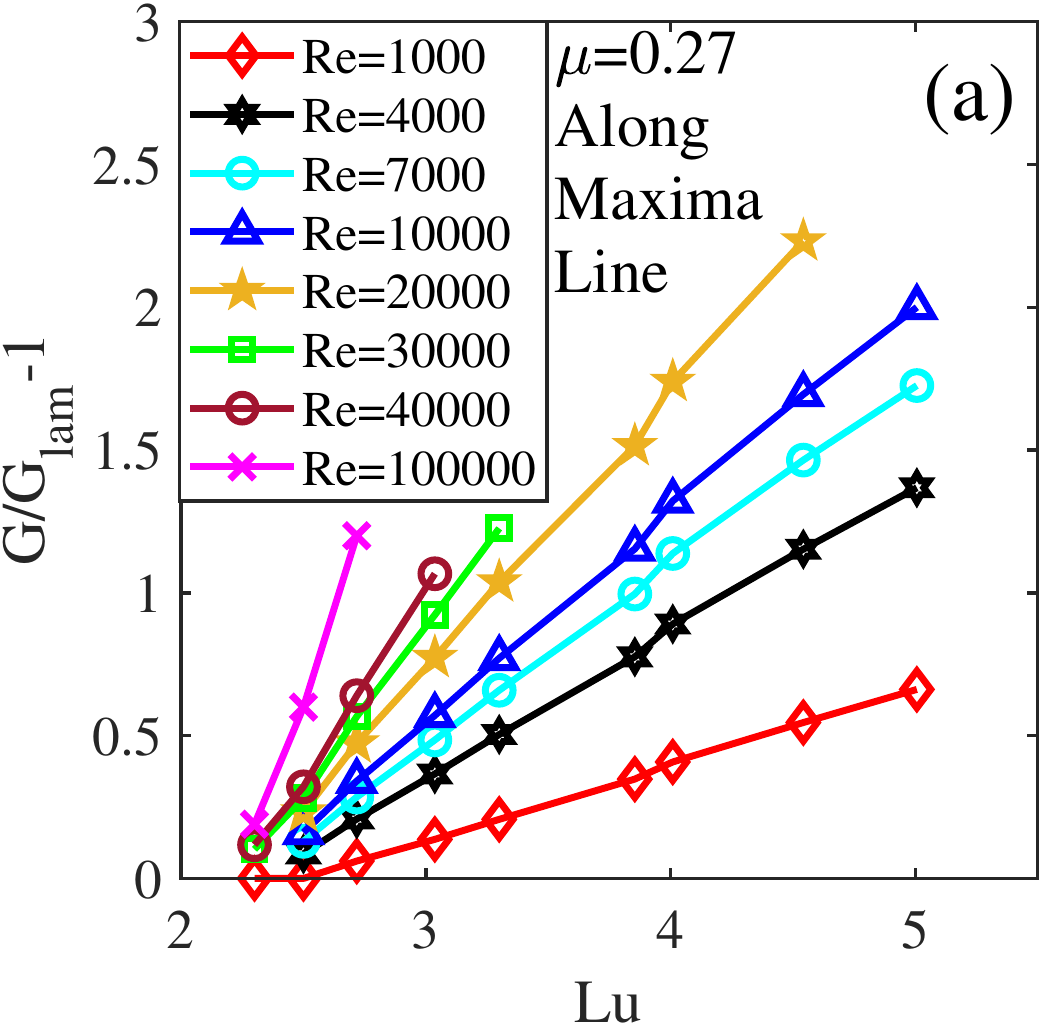}
\hspace{0.6em}
\includegraphics[width=0.27\textwidth]{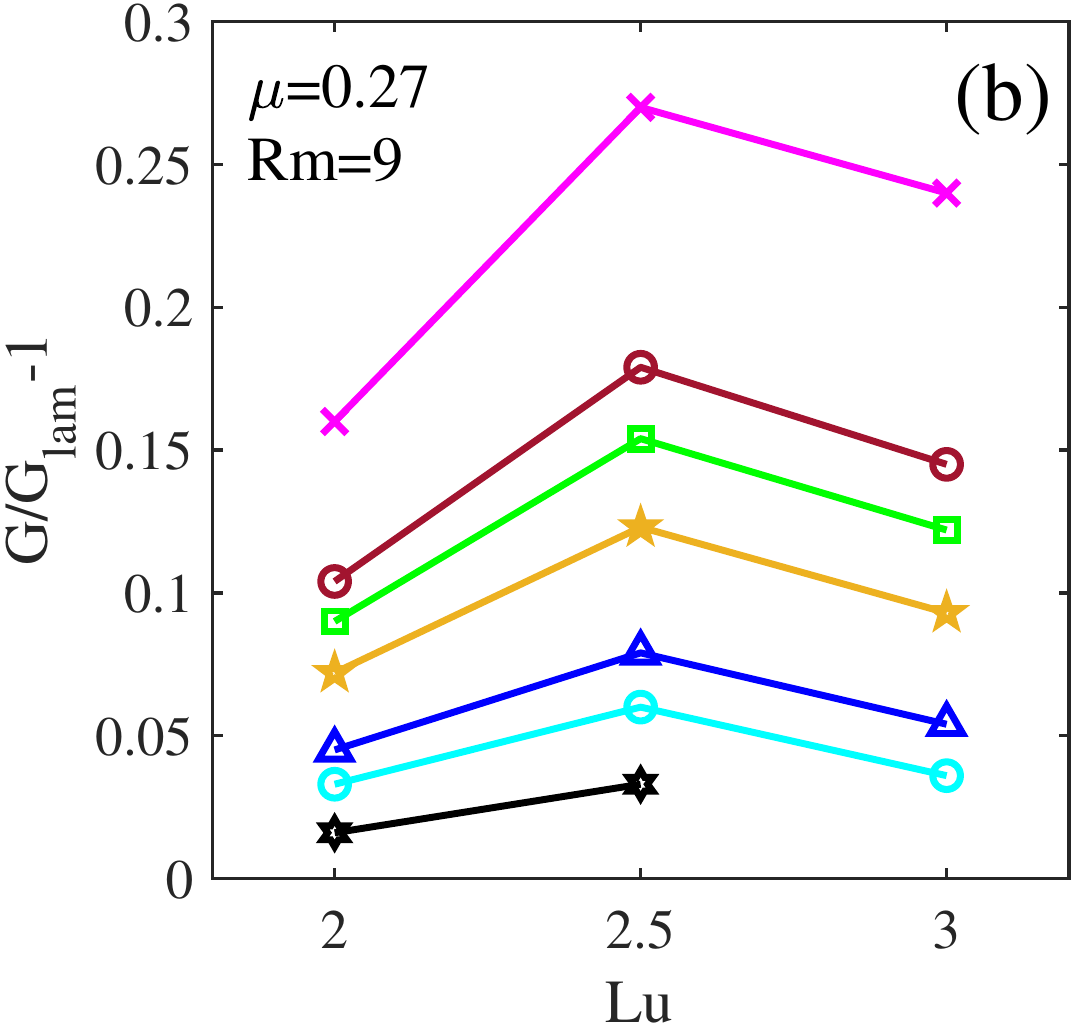}
\hspace{1em}
\includegraphics[width=0.26\textwidth]{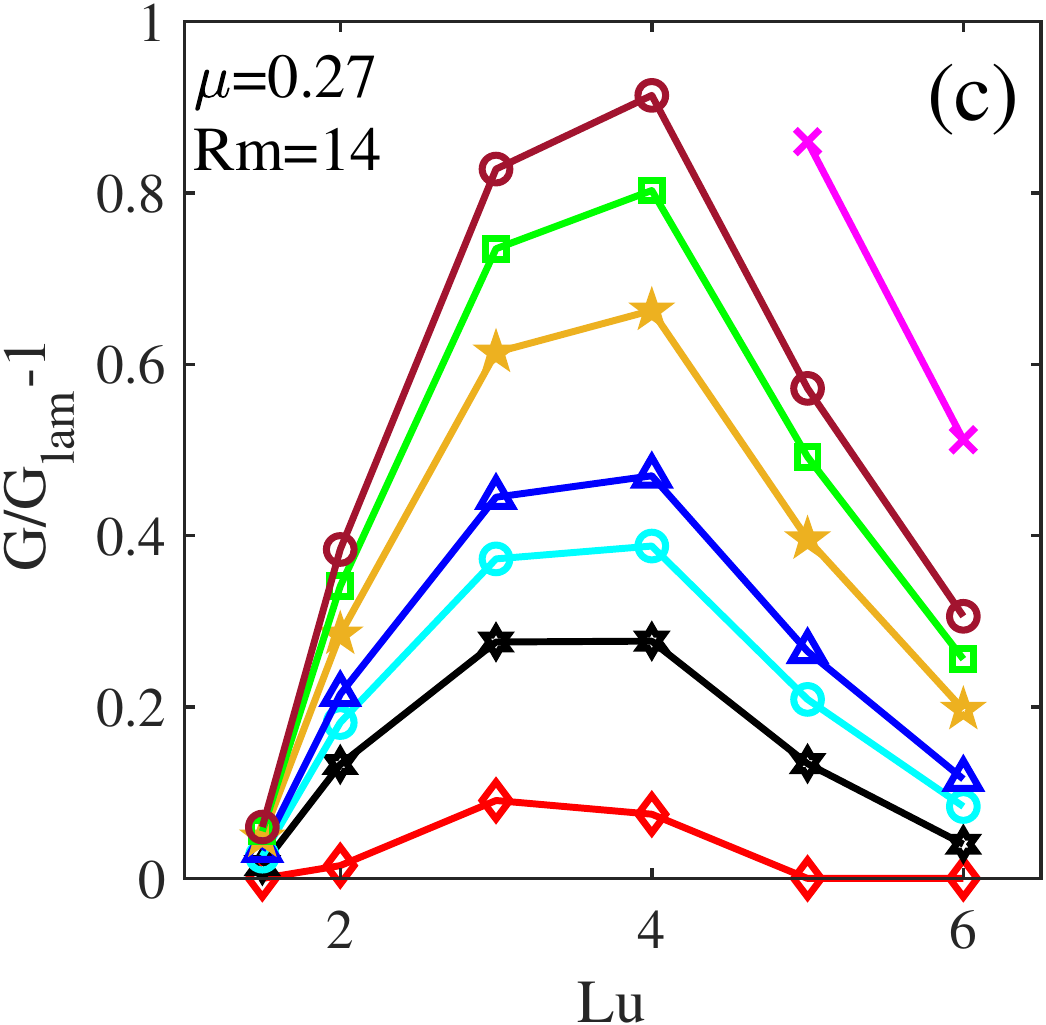}
\hspace{1em}
\includegraphics[width=0.265\textwidth]{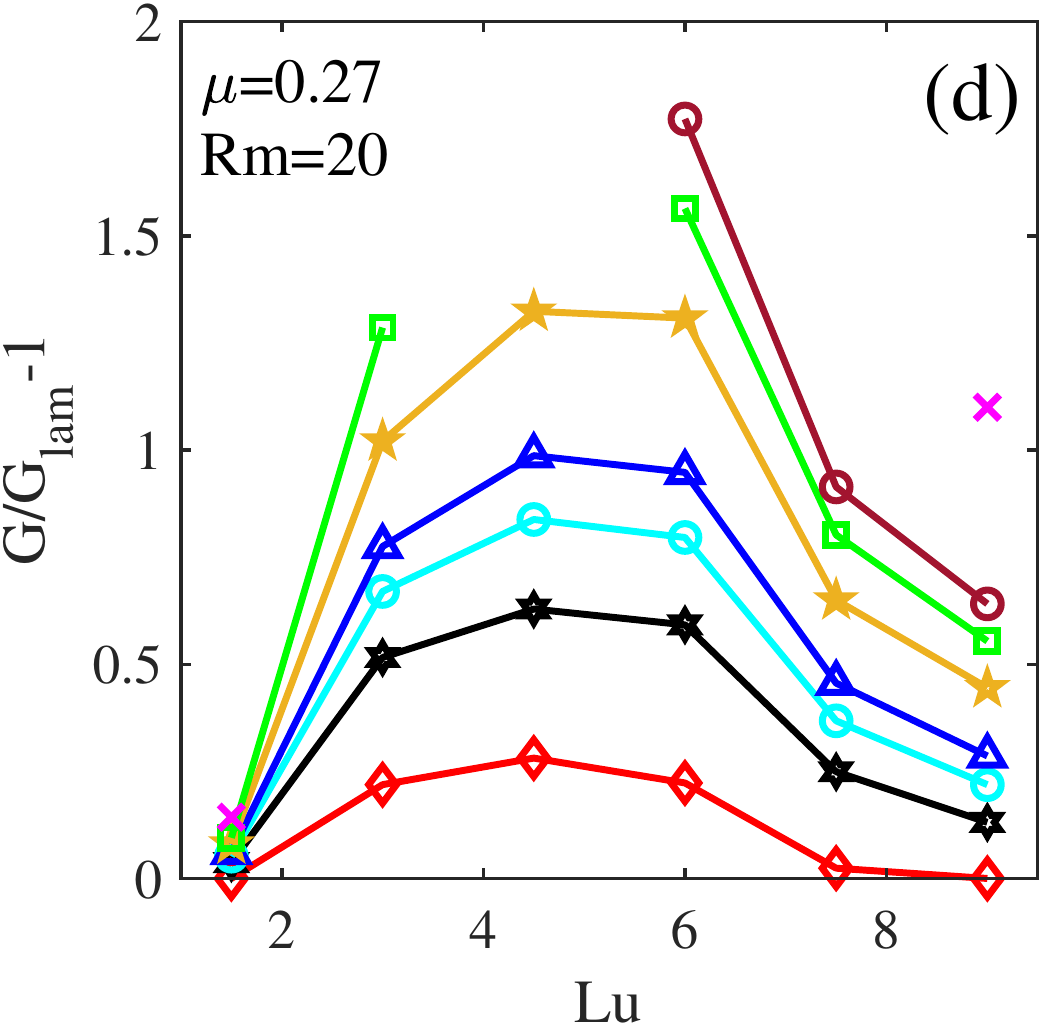}
\hspace{1em}
\includegraphics[width=0.27\textwidth]{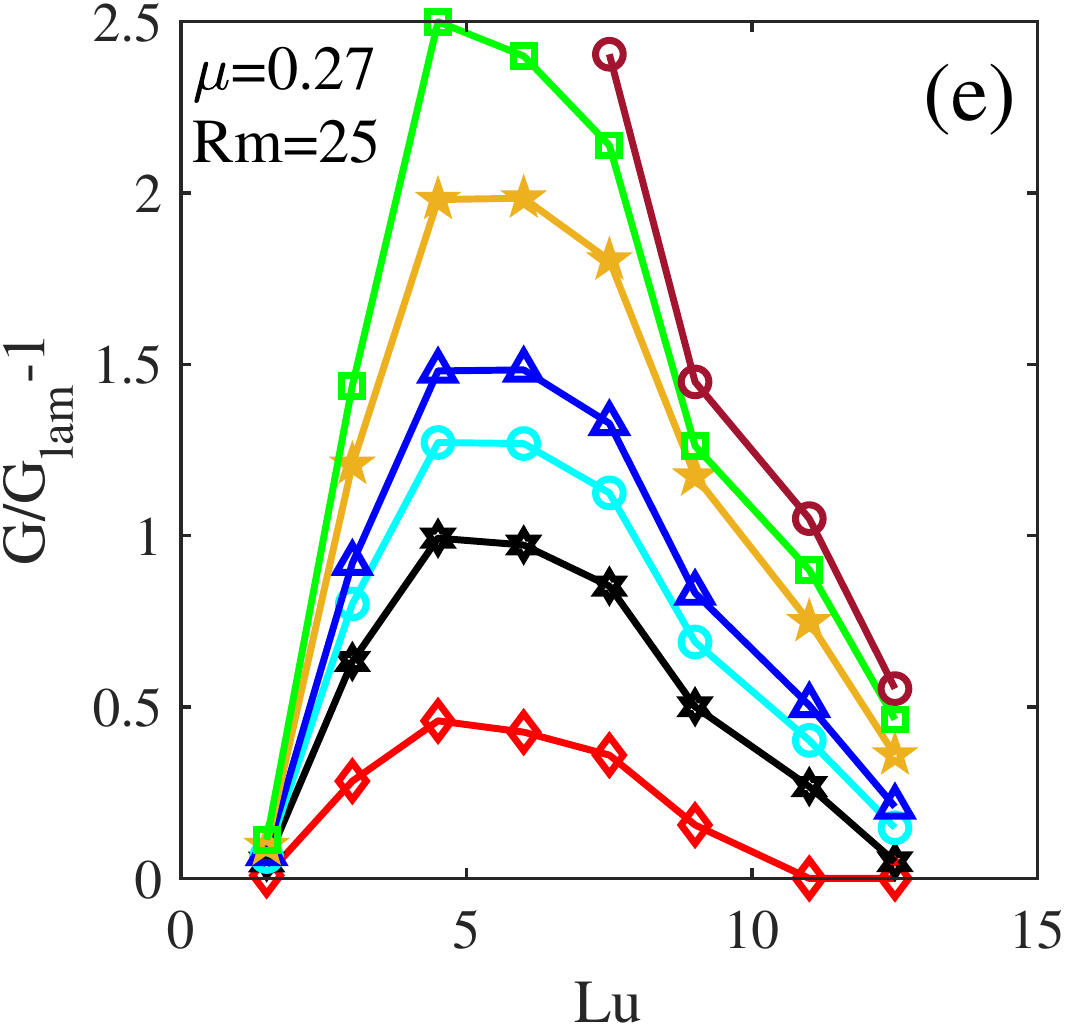}
\hspace{1em}
\includegraphics[width=0.26\textwidth]{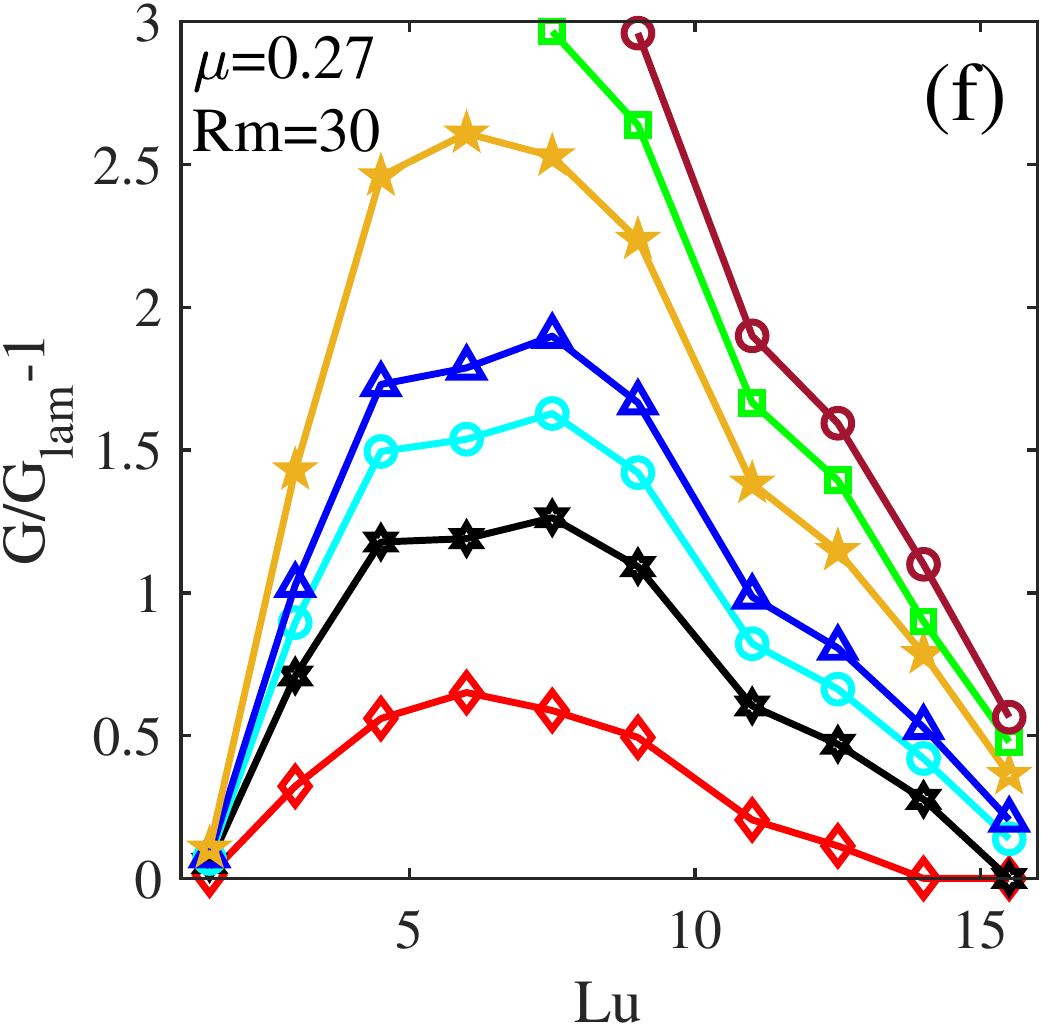}
\caption{The torque $G/G_{lam}-1$ as a function of $Lu$ for fixed $\mu=0.27$ and different $Re=1000$ (red), 4000 (black), 7000 (cyan), 10000 (blue), 20000 (yellow), 30000 (green), 40000 (brown), 100000 (magenta) (a) along the line of the largest growth as well as for different (b) $Rm=9$, (c) $Rm=14$, (d) $Rm=20$, (e) $Rm=25$, (f) $Rm=30$.}
\label{fig:Torque_Sat_vs_Lu}
\end{figure*}

\begin{figure*}
\centering
\includegraphics[width=0.275\textwidth]{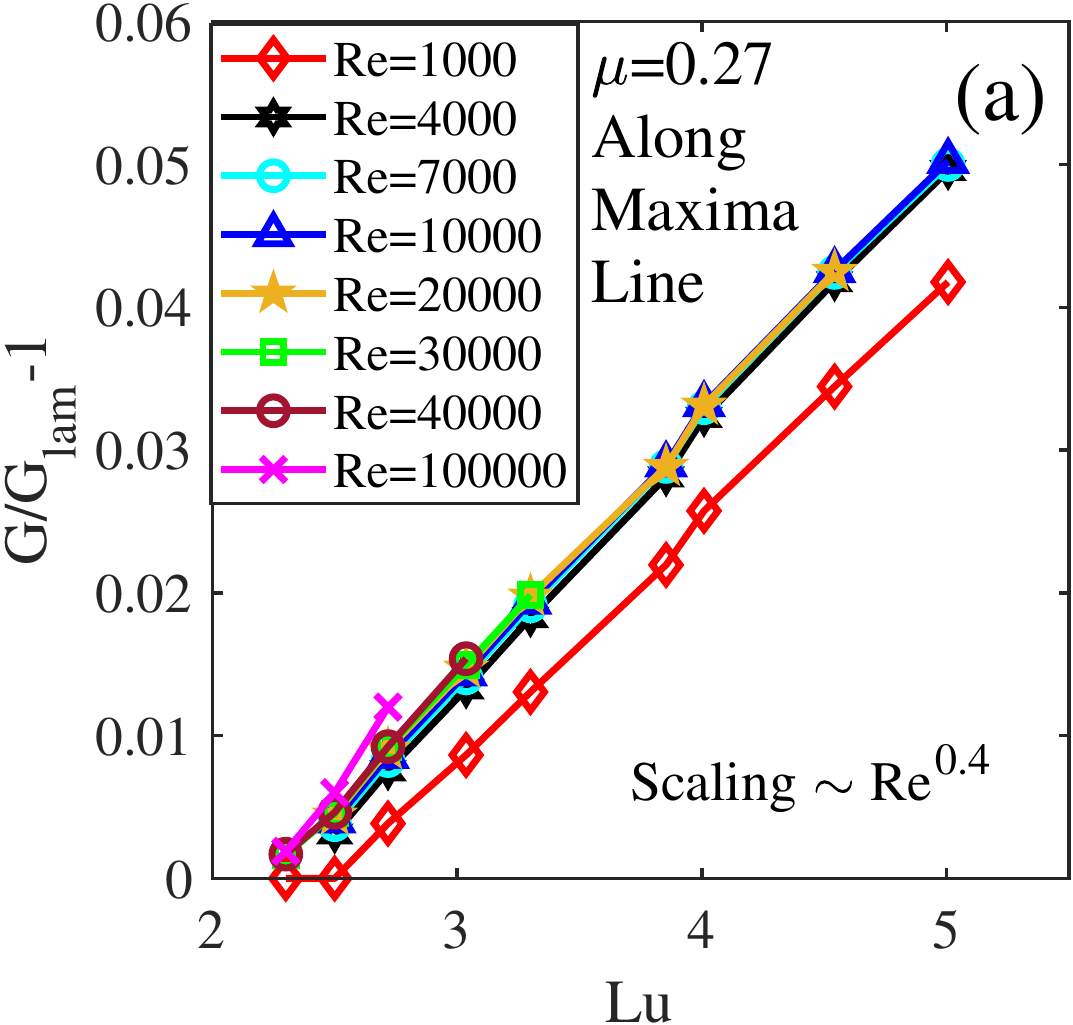}
\hspace{0.5em}
\includegraphics[width=0.27\textwidth]{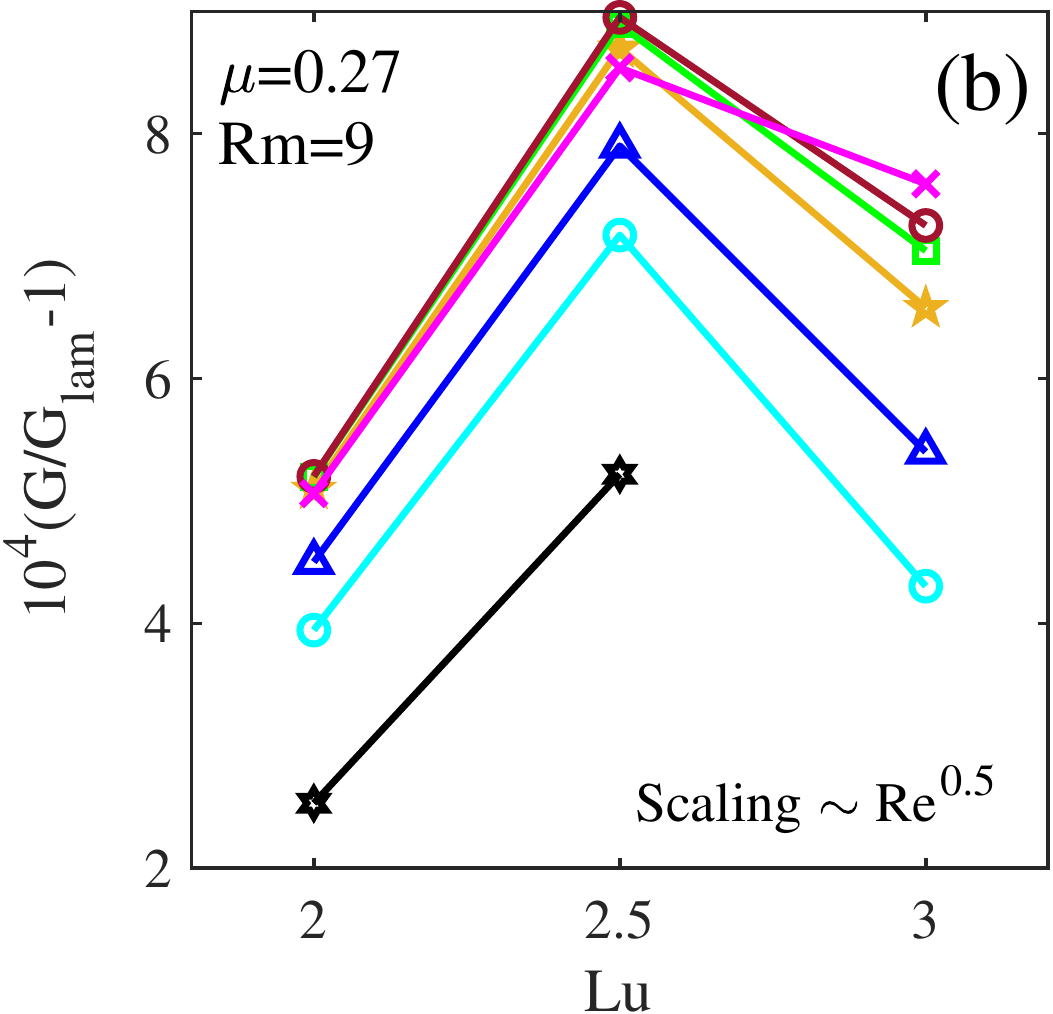}
\hspace{1.5em}
\includegraphics[width=0.26\textwidth]{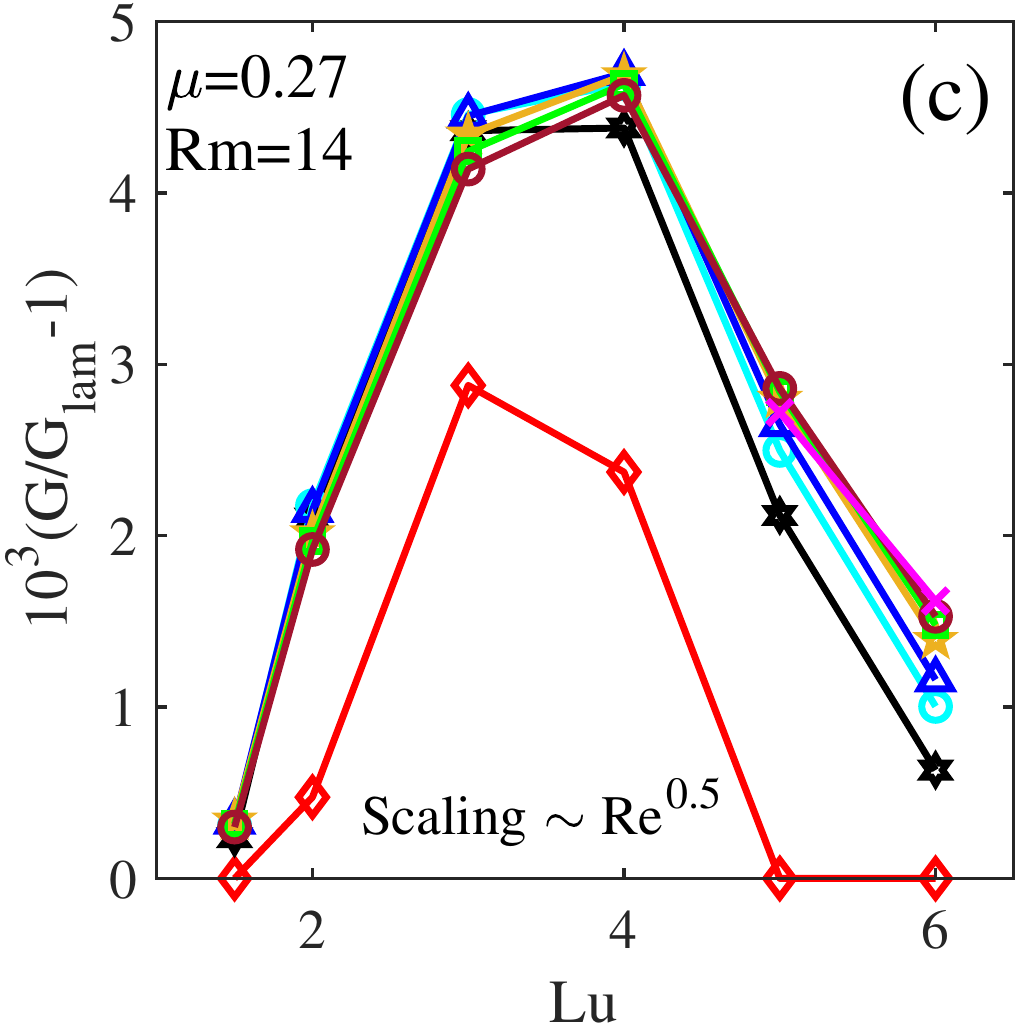}
\hspace{1em}
\includegraphics[width=0.27\textwidth]{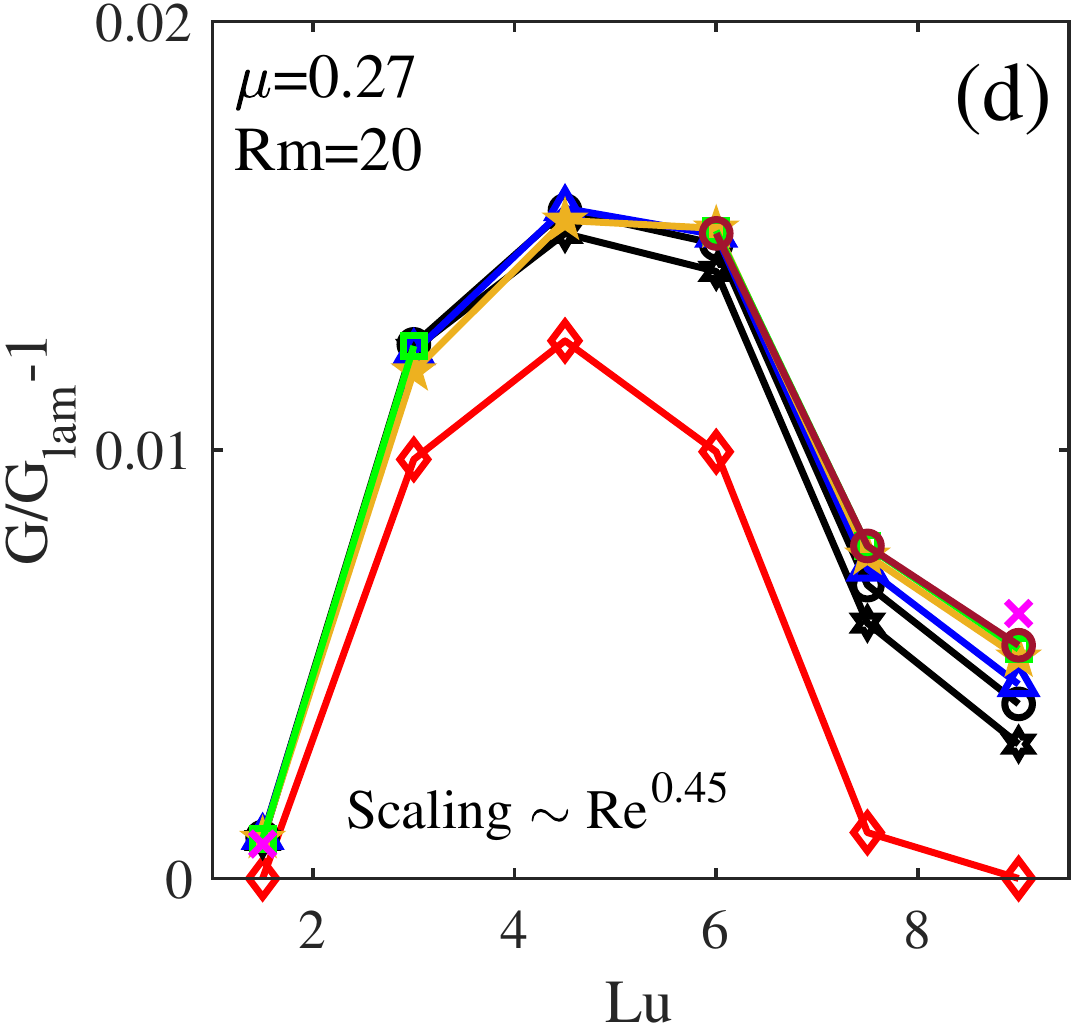}
\hspace{1em}
\includegraphics[width=0.275\textwidth]{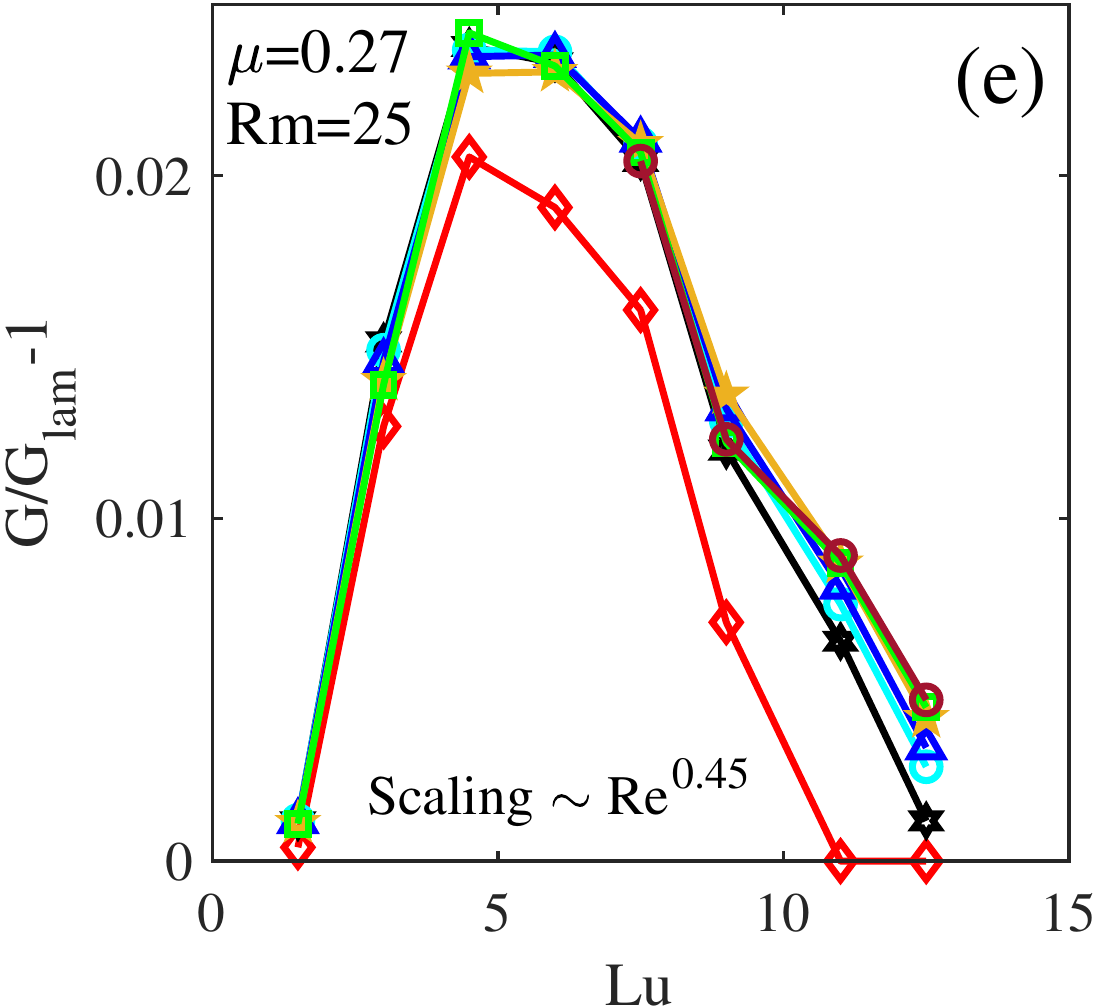}
\hspace{1em}
\includegraphics[width=0.27\textwidth]{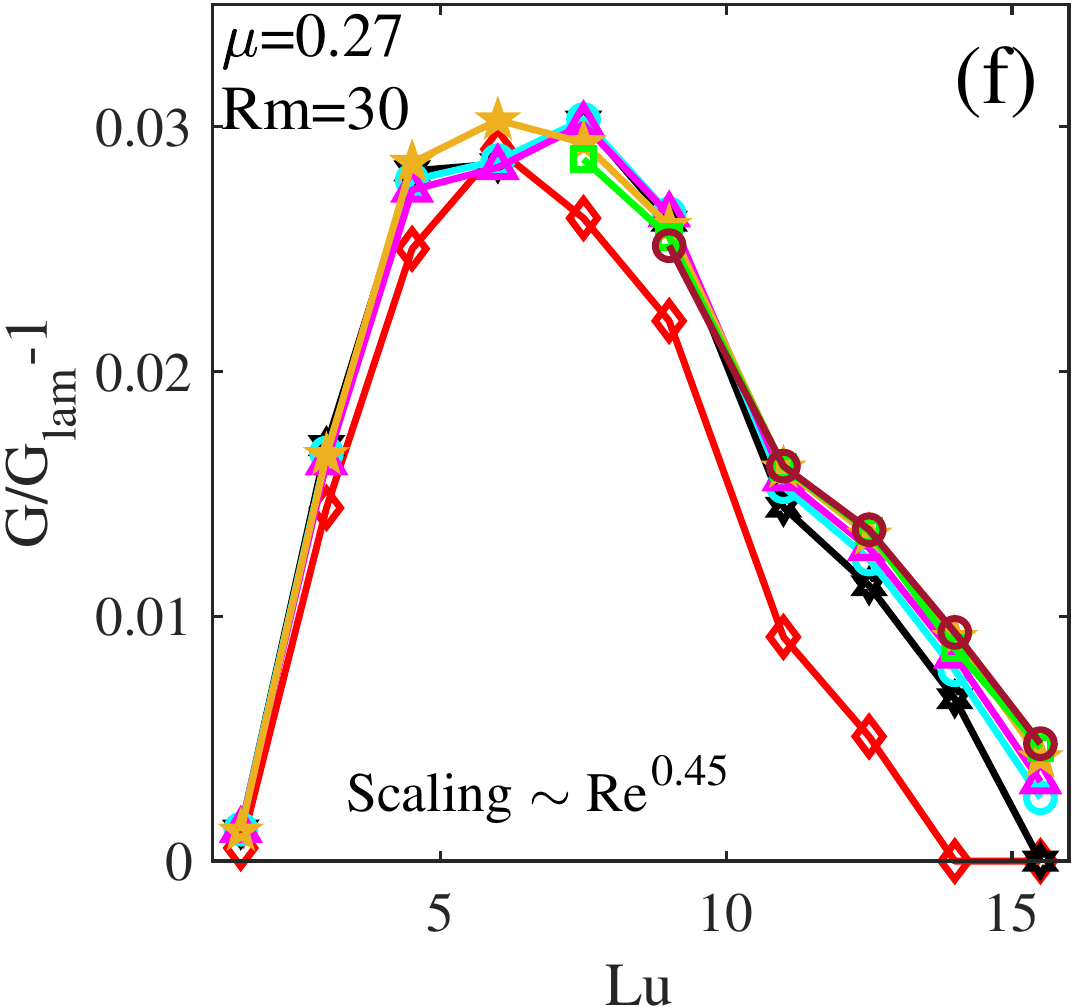}
\caption{Same as in Fig. \ref{fig:Torque_Sat_vs_Lu} except that the curves are scaled with the power law $Re^b$ where $b$ is the scaling exponent. For smaller $Rm$, $b \approx 0.5$ while for larger $Rm$, $b \approx 0.45$. With these scalings, the curves well collapse compared to those in Fig. \ref{fig:Torque_Sat_vs_Lu}.}
\label{fig:Torque_Scaling}
\end{figure*}

Finally, comparing Figs. \ref{fig:Scal_Sat_Mag_Energy_Vs_Lu} and \ref{fig:Torque_Scaling}, we note an interesting approximate relationship between the scaling exponents $a$ for the saturated magnetic energy and $b$ for the torque, $b-a\approx 1$, which appears to hold for all considered $Rm={9, 14, 20, 25, 30}$ as well as along the maximum growth line and can be generalized to other $Rm$.

\subsubsection{Analytical derivation of the scaling relations}

In this subsection, we analytically derive the above scaling relations for the magnetic field and torque with $Re$, which are the most important findings of this paper, based on force balances. We show that these relations follow from the interplay of the dynamics in the boundary layer near the inner cylinder and in the magnetic reconnection region. In particular, these derivations yield the relation $b-a \approx 1$ deduced above empirically. 

\begin{figure}
\centering
\includegraphics[width=\columnwidth]{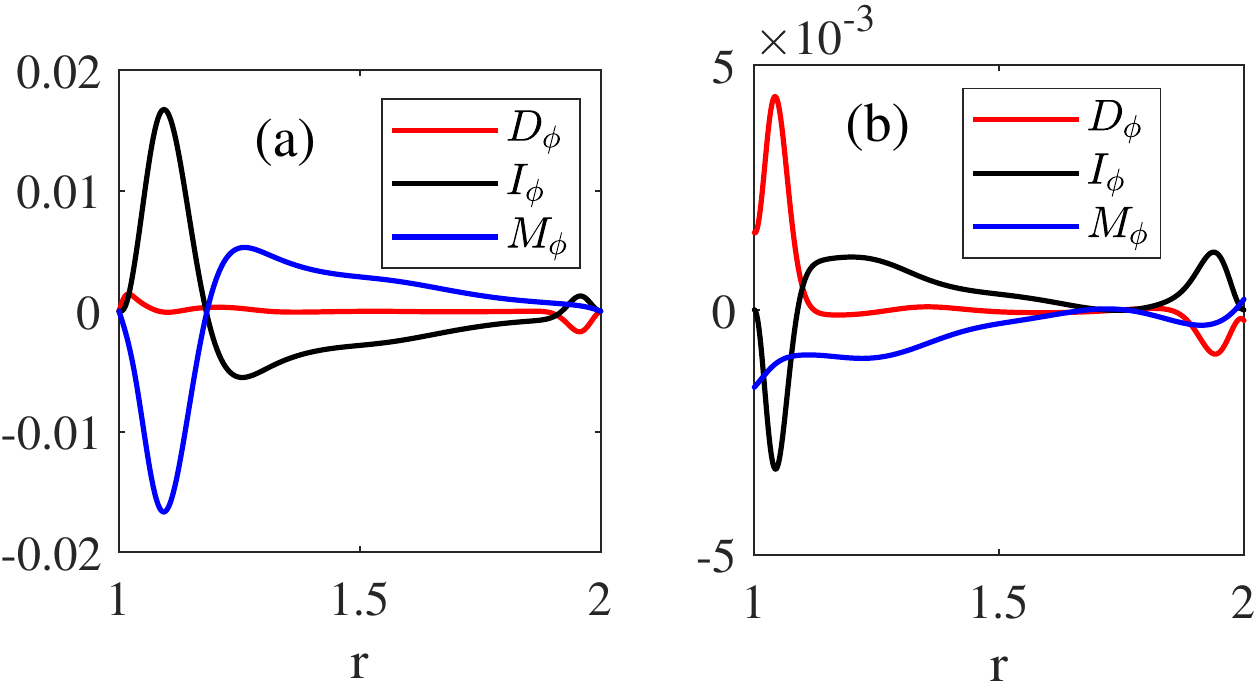}
\caption{Radial profiles of inertial $I_\phi$ (black), Lorentz $M_\phi$ (blue) and viscous $D_\phi$ (red) forces in the azimuthal component of Eq. \ref{moment} at (a) $z=5$ across the reconnection layer and (b) $z=2.5$ across the boundary layer of the saturated state in Fig. \ref{fig:r_z_slice} for $Lu=9$, $Rm=30$ and $Re = 40000$.}\label{fig:force_balance}
\end{figure}

\begin{figure}
\centering
\includegraphics[width=0.75\columnwidth]{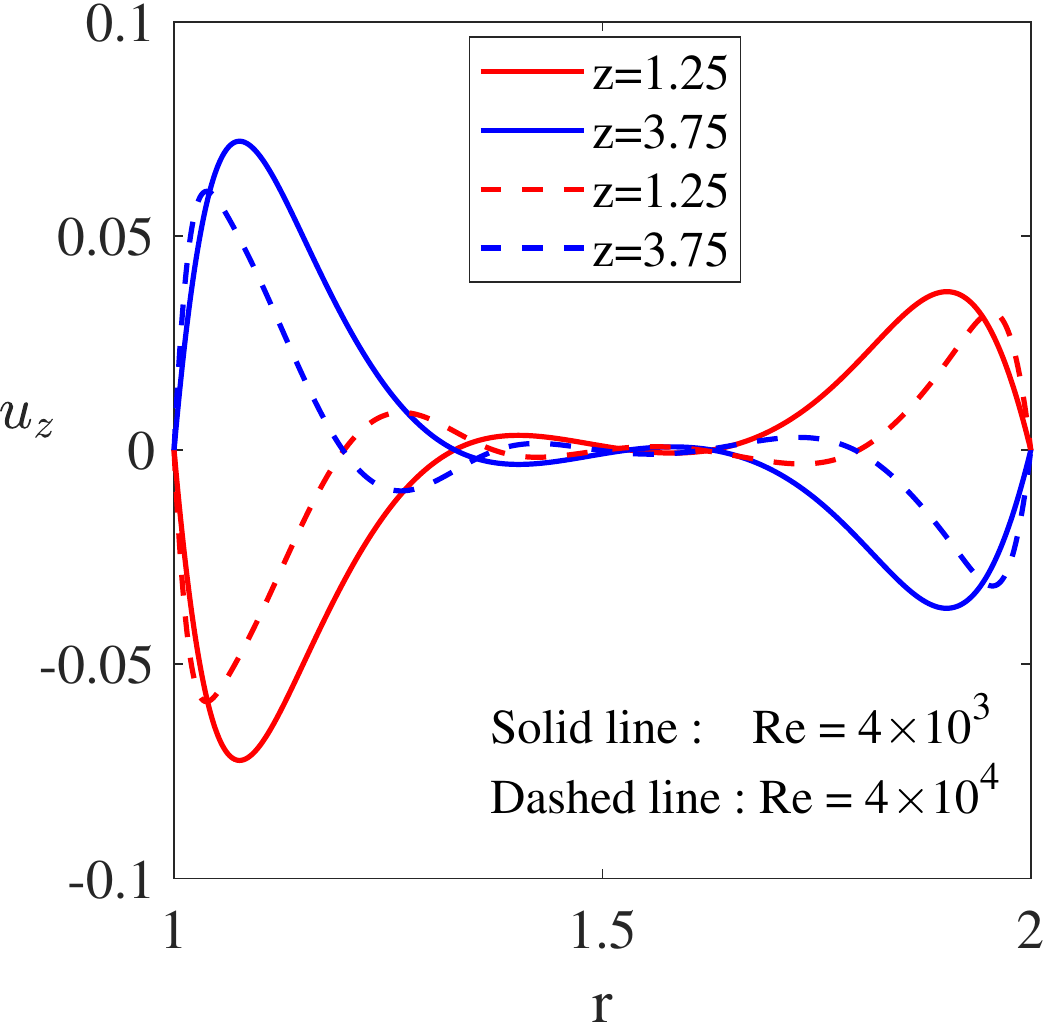}
\caption{Radial profiles of the axial velocity $u_z$ at different $z$-coordinates along the visocus boundary layer for the same parameters as in Fig. \ref{fig:force_balance} but at different $Re=4000$ (solid) and  $Re=40000$ (dashed). Note that the magnitude of $u_z$ does not change much with $Re$ (which vary by 10 times).}
\label{fig:uz_Re4k_40k}
\end{figure}
The azimuthal $\phi$-component of momentum equation Eq. (\ref{moment}) contains inertial ($I_{\phi}$), Lorentz ($M_{\phi}$) and viscous ($D_{\phi}$) forces, which for axisymmetric ($\partial /\partial\phi=0$) perturbations considered here take the following non-dimensional form,
\begin{equation}\label{inertial_phi_force_equation}
    I_{\phi}=-u_r\frac{\partial u_{\phi}}{\partial r}-u_z\frac{\partial u_{\phi}}{\partial z}-\frac{u_{\phi} u_r}{r}, 
\end{equation}
\begin{equation}\label{magnetic_phi_force_equation}
    M_{\phi}=\frac{Lu^2}{Rm^2}\left(B_r\frac{\partial B_{\phi}}{\partial r}+B_z\frac{\partial B_{\phi}}{\partial z}+\frac{B_{\phi} B_r}{r}\right), 
\end{equation}
\begin{equation}\label{viscous_phi_force_equation}
    D_{\phi}=\frac{1}{Re}\left( \frac{1}{r}\frac{\partial}{\partial r}\left(r\frac{\partial u_{\phi}}{\partial r}\right) +\frac{\partial^2 u_{\phi}}{\partial z^2}-\frac{u_\phi}{r^2} \right).
\end{equation}
The saturated state is steady in time and therefore these three forces should be in balance. Figure \ref{fig:force_balance} shows the radial profiles of $I_{\phi}$, $M_{\phi}$ and $D_{\phi}$ in the saturated state shown in Fig. \ref{fig:r_z_slice} across the reconnection region at $z=5$ (a) and in the viscous boundary layer at $z=2.5$ adjacent to the inner cylinder (b). In the reconnection layer, where the most of resistive dissipation occurs [Fig. \ref{fig:mag_kin_dissipation}(a)], and outside the viscous boundary layer, predominantly magnetic and inertial forces balance each other, while viscous force is negligible. On the other hand, in the inner boundary layer, where viscous dissipation is largest [Fig. \ref{fig:mag_kin_dissipation}(b)], viscous and inertial forces balance each other, with the Lorentz force being relatively small (since the boundary conditions are insulating). This balance of forces allows us to estimate the radial thickness $\delta_1$ of the boundary layer and axial width of the magnetic reconnection region, $\delta_2$ (which is assumed to be of the same order as its radial length) denoted in Fig. \ref{fig:mag_kin_dissipation}. It is seen from the structure of streamlines in the saturated state [Fig. \ref{fig:reconnection}(m)] that, because of mass conservation, the axial velocity of the fluid flux entering the magnetic reconnection region from the boundary layer and the radial velocity of the jet outflowing from that region satisfy
\begin{equation}\label{flux_cont}
    u_z\delta_1 \sim u_r \delta_2.
\end{equation} 
Balance of inertial and Lorentz forces in the reconnection region, $I_{\phi}\sim M_{\phi}$ [Fig. \ref{fig:force_balance}(a)], gives to leading order
\begin{equation*} 
\frac{u_ru_{\phi}}{\delta_2} \sim  \frac{Lu^2}{Rm^2}\frac{B_{z}B_\phi}{\delta_2}.
\end{equation*}
Since in the saturated state, the azimuthal velocity and axial magnetic field perturbations are small, we can write, respectively, $u_{\phi}\approx \Omega r\sim 1$ and $B_z\approx B_{0z}=1$ in this equation 
\begin{equation}\label{iner_mag_balance}
    u_r \sim \frac{Lu^2}{Rm^2} B_\phi.
\end{equation} 
\noindent
Similarly, from the $\phi$-component of induction Eq. (\ref{induc}) we obtain to leading order  
\begin{equation*}
   \frac{u_\phi}{\delta_2} \sim \frac{1}{Rm}\frac{B_\phi}{\delta_2^2},
\end{equation*}
or, equivalently
\begin{equation}\label{recon_ind_eq}
B_{\phi} \sim Rm\cdot \delta_2.
\end{equation}
\noindent
Similar analysis can be done for the viscous boundary layer at the inner cylinder wall where the inertial force balances the viscous force, $I_{\phi}\sim D_{\phi}$ [Fig. \ref{fig:force_balance}(b)] and fluid motion in the $(r,z)$-plane is mostly axial $u_r \ll u_z$ (Fig. \ref{fig:reconnection}), yielding
\begin{equation*}
    u_z\frac{u_{\phi}}{\lambda_z} \sim \frac{1}{Re}\frac{u_{\phi}}{\delta_1^2},
\end{equation*}
which can be written as
\begin{equation}\label{iner_visc_balance}
\delta_1^2\sim \frac{1}{Re}\frac{\lambda_z}{u_z},
\end{equation}
\noindent
where $\lambda_z$ is the axial length of large-scale structures (rolls) in the saturated SMRI state (Fig. \ref{fig:r_z_slice}), which is comparable to the gap width $\lambda_z \sim d=1$ and does not depend on $Re$. It also follows from the simulations (Fig. \ref{fig:uz_Re4k_40k}) that the magnitude of the axial velocity $u_z$ is nearly independent of $Re$ and determined rather by the global SMRI mode; only the radial profile of $u_z$ becomes more concentrated towards the cylinder walls as $Re$ increases. As a result, Eq. (\ref{iner_visc_balance}) gives the scaling of the boundary layer thickness with $Re$,
\begin{equation}\label{bl_thickness}
\delta_1\sim Re^{-0.5},
\end{equation}
which coincides with the classical one for a hydrodynamical laminar boundary layer \cite{Landau}. 
Combining Eqs. (\ref{flux_cont})-(\ref{iner_visc_balance}), one can obtain the relation between the boundary layer thickness and reconnection region width,
\begin{equation}\label{delta_re_relation}
\delta_1 \delta_2^2 \sim \frac{1}{Re}\frac{Rm}{Lu^2},
\end{equation}
which using Eq. (\ref{bl_thickness}) yields
\begin{equation}\label{delta2_scaling}
\delta_2\sim \left(\frac{Rm}{Lu^2}\right)^{0.5}(Re\cdot \delta_1)^{-0.5} \sim Re^{-0.25}.
\end{equation}
Note that the magnetic reconnection width, which scales with $Re^{-0.25}$, is also proportional to the square root of the inverse of the interaction parameter $Lu^2/Rm$. 

From induction Eq. (\ref{induc}), one can show that $B_\phi$ is of the same order as $B_r$ and $B_z$, and hence from Eqs. (\ref{recon_ind_eq}) and (\ref{delta2_scaling}), we obtain the same scaling for the magnetic energy 
\begin{equation} \label{mag_energy_scaling}
\mathcal{E}_{mag} \sim Rm^2 \delta_2^2 \sim Re^{-0.5},
\end{equation} 
which we have deduced above empirically based on the simulation data. Note that this central scaling law for the magnetic energy in this paper essentially follows from Eq. (\ref{delta_re_relation}), which thus has an important role to couple the boundary layer dynamics, dominated by viscosity (characterized by $Re$), to the reconnection region dynamics, dominated by inertial and Lorentz forces and resistive dissipation. Consequently, the magnitude of the saturated magnetic field perturbation due to nonlinear SMRI appears to be dependent on Reynolds number.

We derive now the scaling for the torque at the cylinder walls, which is determined by the boundary layer properties. Using Eqs. (\ref{torque_at_cylinders}) and (\ref{bl_thickness}), the normalized perturbation torque at the inner cylinder is given by order of magnitude as: 
\begin{equation}\label{tor_delta}
    G/G_{lam}-1 \sim \frac{u_{\phi}}{r_{in}\delta_1}\sim \delta_1^{-1}\sim Re^{0.5},
\end{equation}
which together with the magnetic energy scaling (Eq. \ref{mag_energy_scaling}) satisfies
\begin{equation}\label{scaling_equation_energy_torque}
    \mathcal{E}_{mag}^{-1}\left(G/G_{lam}-1\right) \sim Re,
\end{equation}
recovering the above relation between the scaling exponents $b-a \approx 1$. Slight deviations from the individual scalings (\ref{mag_energy_scaling}) and (\ref{tor_delta}) at different $Lu$ and $Rm$ that we observe in Figs. \ref{fig:Scal_Sat_Mag_Energy_Vs_Lu} and \ref{fig:Torque_Scaling} could be due to the small contribution of Lorentz force in the dynamical balance within the boundary layer, which we have ignored in the above analysis. Despite these deviations the relation (\ref{scaling_equation_energy_torque}) is still satisfied, as evident from Figs. \ref{fig:Scal_Sat_Mag_Energy_Vs_Lu} and \ref{fig:Torque_Scaling}.

\subsubsection{Relevance to DRESDYN-MRI Experiment}

\begin{table}
\scriptsize
\centering
\resizebox{\columnwidth}{!}{
\begin{tabular}[b]{c|c|c|c|c}	
	\hline
	$ (Lu, Rm)$ & $u$  & $ u $ & $ b $ & $b $ \\
		 &  $(\Omega_{in}r_{in})$ &  $(m/s)$ & $ (B_0)$ & $ (mT)$ \\
		\hline
		$(2, 14)$ & $0.0138$ & $0.0808$ & $0.0078$ & $0.222$ \\
		$(4, 14)$ & $0.0437$ & $0.2560$ & $0.0068$ & $0.387$ \\
		$(1.5, 20)$ & $0.0053$ & $0.0443$ & $0.0085$ & $0.181$ \\
		$(6, 20)$ & $0.0745$ & $0.6234$ & $0.0079$ & $0.675$ \\
		$(1.5, 30)$ & $0.0059$ & $0.0740$ & $0.0125$ & $0.267$ \\
		$(9, 30)$ & $0.1047$ & $1.3142$ & $0.1020$ & $13.073$ \\
  \hline
\end{tabular}
}
\caption{The rms of the velocity and magnetic field perturbations in the nonlinear saturated state of SMRI expected in DRESDYN-MRI experiment. These rms values have been estimated by extrapolating the simulation results for different $Lu$ and $Rm$ at $\mu=0.27$ down to $Pm=7.77\times 10^{-6}$ of liquid sodium using the scaling relations given in subsection B.4.}\label{Table2:RMS_value_Table}
\end{table}
	
In Paper I, we explored the parameter regime of linear SMRI in the context of upcoming DRESDYN-MRI experiment and showed that it can be detected convincingly in the parameter regime achievable in this experiment. However, 1D linear stability analysis only allowed us to characterize the SMRI properties in the growth phase of the instability, while in the experiment measurements are usually conducted after the saturation has been reached. Given the great importance of SMRI, it would be interesting to estimate the magnitude of velocity and magnetic field perturbations expected in the DRESDYN-MRI experiment for $\mu=0.27$ and a realistic value $Pm=7.77 \times 10^{-6}$ of liquid sodium, using the simulation results and the scaling behaviour discussed in the previous subsection.

Table \ref{Table2:RMS_value_Table} shows the calculated rms of the velocity and magnetic field for different sets of $(Lu, Rm)$ by extrapolating the scaling laws obtained above for the considered range $Pm \in [8.5\times 10^{-5}, 0.0327]$ down to even smaller $Pm$ of sodium. Note that these rms values are larger for larger $Lu$ and $Rm$. It is seen from Table \ref{Table2:RMS_value_Table} that for the chosen values of $(Lu, Rm)$, the rms velocity perturbations vary from the smallest $0.07~{\rm ms^{-1}}$ to the highest $1.31~{\rm ms^{-1}}$ and the rms magnetic field perturbations from $0.18~{\rm mT}$ to as high as $13.07~{\rm mT}$. These estimated rms values in the saturated SMRI state (which is actually present in the experiments) are encouraging as they will facilitate detection of SMRI, although should be viewed with caution, because at such high $Re$ the saturated state and, especially the boundary layers, can become turbulent.  
	

\begin{figure*}
\centering
\includegraphics[width=0.35\textwidth]{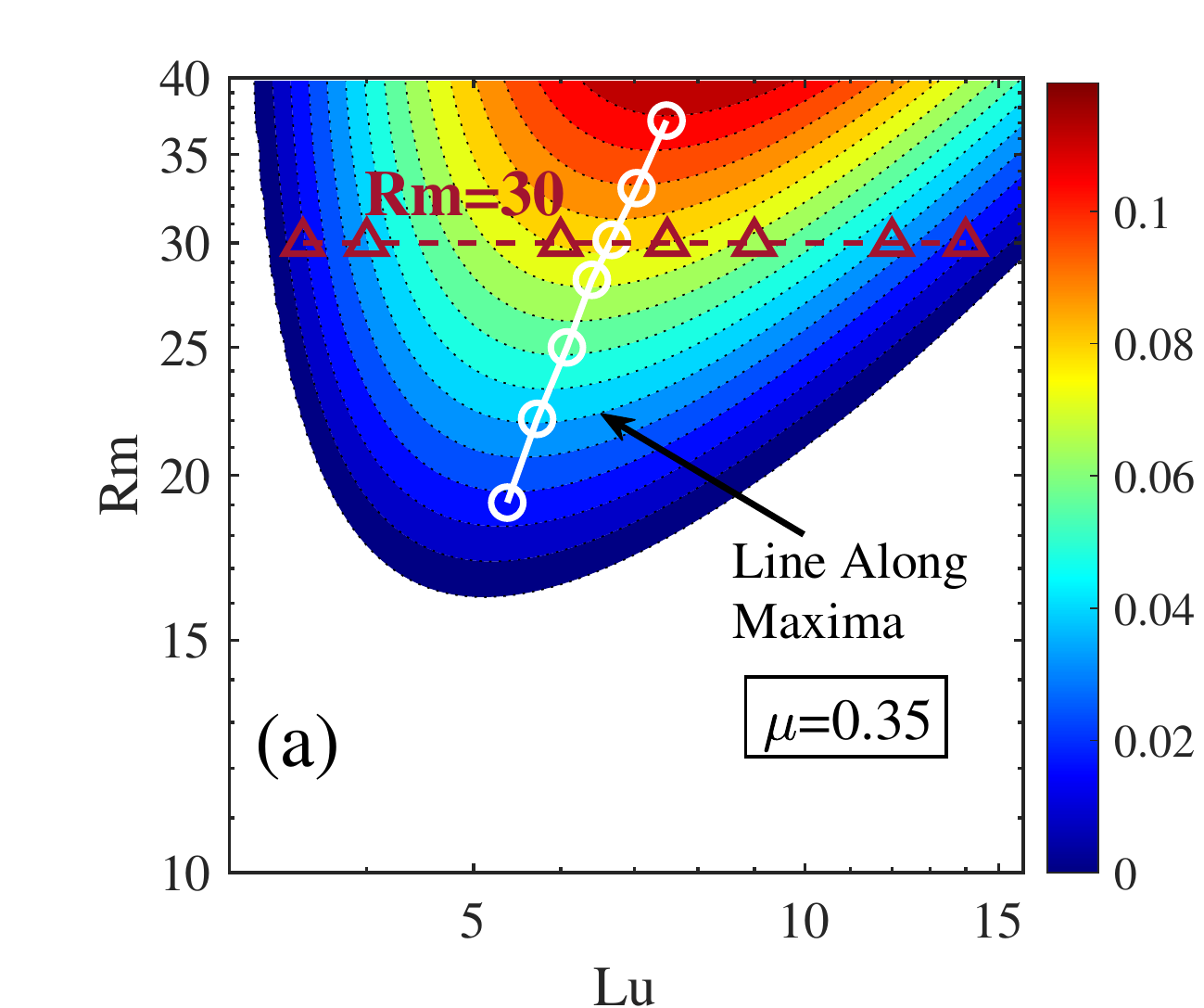}
\includegraphics[width=0.28\textwidth]{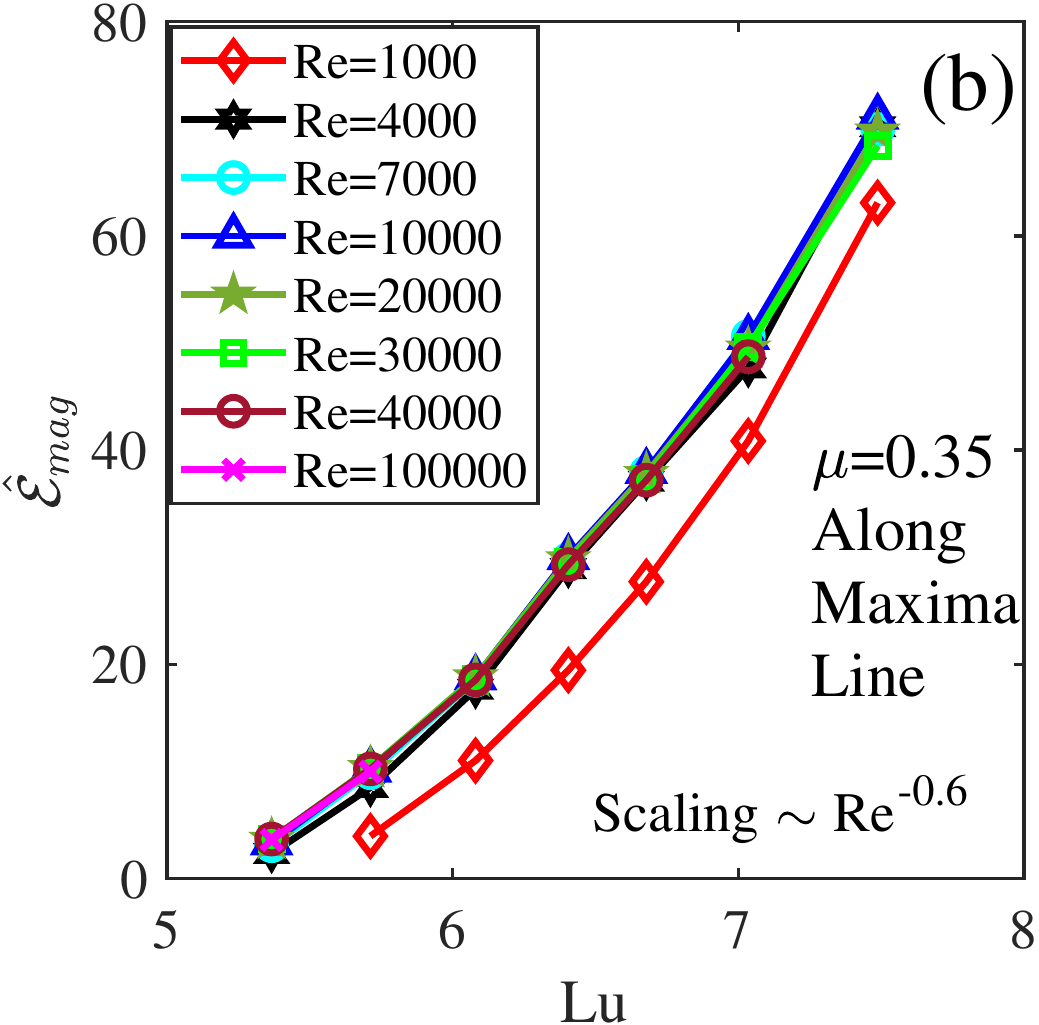}
\includegraphics[width=0.28\textwidth]{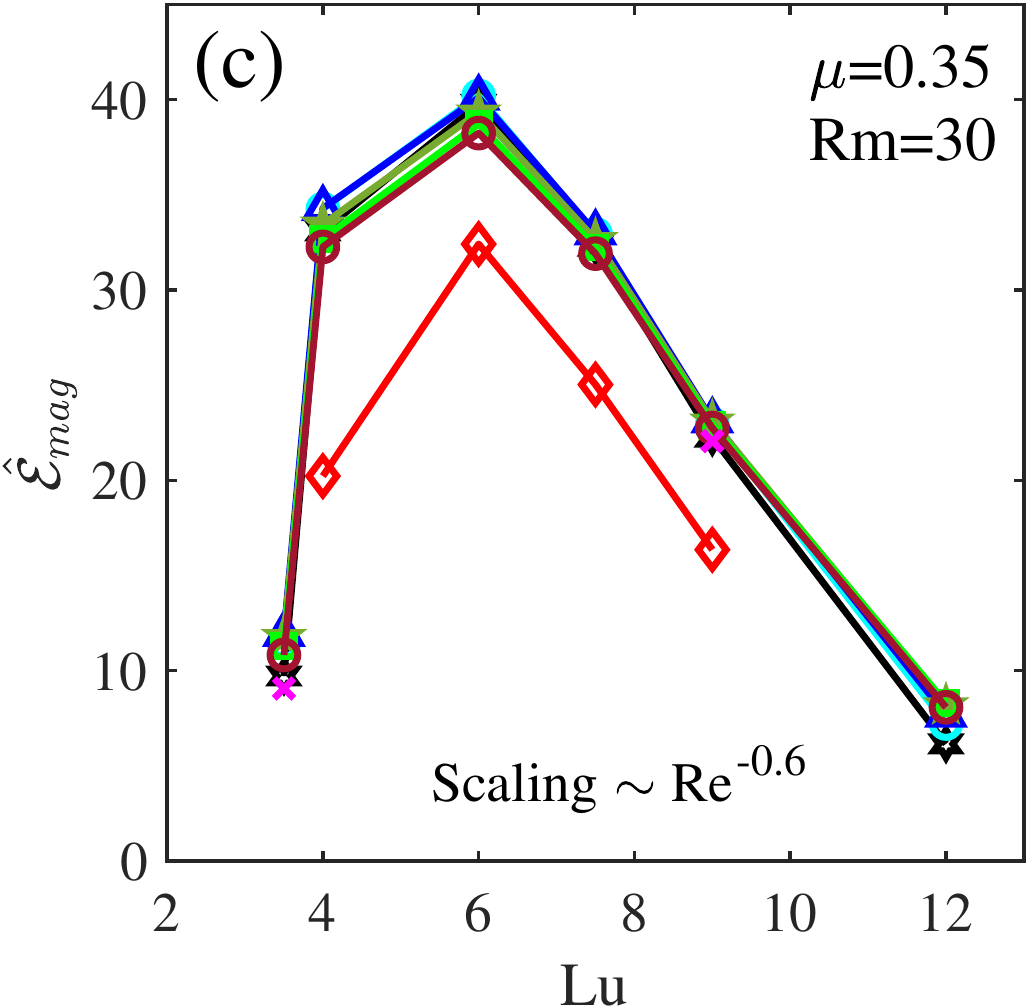}
\includegraphics[width=0.31\textwidth]{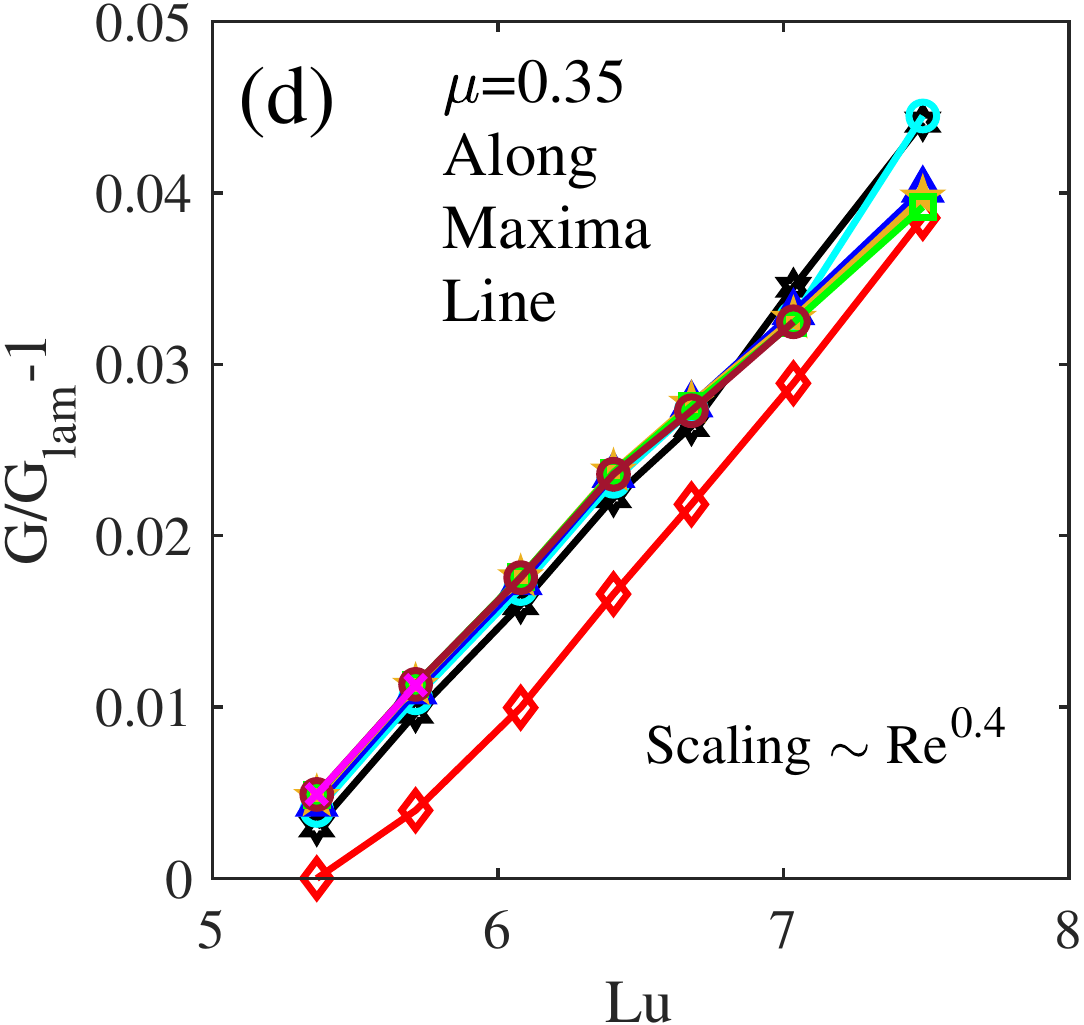}
\includegraphics[width=0.31\textwidth]{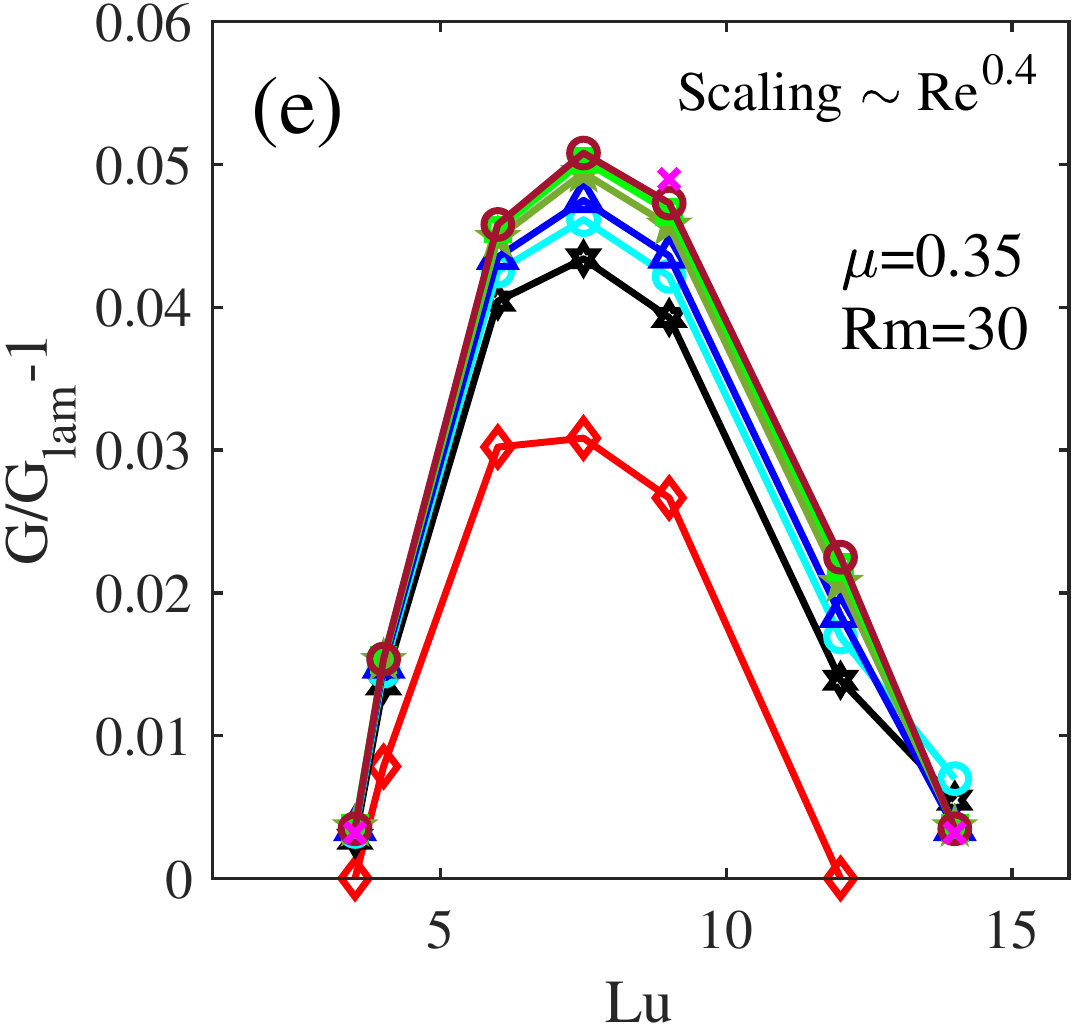}
\includegraphics[width=0.3\textwidth]{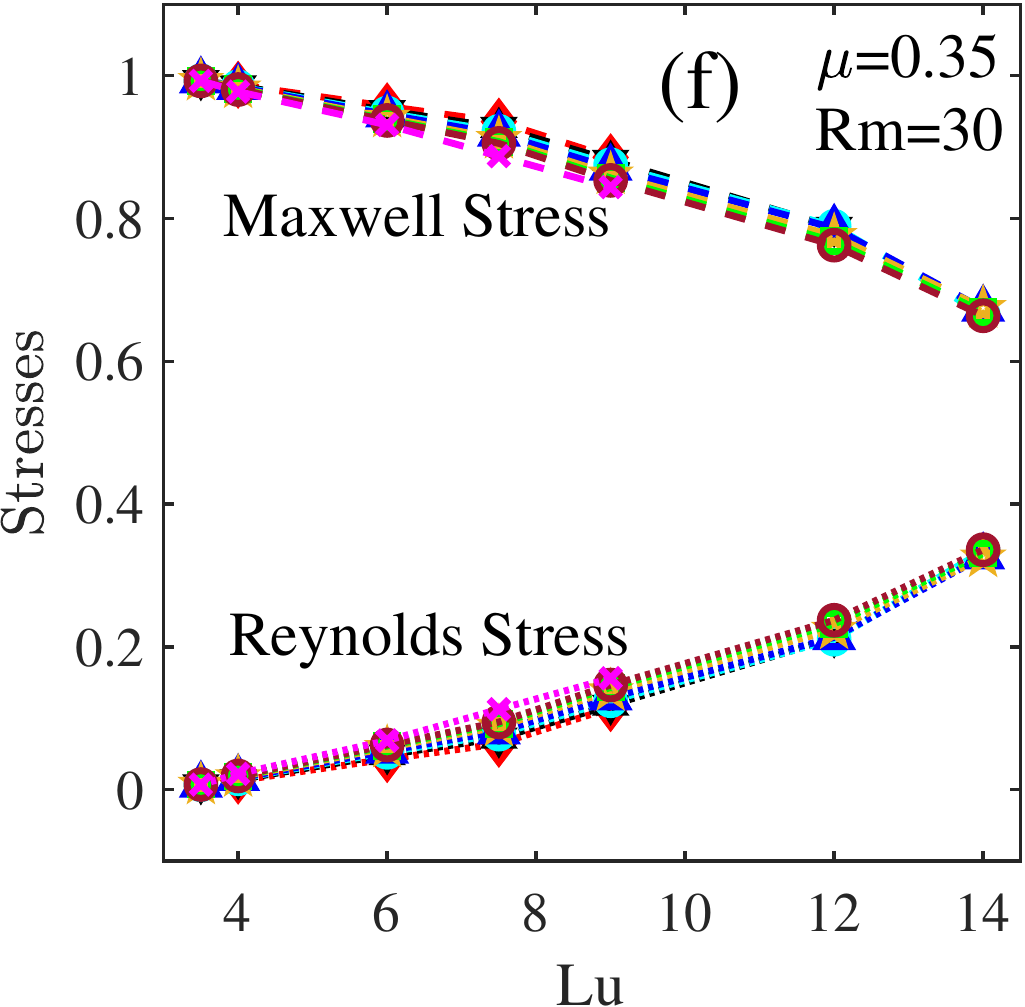}
\caption{(a) The growth rate of SMRI from 1D linear stability analysis for $\mu=0.35$ and $Pm=7.77\times 10^{-6}$ \cite{Mishra_mamatsashvili_stefani_2022PRF}. The nonlinear simulations are performed for $Lu$ and $Rm$ lying along the line of the maximum growth (circles) as well as for fixed $Rm=30$ and varying $Lu$ (triangles). (b) Saturated magnetic energy as a function of $Lu$ at different $Re \in \{1, 4, 7, 10, 20, 30, 40, 100\} \times 10^3$. The curves are scaled by $Re^a$ with $a \approx -0.6$, for which they well collapse. (c) Same as in (b) but for fixed $Rm=30$ with the same scaling exponent $a\approx -0.6$. (d) The torque $G/G_{lam}-1$ as a function of $Lu$ for the same set of $Re$ scaled by $Re^b$ with $b\approx 0.4$. (e) Same as in (d) but for fixed $Rm=30$ and the scaling exponent $b\approx 0.4$. (f) Normalized Maxwell stress (dashed) and Reynolds stress (dotted) for varying $Lu$ and $Re$ at $Rm=30$.}\label{fig:mu_35_scaling_vs_Lu}
\end{figure*}

 \subsection{Quasi-Keplerian profile with $\mu=0.35$}
	
The above results on the nonlinear evolution of SMRI in the magnetzed TC flow for $\mu=0.27$ together with the linear stability analysis presented in Paper I supports our assertion that the SMRI can be unambiguously captured in the upcoming DRESDYN-MRI experiment with detectable rms values of velocity and magnetic field perturbations. In this regard, it is now interesting to quantify the characteristics of the saturated SMRI state for the quasi-Keplerian profile due to its importance for astrophysical disks. In the context of TC flow, quasi-Keplerian rotation is defined by assuming that the cylinders rotate according to Kepler's law $\Omega \propto r^{-3/2}$ \cite{Ruediger_etal_2018_PhysRepo} and therefore the ratio of their angular velocities $\mu=\Omega_{out}/\Omega_{in}=(r_{in}/r_{out})^{3/2}\approx 0.35$ at $r_{in}/r_{out}=0.5$ in DRESDYN-MRI device.

Figure \ref{fig:mu_35_scaling_vs_Lu}(a) shows the linearly unstable region of SMRI in the $(Lu, Rm)$-plane for $\mu=0.35$ and those wavenumbers $k_z\geq 2\pi/L_z$ which fit in the device. This instability region is well within the achievable parameter regime of the DRESDYN-MRI experiment. Like the $\mu=0.27$ case above, here we perform nonlinear simualations for those pairs of $(Lu, Rm)$ that lie on the maximum growth line (white) of the instability, where $Rm$ increases approximately linearly with $Lu$ as $Rm\approx A\cdot Lu+B$ at $A=8.46$ and $B=-26.3$, and the minimum and maximum sets of $(Lu, Rm)$ being (5.4, 19) and (7.5, 37.1), respectively, as well as along the $Rm=30$ line across the linearly unstable region in Fig. \ref{fig:mu_35_scaling_vs_Lu}(a). Although this parameter space is smaller than that at $\mu=0.27$ in Fig. \ref{fig:Analysis_points}(b), it provides enough insight into the general saturation behaviour and properties of SMRI for the quasi-Keplerian flow profile.

Figures \ref{fig:mu_35_scaling_vs_Lu}(b)- \ref{fig:mu_35_scaling_vs_Lu}(e) show the scaling behavior of the magnetic energy and torque in the saturated state both along the line of the maximum growth and at $Rm=30$, which exhibit overall a similar qualitative dependence on $Lu$ and $Re$ as that for $\mu=0.27$ above. The best fit for the scaling behavior with $Re$ is achieved with the averaged exponents $a\approx -0.6$ for the energy [Figs. \ref{fig:mu_35_scaling_vs_Lu}(b) and \ref{fig:mu_35_scaling_vs_Lu}(c)] and $b\approx 0.4$ for the torque [Figs. \ref{fig:mu_35_scaling_vs_Lu}(d) and \ref{fig:mu_35_scaling_vs_Lu}(e)], so the relation $b-a\approx 1$ holds again. For these exponents the curves well collapse to a single line for all	$Re \geq 4000$, while a little deviation is observed for $Re=1000$, which is not yet in the asymptotic regime. These scaling are also confirmed in Figs. \ref{fig:mu_35_scaling_vs_Re}(a) and \ref{fig:mu_35_scaling_vs_Re}(b), which show $Re$-dependence of these variables for certain chosen values of $Lu$, although the scaling exponents are a bit different: $a\approx -0.5$ and $b\approx 0.5$, but similar to those obtained for $\mu=0.27$ in Figs. \ref{fig:Scaling_energy}(b) and \ref{fig:torque_energy}(b). This implies that the scaling of the saturated torque and magnetic energy remain almost the same irrespective of the ratio of the cylinders' angular velocities, $\mu$, in the Rayleigh-stable regime.

\begin{figure*}[t!]
\centering
\includegraphics[width=0.31\textwidth]{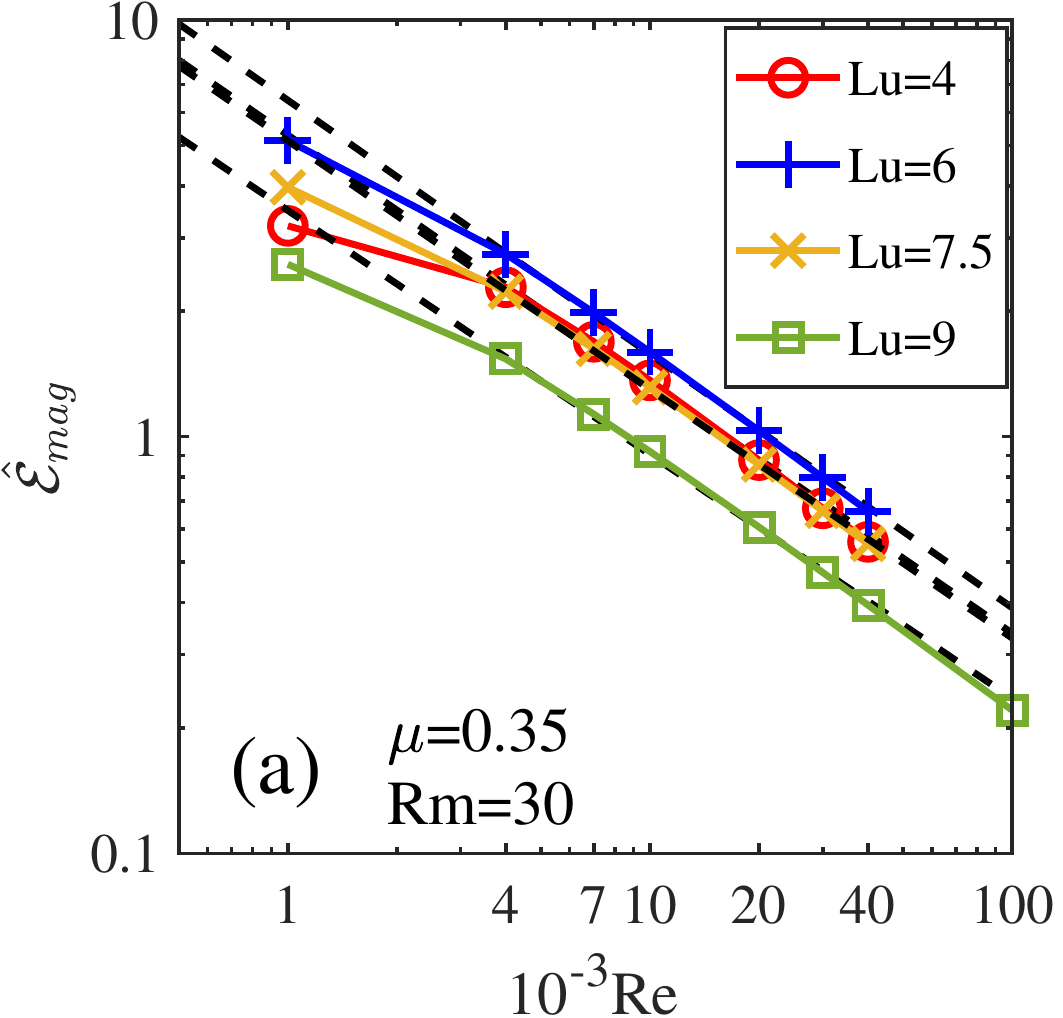}
\includegraphics[width=0.305\textwidth]{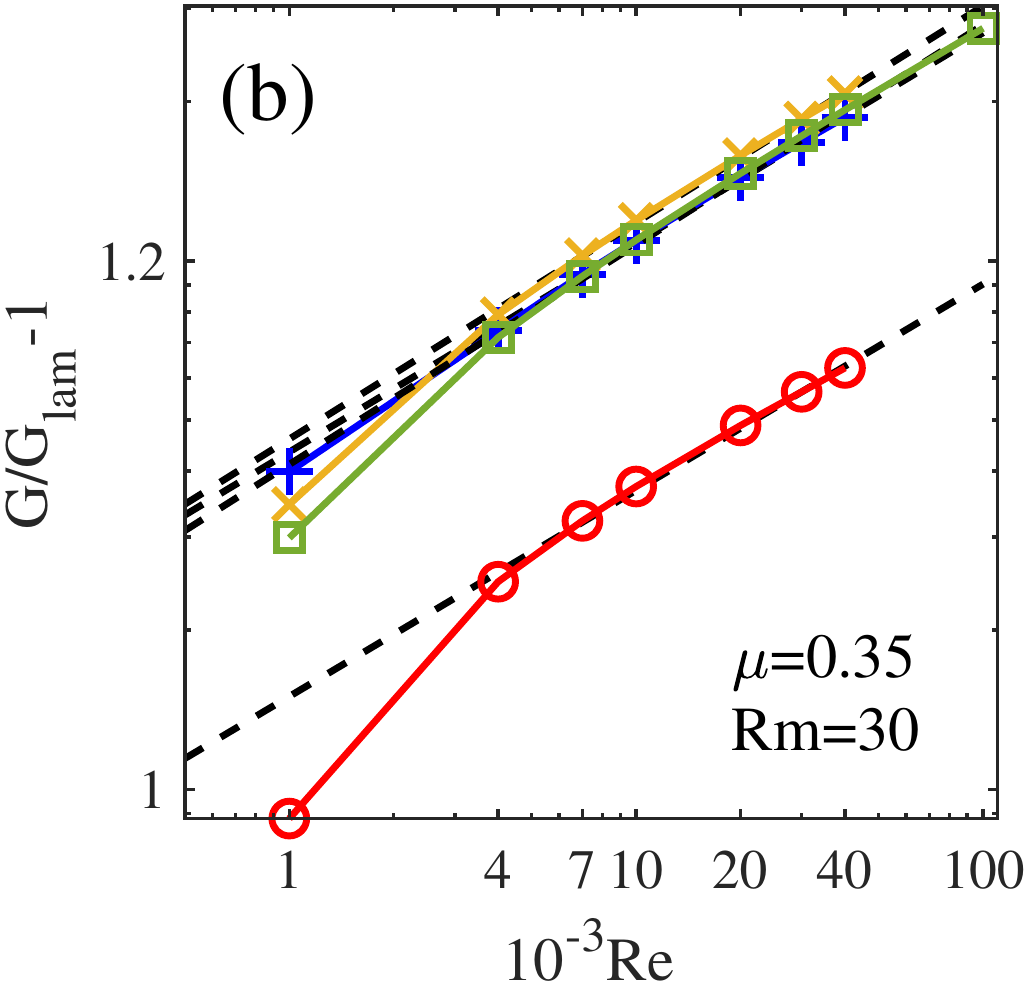}
\includegraphics[width=0.3\textwidth]{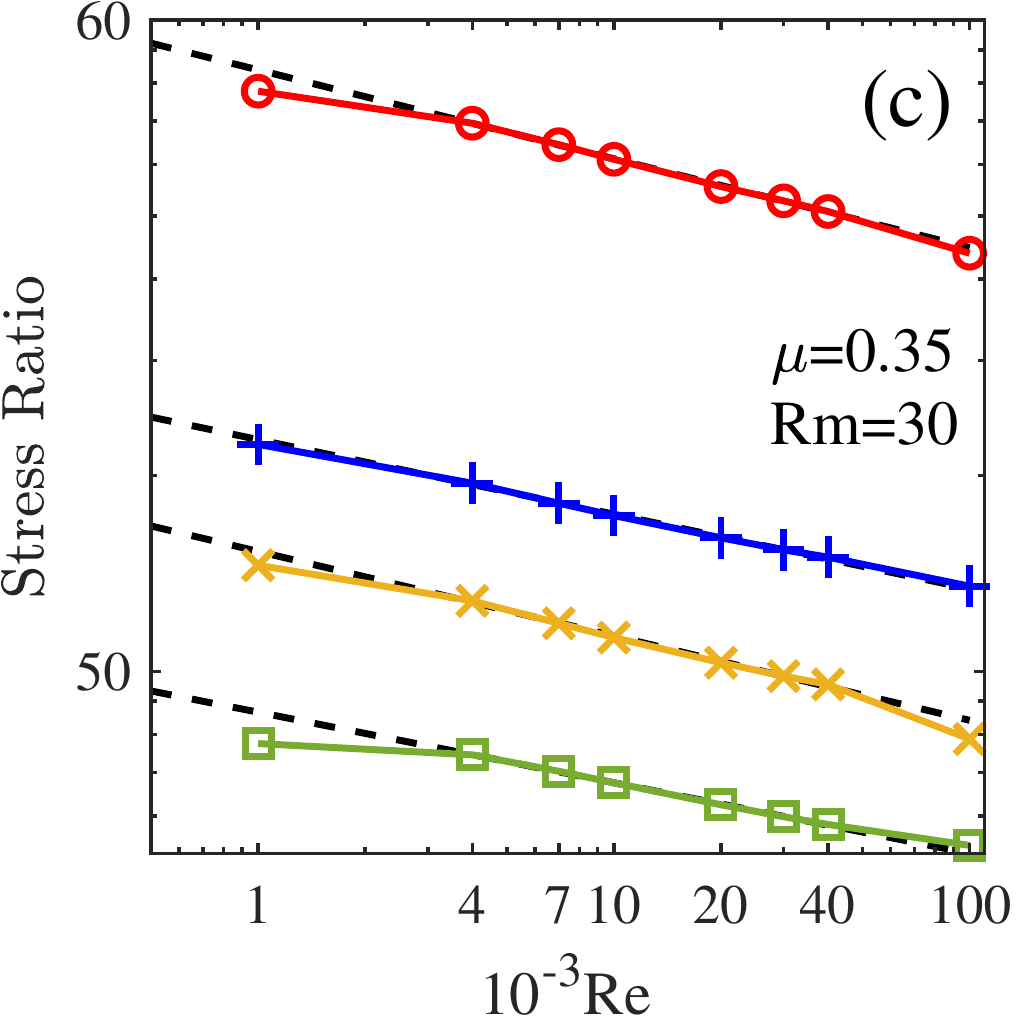}
\caption{(a) Saturated magnetic energy $\mathcal{\hat{E}}_{mag}$ as a function of $Re$ for $\mu=0.35$, $Rm=30$ and different $Lu=4$ (red), 6 (blue), 7.5 (yellow), 9 (green). Black dashed lines show the power law function $Re^a$ with the average scaling exponent tending towards $a\approx -0.5$. (b) The torque $G/G_{lam}-1$ as a function of $Re$ for the same parameters as in (a) which follows the power law $Re^b$ with scaling exponent tending towards $b\approx 0.5$ (dashed black). (c) Maxwell to Reynolds stress ratio, which decreases with $Re$ also as a power law $Re^c$ with $c\approx-0.12$ (dashed black).}\label{fig:mu_35_scaling_vs_Re}
\end{figure*}

Figure \ref{fig:mu_35_scaling_vs_Lu}(f) shows the maximum value of the Maxwell and Reynolds stresses over radius normalized by their sum as a function of $Lu$ and fixed $Rm=30$ in the saturated state. The Maxwell stress is much larger than that the Reynolds one for all $Re$. The contribution of Maxwell stress decreases and that of Reynolds stress increases with $Lu$, while their ratio for fixed $Lu$ follows a power law $Re^c$ with $c\approx-0.12$ [Fig. \ref{fig:mu_35_scaling_vs_Re}(c)]

\begin{figure*}
\centering
\includegraphics[width=0.31\textwidth]{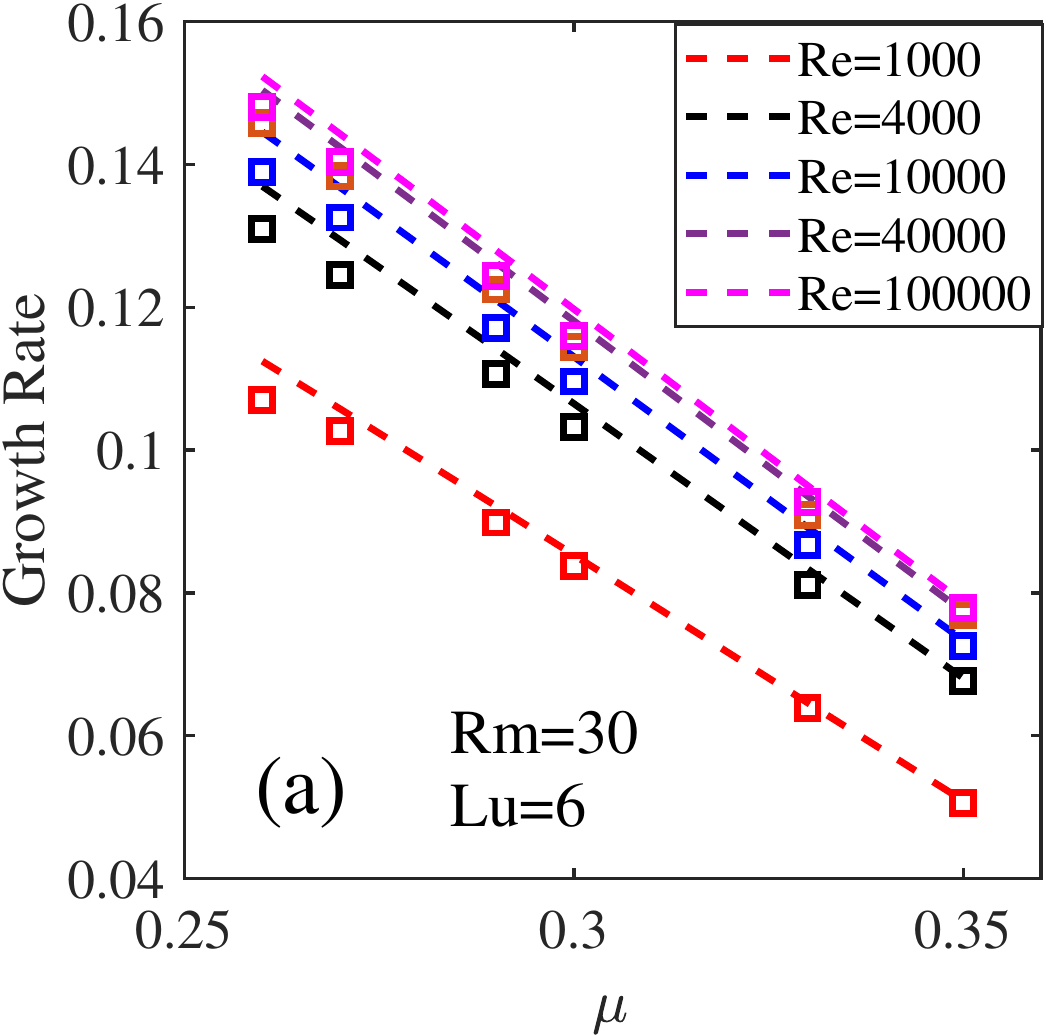}
\hspace{0.4em}
\vspace{1em}
\includegraphics[width=0.30\textwidth]{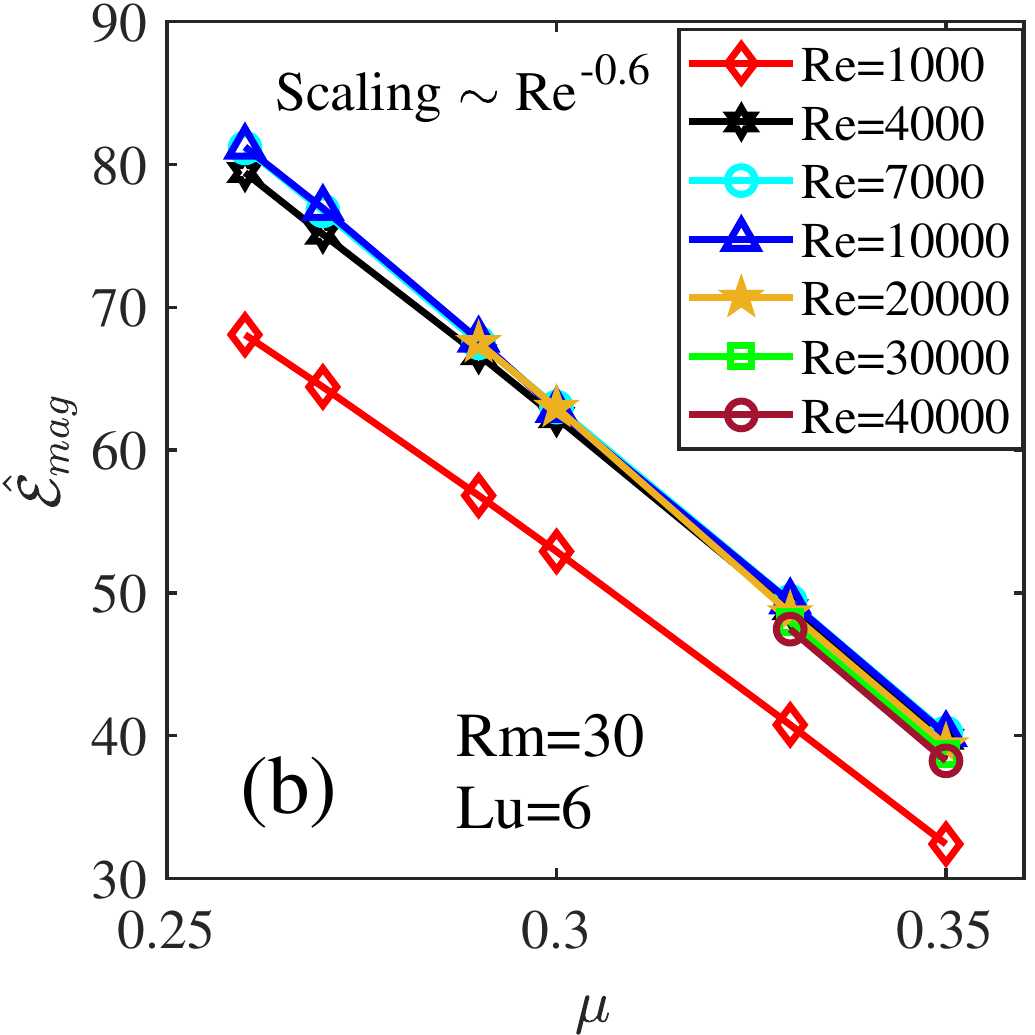}
\hspace{0.3em}
\includegraphics[width=0.31\textwidth]{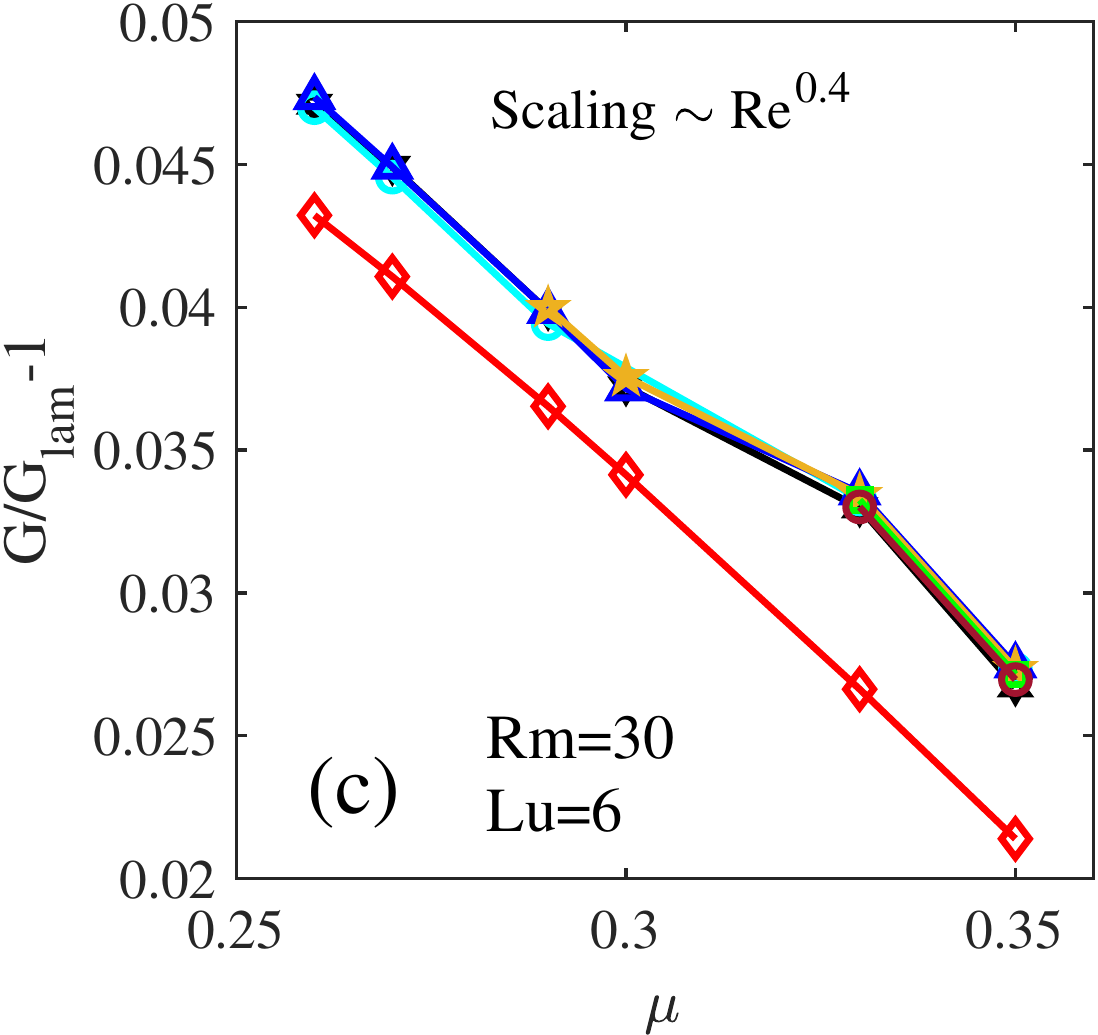}
\hspace{1em}
\includegraphics[width=0.30\textwidth]{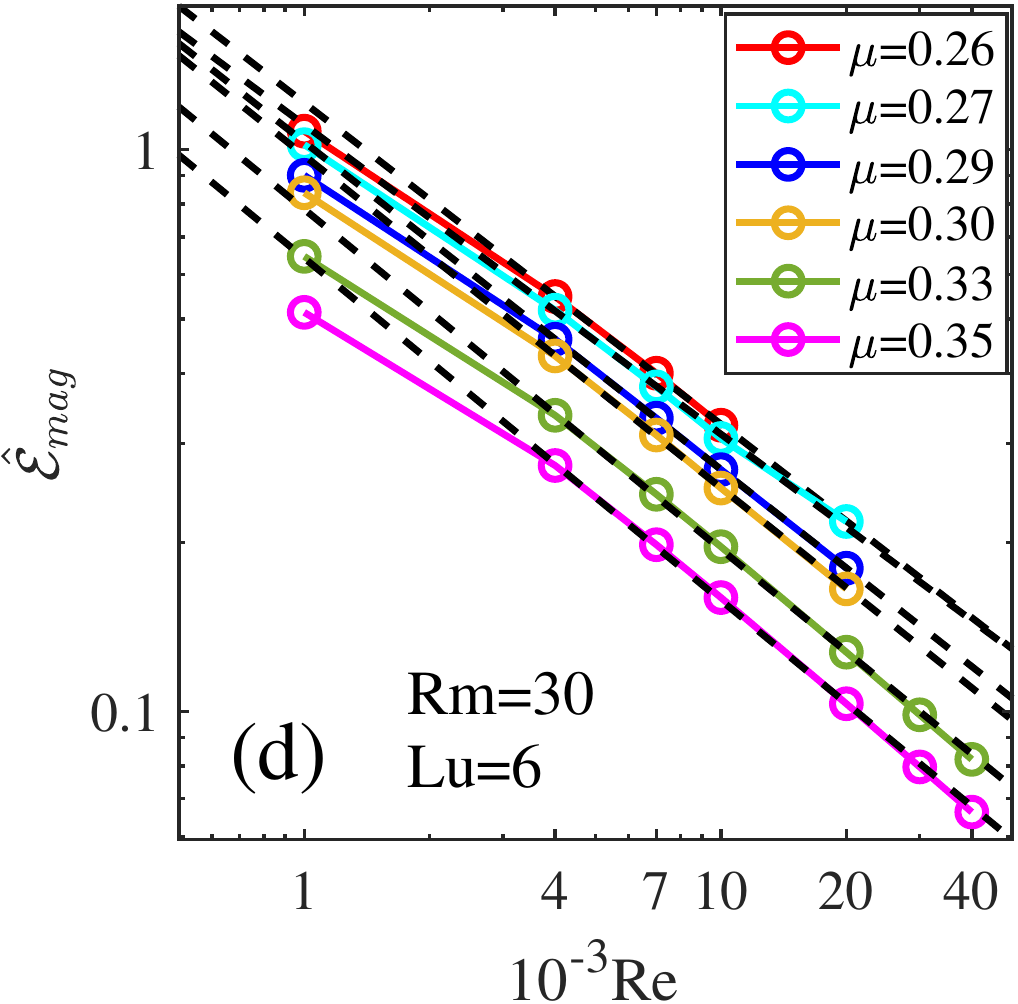}
\hspace{1em}
\includegraphics[width=0.29\textwidth]{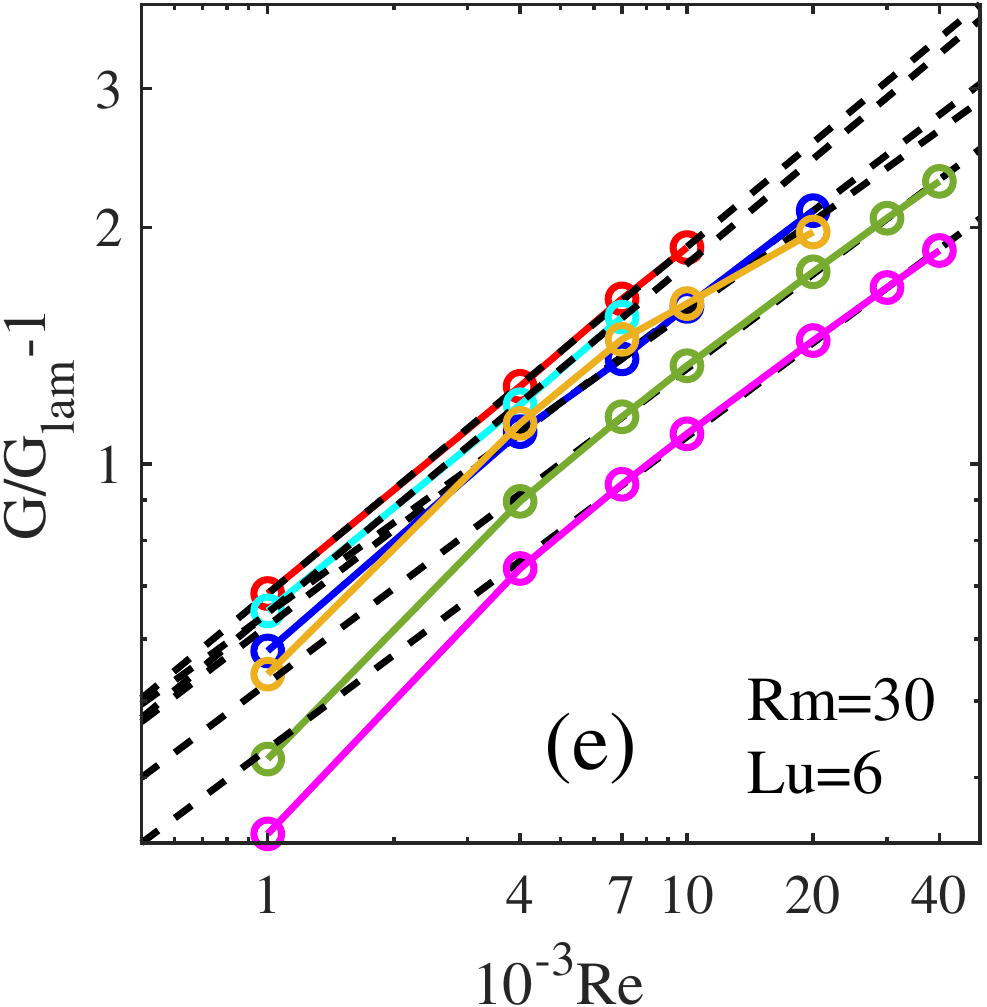}
\hspace{2em}
\includegraphics[width=0.28\textwidth]{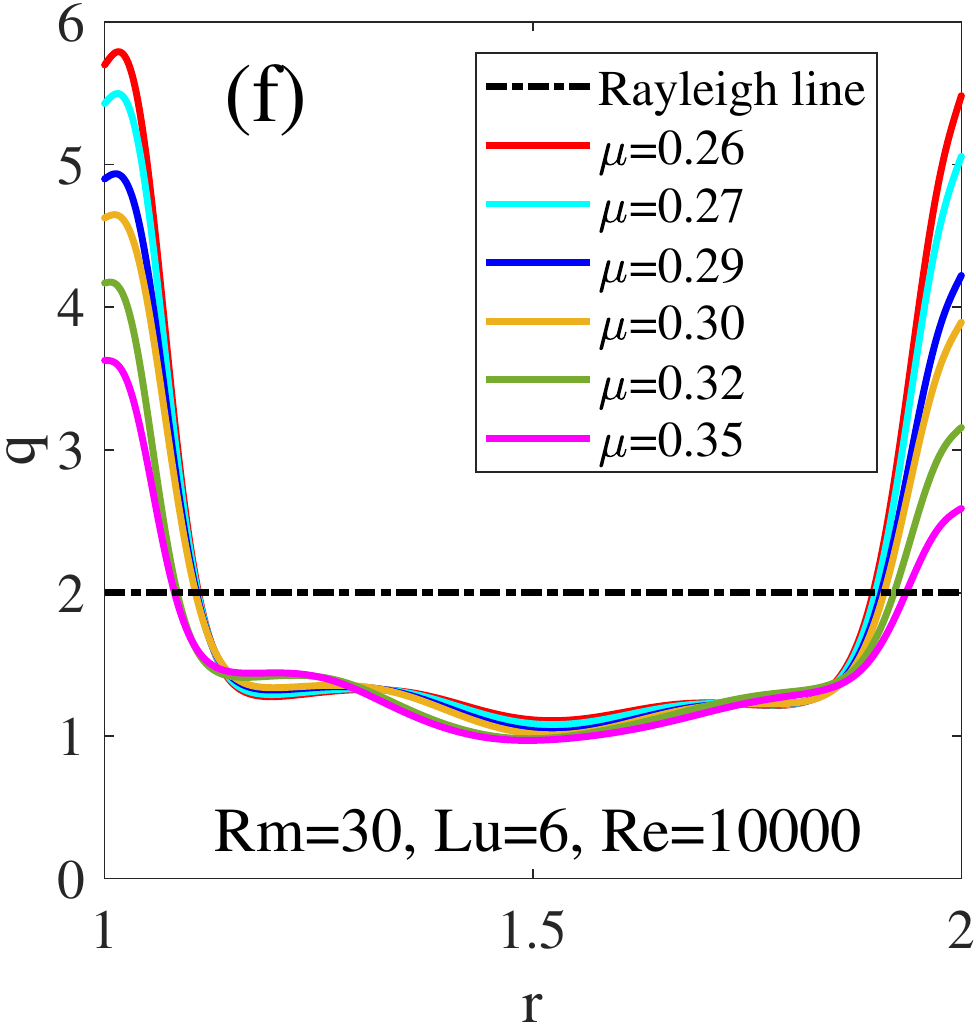} 
\hspace{1em}
\caption{(a) Comparison between the growth rates in the linear regime derived from the nonlinear code (squares) and from the linear 1D code (dashed) at different $Re$ as a function of $\mu$. (b) The saturated magnetic energy $\mathcal{\hat{E}}_{mag}$ and (c) torque $G/G_{lam}-1$ as a function of $\mu$ for different $Re=$ 1000 (red), 4000 (black), 7000 (cyan), 10000 (blue), 20000 (yellow), 30000 (green), 40000 (violet) and 100000 (magenta). The curves are scaled by the power laws $Re^a$ with $a\approx -0.6$ for the magnetic energy and $Re^b$ with $b\approx 0.4$ for the torque. (d) and (e) show the same quantities as a function of $Re$ at different $\mu=$ 0.26 (red), 0.27 (cyan), 0.29 (blue), 0.30 (yellow), 0.33 (green) and 0.35 (magenta). Black dashed lines show the power law curves, which are $Re^a$ with $a\approx-0.585$ for the magnetic energy and $Re^b$ with $b\approx 0.4$ for the torque. (f) The radial profile of the shear parameter $q$ for different $\mu$ and $Re=10000$. In all the panels $Lu=6$ and $Rm=30$.} \label{fig:diff_mu}
\end{figure*}

\begin{table}
	\scriptsize
	\centering
	\resizebox{\columnwidth}{!}{%
		\begin{tabular}[b]{c|c|c|c|c}	
		\hline
		$ (Lu, Rm)$ & $ u$  & $ u $ & $ b $ & $b $ \\
		 &   $(\Omega_{in}r_{in})$ &  $(m/s)$ & $ (B_0)$ & $ (mT)$ \\
		\hline
		$(4, 30)$ & $0.0479$ & $0.6013$ & $0.0087$ & $0.4956$ \\
		$(6, 30)$ & $0.0750$ & $0.9414$ & $0.0093$ & $0.7946$ \\
		$(9, 30)$ & $0.0853$ & $1.0707$ & $0.0076$ & $0.9741$ \\
		$(6.08, 25)$ & $0.0731$ & $0.7646$ & $0.0076$ & $0.6580$ \\
		$(5.71, 22.07)$ & $0.0545$ & $0.5033$ & $0.0055$ & $0.4474$ \\
       \hline
	\end{tabular}
		}
\caption{The rms of velocity and magnetic field perturbations estimated from the simulations that can be expected in the DRESDYN-MRI experiment for quasi-Keplerian rotation with $\mu=0.35$ at $Pm=7.77\times10^{-6}$.}\label{Table3:RMS_mu35}
\end{table}

We further use the above-established scaling relations and, extrapolating them down to $Pm=7.77 \times 10^{-6}$ of sodium, calculate the expected rms of velocity and magnetic field perturbations for the quasi-Keplerian flow profile with $\mu=0.35$ in the upcoming DRESDYN-MRI experiment (Table \ref{Table3:RMS_mu35}). For the adopted characteristic values of $Lu$ and $Rm$ in this table, these rms values vary, respectively, from the smallest $0.503~{\rm ms^{-1}}$ and $0.447~{\rm mT}$ corresponding to the lower values $(Lu, Rm)=(5.71, 22.07)$ to the maximum $1.071~{\rm ms^{-1}}$ and $0.974~{\rm mT}$ at higher $(Lu, Rm)=(9, 30)$. Since these sets of $(Lu, Rm)$ are well within the regimes of the DRESDYN-MRI experiment, the obtained rms velocity and magnetic field perturbations will enable the detection of SMRI for the astrophysically important Keplerian flow profile. 


\subsection{Dependence on $\mu$}
	
So far the analysis of the nonlinear saturation properties of SMRI has been carried out for a range of $Lu$ and $Rm$, but only for two values $\mu=0.27$ and quasi-Keplerian $\mu=0.35$. These results suggest that the scaling exponents $a$ and $b$ appear to be nearly independent of $\mu$. In this section, we look into the effect of varying $\mu$ on the nonlinear evolution of SMRI in more detail.

Figure \ref{fig:diff_mu}(a) compares the growth rates of SMRI obtained from the 1D linear stability analysis and from the simulations during the exponential amplification stage of the magnetic energy as a function of $\mu$ for fixed $Lu=6$ and $Rm=30$. The results from these two approaches match very well implying that our code works quite well for all $\mu \in (0.25, 0.35]$ covered here. 

Figures \ref{fig:diff_mu}(b) and \ref{fig:diff_mu}(c) show, respectively, the saturated magnetic energy and torque as a function of $\mu$ for the same  $Lu=6$, $Rm=30$ and different $Re$. Similar to the above cases with $\mu=0.27$ and $0.35$, they scale again as power laws  $Re^a$ with $a \approx -0.6$ for the magnetic energy and $Re^b$ with $b\approx 0.4$ for the torque, which hold for all $Re \geq 4000$ (with some deviation at $Re=1000$) and $\mu$. Taking into account this scaling, the $Re$-dependence of the energy collapses in a single line, exhibiting a linear decrease with $\mu$ [Fig. \ref{fig:diff_mu}(b)], which can be extrapolated to higher $\mu$. In a similar manner, after scaling the curves for the torque with $Re^{0.4}$, they collapse into a single curve that decreases linearly with $\mu$ except a little deviation from this trend at $\mu \in (0.30,0.33]$, where the decrease is less steep [Fig. \ref{fig:diff_mu}(c)]. This could be related to the changes in the properties (shear profile) of the boundary layers near the cylinder walls, which are Rayleigh-unstable [Fig. \ref{fig:diff_mu}(f)] and can develop small-scale fluctuations directly affecting the torque at the walls. 

These scalings are also supported by Figs. \ref{fig:diff_mu}(d) and \ref{fig:diff_mu}(e) showing, respectively, the variation of the magnetic energy and torque, as a function of $Re$ in the saturated state at $Lu=6,~Rm=30$ and different $\mu$ with the average scaling exponents $a \approx -0.585$ and $b \approx 0.4$. 

Finally, Fig. \ref{fig:diff_mu}(f) shows the radial profile of the shear parameter $q$ in the saturated state for different $\mu$ and $Lu=6, Rm=30$ and $Re=10000$. It is seen that the mean shear remains nearly the same in the bulk of the flow irrespective of $\mu$, while the shear near the boundary decreases with increasing $\mu$. This suggests that nonlinear SMRI present in the bulk of the flow modifies the mean shear to a similar extent irrespective of cylinder rotation rates, or equivalently of the global shear profile of the original TC flow.
	
\section{Conclusion} \label{sec_4_conclusion}

In this paper, we studied the nonlinear dynamics of standard MRI in a viscous and resistive Taylor-Couette flow with infinite height and an imposed constant background axial magnetic field using direct numerical simulations. This study, as our previous one in Paper I, is intended as preparatory, forming a theoretical basis for upcoming large-scale DRESDYN-MRI experiments with liquid sodium, which aim to detect and further study MRI in laboratory. In this analysis, we focused exclusively on axisymmetric ($m=0$) modes of SMRI, which are the dominant unstable ones in the flow. We investigated in detail the nonlinear saturation process of SMRI and its scaling properties in the final saturated state for a wide range of the system parameters: Lundquist ($Lu$), Reynolds ($Re$), magnetic Reynolds ($Rm$) numbers and the ratio of cylinders' angular velocities ($\mu$), including the Keplerian rotation regime relevant to astrophysical disks. In all these cases, the magnetic Prandtl number ($Pm$) has been kept small $10^{-4} \lesssim Pm \lesssim 0.03$ as is typical of liquid metals. In this low-$Pm$ regime, the linear behavior of SMRI is determined by $Lu$ and $Rm$, so the simulation points in the $(Lu,Rm)-$plane were chosen from the 1D linear stability analysis of Paper I where the instability has maximum (or near to maximum) growth rates. In this work, as distinct from previous ones mostly focusing on the weakly nonlinear analysis of SMRI, we explored the regime far from the marginal stability where the subsequent development of the instability is strongly nonlinear.  It was shown that the saturation process of SMRI from exponential growth phase to the nonlinear saturated state proceeds via magnetic reconnection and consists of four main -- growth, decay, restructuring and the final saturated -- stages. In the initial \textit{growth stage}, the instability, being in the linear regime, increases exponentially until it attains a certain peak. At this moment the (azimuthal) current becomes concentrated near the inner cylinder wall, forming a current sheet there. At this location, magnetic reconnection occurs which allows the system to release energy through the formation of the radially outflowing jet and as a consequence the magnetic energy decreases. During this \textit{decay stage}, magnetic energy is dissipated due to magnetic reconnection and partly converted in the kinetic energy of the jet. The decay stage is followed by the restructuring stage, when the radial jet motion at the inner cylinder rearranges the current and hence magnetic field distribution to achieve a steady magnetic configuration. As a result, SMRI reaches \textit{saturation} and the nonlinear state no longer changes in time. In this final state, the mean angular velocity profile of the flow deviates from the initial TC profile, i.e., angular velocity decreases near the inner cylinder and increases near the outer cylinder, resulting in the reduction of the overall shear (that feeds SMRI) compared to that of the TC flow. The mean axial magnetic field in the saturated state is also reduced, but only very little, about $0.5 \%$ of the imposed field. 
    
We explored the magnetic energy and angular momentum transport in the nonlinear saturated state of SMRI for a wide range of the main parameters ($Lu$, $Rm$, $Re$ and $\mu$) of the basic TC flow, allowing us to systematize the parameter dependence of these quantities. We found that for fixed $Rm$ and $\mu$, SMRI operates in a certain range of $Lu$, which is broader the higher $Rm$ is. The scaling of the saturated magnetic energy with $Re$ at different $Lu$ depends on $Rm$ (Fig. \ref{fig:Scal_Sat_Mag_Energy_Vs_Lu}). For smaller $Rm\leq 9$, the scaling is close to $Re^{-0.5}$, between $14\leq Rm \leq 20$ to $Re^{-0.55}$ and for large $Rm\geq 25$ it is $Re^{-0.6}$. Hence, the scaling of the magnetic energy of the saturated SMRI varies in the range $Re^{-0.6..-0.5}$. A similar trend is also seen for the scaling of the normalized perturbation torque with respect to $Re$ at different $Lu$, $Rm$ and $\mu$, which varies in the range $Re^{0.4..0.5}$ (Fig. \ref{fig:Torque_Scaling}). These results also carry over to the quasi-Keplerian rotation with $\mu=0.35$. We presented the analytical derivation of these scaling behavior based on the dynamical approach, i.e., looking at balances among three main -- inertial, viscous and Lorentz -- forces in the boundary layer near the cylinder walls and the reconnection region. These relations result from the interplay between the dynamics of the boundary layer characterized by viscosity ($Re$) and of the magnetic reconnection region, where inertial and Lorentz forces are in balance and most of magnetic dissipation takes place therein. Consequently, the magnetic field and hence magnetic energy of perturbations are related to the characteristics of the boundary layer -- viscous torque (Eq. \ref{scaling_equation_energy_torque}) and therefore Reynolds number (Eq. \ref{mag_energy_scaling}). Notably, as we checked, these scaling relations hold also in the 3D case including non-axisymmetric $|m|\geq1$ modes, which will be discussed elsewhere.

Another important goal was to quantify the magnitude of velocity and magnetic field perturbations in the nonlinear saturated state of SMRI that can be expected in upcoming DRESDYN-MRI experiment using liquid sodium. However, given extremely small $Pm=7.77\times 10^{-6}$ (at $T=130^{o}C$) of liquid sodium, very high Reynolds numbers $Re \gtrsim 10^6$ should be achieved in these experiments in order to activate SMRI, which are still beyond the capabilities of present numerical codes. In this regard, the above scaling relations are the most important and useful result of this study, because they allow us to estimate the characteristic quantities (diagnostics) of the nonlinear SMRI state at such higher Reynolds numbers. For this reason, we extrapolated the scalings obtained here at lower $Re\leq 10^5$ (higher $Pm \gtrsim 10^{-4}$) to higher $Re \gtrsim 10^6$ (smaller $Pm\lesssim 10^{-5}$) of liquid sodium for different pairs of $(Lu, Rm)$. As a result, we obtained the rms of velocity and magnetic field perturbations, which generally increase  with increasing $Lu$ and $Rm$ (see Tables I and II). For typical values $Lu \leq 9$ and $Rm \leq 30$ attainable in DRESDYN-MRI experiments, the rms of velocity perturbations vary from $0.074~{\rm mms^{-1}}$ to $1.314~{\rm ms^{-1}}$ and magnetic field rms from $0.22~{\rm  mT}$ to $13.07~{\rm mT}$ at $\mu=0.27$ (see Table I), while  for the quasi-Keplerian rotation $\mu=0.35$ they vary, respectively, from $0.5~ {\rm ms^{-1}}$ to $0.94~{\rm ms^{-1}}$ and from $0.45~{\rm mT}$ to $0.97~{\rm mT}$ (Table II). These magnitudes of the velocity and magnetic field perturbations are in fact sufficient for an unambiguous detection of SMRI in DRESDYN experiment. They also serve as reference values for designing and preparing these experiments with appropriate measuring techniques. Still, these numbers should be viewed with caution, because at such high $Re$, the nonlinear SMRI dynamics can become more complex: small-scale fluctuations (turbulence) can develop on top of the large-scale saturated state, especially near the inner cylinder wall, affecting the torque there and possibly modifying the scaling relations. Therefore, high-resolution and high Reynolds number $Re \gtrsim 10^6$ simulations are necessary to deeper understand the nonlinear dynamics of SMRI under conditions as close as possible to experimental ones and, in particular, to test the scaling laws obtained here at $Re \lesssim 10^5$.

\begin{acknowledgments}
We thank R. Hollerbach for providing the linear 1D code and A. Guseva for the nonlinear code used in this paper. FS acknowledges useful discussions with G. R\"udiger and AM critical discussions with Vivaswat Kumar. We also thank anonymous Referees whose useful comments improved the presentation of our results. This work received funding from the European Union's Horizon 2020 research and innovation programme under the ERC Advanced Grant Agreement No. 787544.
\end{acknowledgments}





\bibliography{ref}
\end{document}